\documentclass[a4paper,11pt]{article}
\usepackage[english]{babel}
\usepackage{jheppub}
\usepackage{amsfonts}
\usepackage{amsmath,bm}
\usepackage{amssymb}
\usepackage[titletoc,toc,title]{appendix}
\usepackage{braket}
\usepackage{cancel}
\usepackage[font={footnotesize,it}]{caption}
\usepackage{color}
\usepackage{comment}
\usepackage{enumitem}
\usepackage{epsfig}
\usepackage{epstopdf}
\usepackage{float}
\usepackage{graphicx}
\usepackage{hyperref}
\usepackage{cleveref}
\usepackage{physics}
\usepackage{mathtools}
\usepackage{ragged2e}
\usepackage{subcaption}
\usepackage{tikz}

\usepackage{colortbl}
\definecolor{LouisBlue}{RGB}{55, 114, 202}
\definecolor{LouisOrange}{RGB}{180, 54, 22}
\definecolor{LouisColor1}{RGB}{0, 118, 63}
\definecolor{LouisColor2}{RGB}{111, 73, 189}
\definecolor{lightblue}{RGB}{135, 206, 250}
\definecolor{yellowr}{RGB}{250, 250, 210}
\def\D{\mathrm{d}}

\newcommand{\figref}[1]{Fig.\ref{#1}}

\def\D{\mathrm{d}}
\def\is{\mathrm{Is}}

\hypersetup{
	citecolor=red,
	colorlinks=true,
	filecolor=red,
	linkcolor=blue,
	linktocpage=true,
	urlcolor=blue
}

\begin{document}
	
	\title{Entanglement islands and cutoff branes from path-integral optimization }
	
	\author[a,b]{Ashish Chandra,}
	\author[c]{Zhengjiang Li,}
	\author[a]{Qiang Wen}

		\affiliation[a]{ Shing-Tung Yau Center and School of Physics,\\
		Southeast University,\\
		Nanjing 210096, China
		\bigskip
	}
	\affiliation[b]{
		Department of Physics,\\
		Indian Institute of Technology,\\
		Kanpur 208 016, India
		\bigskip
	}
        \affiliation[c]{
		School of Physics and Astronomy,\\
		Sun Yat-sen University,\\
		Zhuhai 519082, China
		\bigskip
	}

	\emailAdd{achandra@iitk.ac.in}
	\emailAdd{lizhj66@mail2.sysu.edu.cn}
	\emailAdd{wenqiang@seu.edu.cn}

	\abstract{Recently it was proposed that, the AdS/BCFT correspondence can be simulated by a holographic Weyl transformed CFT$_2$, where the cut-off brane plays the role of the Karch-Randall (KR) brane \cite{Basu:2022crn}. In this paper, we focus on the Weyl transformation that optimizes the path integral computation of the reduced density matrix for a single interval in a holographic CFT$_2$. When we take the limit that one of the endpoint of the interval goes to infinity (a half line), such a holographic Weyl transformed CFT$_2$ matches the AdS/BCFT configuration for a BCFT with one boundary. Without taking the limit, the induced cutoff brane becomes a circle passing through the two endpoints of the interval. We assume that the cutoff brane also plays the same role as the KR brane in AdS/BCFT, hence the path-integral-optimized purification for the interval is in the island phase. This explains the appearance of negative mutual information observed in \cite{Camargo:2022mme}. We check that, the entanglement entropy and the balanced partial entanglement entropy (BPE) calculated via the island formulas, exactly match with the RT formula and the entanglement wedge cross-section (EWCS), which are allowed to anchor on the cutoff brane.
	
		\noindent
	}
	
	
	\maketitle
	
	\clearpage

\section{Introduction}
Recently, a new rule to compute the entanglement entropy in gravitational systems, namely the \textit{island formula} \cite{Penington:2019npb,Almheiri:2019psf,Almheiri:2019qdq,Penington:2019kki,Almheiri:2019hni}, has led to a remarkable new perspective to understand the long standing puzzle of the black hole information paradox \cite{Hawking:1976ra}. The development of the \textit{island formula} originates from the celebrated Ryu-Takayanagi (RT) formula \cite{Ryu:2006bv,Hubeny:2007xt}, which relates the entanglement entropy in holographic CFT to the minimal surfaces in the dual AdS bulk. Then the RT formula was refined to the quantum extremal surface (QES) formula \cite{Lewkowycz:2013nqa,Faulkner:2013ana,Engelhardt:2014gca} with the quantum correction taken into account. In \cite{Penington:2019npb,Almheiri:2019psf} the QES formula was used to compute the entanglement entropy for the Hawking radiation of an evaporating black hole, and the result is consistent with the Page curve. Remarkably, the QES formula suggests that, a region in black hole interior $I$, which we call the \textit{entanglement island}, should be considered as a part of the Hawking radiation $R$. And the entanglement entropy for $R$ is given by the island formula,
\begin{align}\label{islandf1}
    S_R=\text{min}~\text{ext}_I\left\{\frac{\text{Area}(\partial I)}{4G}+\tilde{S}_{bulk}(R\cup I)\right\}\,,
\end{align}
which contains an optimization for all possible $I$. 
Soon, it was argued that when we apply the replica trick in gravitational theory in a path-integral representation, we should take into account the new geometric configurations with wormholes in the replica manifold. When the replica wormhole configuration dominate the path-integral, the entanglement entropy should be computed by the island formula \cite{Penington:2019kki,Almheiri:2019hni}. Note that, replica wormhole arguments apply to generic gravitational theories and do not rely on the existence of holography. 

{When a gravitational system is in island phase, it is interesting to note that for given a region $R$ and its island region $Is(R)$, the state of the island region $\text{Is}(R)$ can be reconstructed from the information in the $R$ region. The existence of such a reconstruction is implied by the island formula \eqref{islandf1} following the same line of argument for entanglement wedge reconstruction \cite{Dong:2016eik,Harlow:2016vwg,Faulkner:2017vdd,Penington:2019kki}. And an explicit scheme of reconstruction in some simple models of quantum gravity was given in \cite{Penington:2019kki} based on the Petz map \cite{Petz:1986tvy,Petz:1988usv,Cotler:2017erl,Chen:2019gbt}. In other words the state of $\text{Is}(R) $ is encoded in the state of $R$, hence the degrees of freedom in $\text{Is}(R)$ are not independent and the Hilbert space of the whole system becomes non-factorizable. In \cite{Basu:2022crn} this property was further refined as the so-called \textit{self-encoding} property between spacelike-separated sub-regions, and a possible generalization of this property to systems beyond gravitational theories was explored in \cite{Basu:2022crn}.}

The AdS/BCFT correspondence \cite{Takayanagi:2011zk,Fujita:2011fp} is a commonly used context under which the entanglement islands emerges. It was found that the AdS/BCFT configuration can be perfectly simulated by models of Weyl transformed holographic CFT$_2$ \cite{Basu:2022crn,Basu:2023wmv,Lin:2023ajt}. The particular Weyl transformations under consideration are special as they depend on the cutoff scale of theory, which effectively introduces finite cutoff scale to the region where entanglement islands can emerge. The Weyl transformations will induce a cutoff brane in the bulk. It was shown in \cite{Basu:2022crn,Basu:2023wmv,Lin:2023ajt} that, the cutoff brane plays the same role as a KR brane where the RT surfaces are allowed to anchor. If we adjust the Weyl transformation such that the cutoff brane overlaps with the KR brane in the AdS bulk, we can exactly reproduce the main features of the AdS/BCFT configuration.

In this paper, we will consider a special Weyl transformation that optimizes the path integral computation of the reduced density matrix of an interval in a holographic CFT$_2$ \cite{Caputa:2017urj,Caputa:2018xuf}, and use it to simulate the AdS/BCFT configuration. This means the path-integral-optimized purification of the interval is in island phase. We give non-trivial consistency check for this simulation by computing the entanglement entropy and the BPE on the field theory side via the island formulas, and find that the results match with the RT surfaces and EWCSs that are allowed to anchor on the cutoff brane.
 
In section \ref{sec2}, we briefly introduce the setup of the AdS$_3$/BCFT$_2$ correspondence and its simulation via a holographic Weyl transformed CFT$_2$. In section \ref{sec3}, we introduce the computation of the reduced density matrix for an interval and the Weyl transformation that optimizes the path integral. Furthermore, we derive the cutoff branes for holographic CFT$_2$ under such Weyl transformations, which coincide with the KR branes in AdS/BCFT. In section \ref{sec4}, assuming that the path-integral-optimized purification for an interval is in island phase, we calculate the entanglement entropy for sub-intervals of the interval from both sides of the holography and find agreement. Also we show that the island phase perspective for this purification solves a puzzle of the existence of negative mutual information in this purification \cite{Camargo:2022mme}. We give a summary in section \ref{secdis}. In the Appendix, assuming the cutoff brane plays the role of the KR brane, we study the EWCS for bipartite sub-interval $AB$ and the corresponding balanced partial entanglement entropy (BPE) in island phases, and test their correspondence. This gives further evidence for our claim that, the path-integral-optimized purification is in island phase.

\section{AdS/BCFT and holographic Weyl transformed CFT}\label{sec2}
In the AdS/BCFT correspondence \cite{Takayanagi:2011zk} (or equivalently the Karch-Randall braneworld \cite{Karch:2000ct,Karch:2000gx}), the boundary theory is a $d$ dimensional CFT$_d$ with boundaries, and the gravity dual is an AdS$_{d+1}$ gravity in $d+1$ dimensions bounded by a co-dimension one KR brane,
\begin{equation}
\begin{aligned}
I_{\text{AdS}} & =\frac{1}{16 \pi G} \int_N \sqrt{-g}(R-2 \Lambda)  +\frac{1}{8 \pi G} \int_{\mathcal Q} \sqrt{-h} (K-T),
\end{aligned}
\end{equation}
where $N$ denotes the bulk AdS spacetime, $\mathcal Q$ denotes the Karch-Randall (KR) brane anchored at the boundary of the CFT and $T$ is the tension of $\mathcal Q$. In the bulk, the von Neumann boundary conditions are imposed on the KR brane. In this paper, we focus on the $d=2$ case and write the bulk metric in terms of the following coordinates,
\begin{align}
\D s^2=&\frac{\ell^2}{z^2}\left(-\D t^2+\D x^2+\D z^2\right)\cr
=&\D \rho^2+\ell^2 \cosh ^2 \frac{\rho}{\ell} \left(\frac{-\D t^2+\D y^2}{y^2}\right)\,,
\end{align}
where $\ell$ is the AdS radius which we will set to be unit in the rest of this paper, and the two sets of coordinates are related by
\begin{align}\label{ztorho}
	z=y\cosh^{-1}{\rho},\quad x=-y\tanh {\rho}\,.
\end{align}
For a BCFT with one boundary settled at $x=0$, the corresponding KR brane locates at $\rho=\rho_0$, where $\rho_0$ is a constant determined by the tension of the brane $\rho_0=\text{arctanh}~ T$.

\begin{figure}[t] 
	\begin{center}
		\begin{tikzpicture}
			\draw[color=red!80!black,line width=1.5pt] (-5,0)--(4,0);
			\draw[color=black,dashed,line width=1.3pt] (-1,0) arc(-90:270:1.2) node[pos=0.5,above,sloped]{$\textcolor{black}{}$};
			\draw[color=blue!90!yellow,line width=1.3pt] (1,0) arc(0:80:1) node[pos=0.5,above,sloped]{$\textcolor{black}{}$};
			\draw[color=blue!90!yellow,dashed,line width=1.3pt] (-1,0) arc(180:80:1) node[pos=0.5,above,sloped]{$\textcolor{black}{}$}; 
			\draw[color=blue!90!yellow,line width=1.3pt] (2,0) arc(0:100:1.5);
			\draw[color=blue!90!yellow,dashed,line width=1.3pt] (-1,0) arc(180:100:1.5);      
			\draw[color=blue!90!yellow,line width=1.3pt] (-3,0) arc(180:100:1);
			\draw[color=blue!90!yellow,dashed,line width=1.3pt] (-1,0) arc(0:100:1);  
			\draw[color=blue!90!yellow,line width=1.3pt] (-4,0) arc(180:80:1.5);
			\draw[color=blue!90!yellow,dashed,line width=1.3pt] (-1,0) arc(0:80:1.5);  
			\draw[fill](-1,0)circle(0.06)node[below,scale=1.3]{$ x_0 $};
		\end{tikzpicture}
	\end{center}   
	\caption{Cut-off sphere at $x=x_0$. For RT surfaces anchored at $x_0$, the part inside the cutoff sphere is excluded, which has the same length $\abs{\phi(x_0)}$}       
	\label{sketch_cutoff}	
\end{figure}
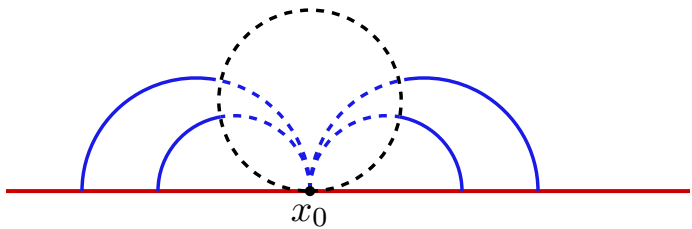

On the other hand, let us consider the vacuum state of a holographic CFT$_2$ in flat background
\begin{equation}
	ds^2=\frac{-dt^2+dx^2}{\epsilon^2},
\end{equation}
where $\epsilon$ represent the UV cutoff of the CFT. Then we perform a Weyl transformation characterized by a scalar field $\varphi(x)$, hence the metric changes to be
\begin{equation}
        ds^2=e^{2 \varphi(x)}\left(\frac{-dt^2+dx^2}{\epsilon^2}\right).
\end{equation}
It can be understood that, the Weyl transformation changes the cutoff scale in a position-dependent way,
\begin{align}
	\epsilon\ \to\ e^{-\varphi(x)}\epsilon\,.
\end{align} 
The scalar $\varphi(x)$ is chosen to be negative hence the Weyl transformation enlarges the cutoff scale.
Accordingly, the entanglement entropy for a single interval $A=[a,b]$ is modified to be \cite{Caputa:2017urj,Caputa:2018xuf,Camargo:2022mme}
\begin{align}\label{twopointf}
S_{A}=\frac{c}{3}\log \frac{(b-a)}{\epsilon}-\frac{c}{6}|\varphi(a)|-\frac{c}{6}|\varphi(b)|\,,
\end{align}
following the transformation rule of the two-point functions of twist operators under Weyl transformations. The above result is just the original entanglement entropy subjecting the absolute value of the scalar field at the endpoints.  Holographically, the constant subjection of the entanglement entropy was understood as inserting cutoff spheres with radius $|\varphi(x)|$ centered at the endpoints \cite{Basu:2022crn}. In other words, when computing the entanglement entropy via the RT formula, we should exclude the portion of the RT surface inside the cutoff spheres, see \figref{sketch_cutoff}. Interestingly, the cutoff sphere centered at $(x,z)=(x_0,\epsilon)$ in AdS$_3$ is just a circle in flat background with radius $\alpha$,  \cite{Wen:2018mev,Basu:2022crn},
	\begin{align}\label{cutoffsph}
		(x-x_0)^2+(z-\alpha)^2=\alpha^2,\qquad \alpha=\frac{\epsilon}{2}e^{\abs{\varphi(x_0)}}\,.
	\end{align}

\begin{figure}[t]  
	\begin{center}
		\includegraphics[width=0.5\textwidth]{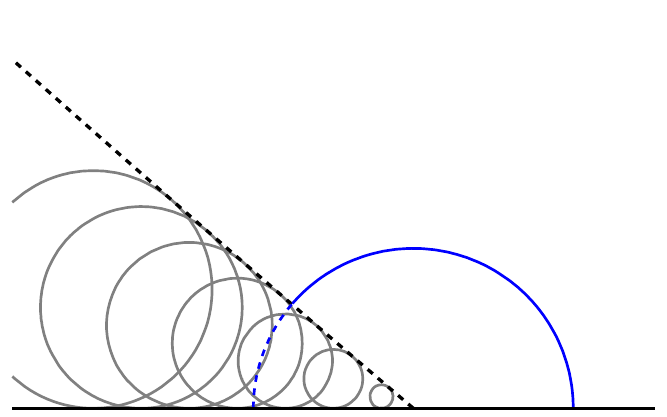}
	\end{center}
	\caption{This figure shows the cutoff spheres and cutoff brane on a time slice for the Weyl transformed CFT with the scalar field \eqref{varphi1}. The radius of the cutoff sphere centered at $x=x_0$ is given by $\alpha=-x_0e^{-\kappa}$.}
	\label{numerical0}
\end{figure}

The particular Weyl transformation, that captures the main features of the AdS/BCFT with the KR brane settled at $\rho=\rho_0$, is given by \cite{Basu:2022crn}\footnote{See \cite{Suzuki:2022xwv} for an earlier discussion.},  
\begin{equation}\label{varphi1}
\varphi(x)= \begin{cases}0, & \text { if } \quad x>0 \\ -\log \left(\frac{2|x|}{\epsilon}\right)+\kappa,  & \ \text {if } \quad x<0\end{cases}\,,
\end{equation}
where $\kappa$ is a constant. The \textit{cutoff brane} \cite{Basu:2022crn} is defined as the common tangent line of all the cutoff spheres, which represent the boundary of the bulk cutoff region (see \figref{numerical0}). In this case the cutoff brane is settled at
\begin{align}\label{cutoffbrane1}
	\rho=\kappa\,,
\end{align}
which can be adjust to exactly overlap with the KR brane by setting
\begin{align}\label{kapparho0}
\kappa=\rho_0\,.
\end{align}

{  To justify the application of the island formula in this Weyl transformed CFT, we need to give an additional assumption, which is that the $x<0$ region was coupled to a gravity. The gravity is an AdS$_2$ gravity as the Weyl transformation changes the metric at the $x<0$ region to AdS$_2$. Later, we will return to this assumption in the discussion section. Then entanglement islands are allowed in this region according to the replica wormhole arguments \cite{Almheiri:2019qdq,Penington:2019kki}. Furthermore, for simplicity we assume the AdS$_2$ gravity is a induced gravity as in \cite{Suzuki:2022xwv}, hence the area term in \eqref{islandf1} will not appear. 

On the field theory side, let us apply the island formula \eqref{islandf1} to calculate $S_A$ for the interval $A=[0,L]$ in this Weyl transformed CFT and assuming the island region $\text{Is}(A)=[-a,0)$, then we have
 \begin{align}
 	S_{A}=\text{min~ext}_a \left[\frac{c}{3}\log \frac{(L+a)}{\epsilon}+\frac{c}{6}\varphi(-a)+\frac{c}{6}\varphi(L)\right]\,.
 \end{align}
 Plugging into \eqref{varphi1}, the above formula is minimized at $a=L$, hence
 \begin{align}\label{ee-wcft0}
 	S_{A}=\frac{c}{6}\log \frac{2L}{\epsilon}+\frac{c}{6}\kappa\,.
 \end{align}
This exactly matches the result in AdS/BCFT with $\rho_0=\kappa$. 

On the gravity side, the holographic entanglement entropy is given by the area of the minimal extreme surface (the RT surface) which is homologous to $A$ and is allowed to anchor at any of the cutoff spheres. As expected \cite{Basu:2022crn}, the RT surface is just the circle emanating from the endpoint $x=L$ on the boundary and anchored on the cutoff brane vertically, see the blue circle in Fig.\ \ref{numerical0}. Provided $\kappa=\rho_0$, this is exactly the RT surface anchored on the KR brane in the AdS/BCFT correspondence. 
 
The above configuration exactly matches with the AdS/BCFT correspondence given $\rho_0=\kappa$. The key for this simulation is that, we should adjust $\kappa_1$ such that the cutoff brane overlaps with the KR brane in the AdS/BCFT configuration which we simulate. Also, the coincidence between the two set-ups can be straightforwardly checked for the calculation of the entanglement entropy for any interval $A=[a,b]$ in the $x>0$ region, with phase transitions for the RT surfaces. }

\section{Weyl transformed CFT from path-integral optimization and AdS/BCFT}\label{sec3}
In the previous section, we simulated the AdS/BCFT configuration via the holographic Weyl transformed CFT$_2$ with the scalar field \eqref{varphi1} and the constant $\kappa$ adjusted to satisfy $\kappa=\rho_0$. Note that, in this case the Weyl transformation is adjusted by hand such that the cutoff brane and the KR brane overlap. Since in the AdS/BCFT correspondence, the location of the KR brane is determined by details of the theory, including the tension and the boundary conditions of the brane. It is intriguing to ask what is special about the corresponding Weyl transformation \eqref{varphi1} from the perspective on the field theory side. One of the main observation in this paper is that, the particular Weyl transformation \eqref{varphi1} that makes the simulation a success is the one that optimizes the path integral computation for the reduced density matrix of the non-gravitational (or bath) region, which satisfies two key points: 1) it preserves the reduced density matrix $\rho_A$ for an interval $A$ at a particular time, and 2) it minimizes the path-integral complexity $C_L[\phi]$ defined later \cite{Caputa:2017urj, Caputa:2017yrh}.
    
Let us start with the path integral computation for a quantum state of the whole system. Consider a 2d CFT on Euclidean flat space $\D s^2=(\D \tau^2+\D \xi^2)/ \epsilon^2=\delta_{ab}/ \epsilon^2$, where $\epsilon$ represents the UV cutoff. Under this metric the ground state $\ket{\Psi}$ wave functional $\Psi[\tilde{\varphi}(\xi)]$ at $\tau=-\epsilon$ is given by the Euclidean path integral on the half plane,
\begin{align}\label{Psi1}
    \Psi_{\delta_{ab}/\epsilon^2}[\tilde{\varphi}(\xi)]=\int\left(\prod_\xi \prod_{-\infty<\tau<-\epsilon}D\varphi(\tau,\xi)\right)e^{-S_{CFT}(\varphi)}\cdot\prod_\xi\delta(\varphi(-\epsilon,\xi)-\tilde\varphi(\xi)).
\end{align}
Here the subscript of the wave function $\Psi$ represents the metric of the Euclidean space where the path integral is performed. One can perform a Weyl transformation, which is a symmetry of the theory, to the metric,
\begin{align}\label{bcbasic}
    &ds^2=e^{2\phi(\tau,\xi)}\frac{d\tau^2+d\xi^2}{\epsilon^2}, \qquad e^{2\phi(\tau=-\epsilon,\xi)}=1.
\end{align}
where the second equation is the boundary condition for the scalar field that characterizes the Weyl transformation. Then the state $\Psi$ computed under the Weyl transformed metric is proportional to \eqref{Psi1},
\begin{align}
   \Psi_{e^{2\phi}\delta_{ab}/\epsilon^2}[(\tilde{\varphi}(\xi))]=e^{C_L[\phi]-C_L[0]}\Psi_{\delta_{ab}/\epsilon^2}[(\tilde{\varphi}(\xi))]\,,
\end{align}
where $C_L[\phi]$ the Liouville action \cite{Polyakov:1981rd},
\begin{align}
    C_L[\phi]=\frac{c}{24\pi}\int_{-\infty}^{\infty}d\xi\int^{-\epsilon}_{-\infty} d\tau\left((\partial_\xi\phi)^2+(\partial_\tau\phi)^2+\mu e^{2\phi} \right).
\end{align}
This means the state $\Psi$ is preserved under the Weyl transformation \eqref{bcbasic}. In \cite{Caputa:2017urj, Caputa:2017yrh}, the Liouville action $C_L[\phi]$ is further related to the complexity functional of the quantum state $\ket{\Psi}$. The path integral optimization then means by computing the path integral under the Weyl transformation that minimizes the Liouville action $C_L[\phi]$. This can be achieved by solving the equation of motion $(\partial_\xi^2+\partial_\tau^2)\phi=e^{2\phi}/\epsilon^2$ of $C_L[\phi]$ (or the Liouville equation), which gives us the following simple solution,
\begin{align}\label{optphi1}
    e^{2\phi}=\frac{\epsilon^2}{\tau^2}.\qquad ds^2=\frac{d\tau^2+d\xi^2}{\tau^2}
\end{align}
It is easy to check that the above solution satisfies the boundary condition in \eqref{bcbasic}. Note that, one can shift the above scalar field by a constant by choosing a different UV cutoff scale $\epsilon$.

The above optimization procedure was generalized to optimizing the path integral computations that preserve the state (or the reduced density matrix) on a single interval $A=[a,b],\,\, t=0$ as described in \cite{Caputa:2018xuf}. This involves the Euclidean path integral over a complex plane $(\eta,\bar{\eta})=(x+it,x-it)$ with the interval $A$ cut open. Interestingly, one can relate this path integral to the one on a half plane by performing a conformal transformation which maps the interval $A$ to an infinitely long line,
\begin{align}
    w=\sqrt{\frac{\eta-a}{b-\eta}}\,,\qquad (\omega,\bar{\omega})=(\xi+i\tau,\xi-i\tau)\,.
\end{align}
Then we can optimize the Euclidean path integral on the half $\omega$ plane whose optimization gives \eqref{optphi1}. Finally, we may obtain the Weyl transformation by mapping \eqref{optphi1} back to the $(t,x)$ plane. Hence, we get the Weyl transformation that optimizes the path integral computation for the reduced density matrix $\rho_{A}$ of the interval $A$.
More explicitly, the relation $\frac{\epsilon^2}{\tau^2} d\omega d\bar{\omega}=e^{2\phi(x)}d\eta d\bar{\eta}$ gives us,
	\begin{equation}\label{varphi2}
		\phi(x)=
		\left\{ 
		\begin{array}{ c l }
			0 & \qquad  a<x<b \\
			\log\left[\frac{\epsilon(b-a)}{2(x-a)(x-b)}\right] +\kappa                & \qquad x>b \ or \ x<a
		\end{array}
		\right\}\,,
	\end{equation}
 where $\kappa$ is a constant which depends on the choice of $\epsilon$.

\begin{figure}[t]  
		\begin{center}
			\includegraphics[width=0.45\textwidth]{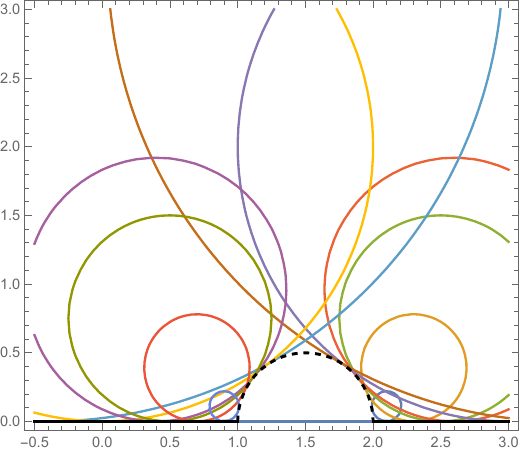}\ 
			\includegraphics[width=0.45\textwidth]{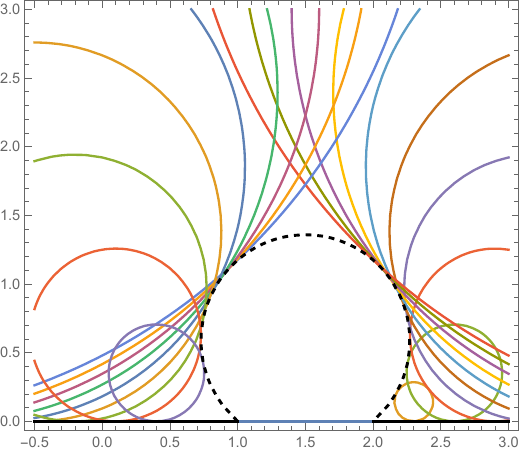}
		\end{center}
		\caption{This figure shows the numerical results of to series of cut-off spheres and the corresponding cutoff brane (dashed line). We have set  a=1, b=2 and $\kappa=0, 1$ in the left and right figures respectively.}
		\label{numerical1}
	\end{figure}

Now we are ready to derive the cutoff brane induced by the above Weyl transformation, which is again the common tangent line of all the cutoff spheres. Firstly, according to \eqref{cutoffsph} the cutoff spheres are circles on a time slice described by
	\begin{align}\label{branebcft2}
		(x-x_0)^2+(z-\alpha)^2=\alpha^2,\quad \alpha=\frac{(x_0-a)(x_0-b)}{b-a}e^{-\kappa}.
	\end{align}
	Then it is easy to check that the common tangent line of all the spheres is also a part of a circle which passes the two endpoints of the interval (see \figref{numerical1}),
	\begin{align}\label{cutoffbrane2}
	\text{cutoff brane:}\quad	(x-\frac{a+b}{2})^2+(z-z_0)^2=z_0^2+\frac{(b-a)^2}{4},
	\end{align}
	where $z_0=\frac{b-a}{4}(e^\kappa-e^{-\kappa})$. Changing the constant $\kappa$ will change the intersection angle between the cutoff brane and the asymptotic boundary. In \figref{numerical1}, we give the cutoff brane for the same interval with different $\kappa$. 

	When we send one of the endpoint to infinity by taking the limit $b\to \infty$, we are back to the BCFT with one boundary discussed in the previous section. Interestingly, under this limit the cutoff brane \eqref{cutoffbrane2} becomes,
 \begin{align}
     z=(a-x) \csch(\kappa)\,.
     \end{align}
     According to the coordinate transformation \eqref{ztorho}, the above equation is just $\rho=\kappa$ if we put the other endpoint at the origin by setting $a=0$. This exactly coincides with the cutoff brane \eqref{cutoffbrane1}, which is adjusted by hand to match the KR brane in the AdS/BCFT correspondence.    
     Note that, if we choose an arbitrary smooth function $\phi(x)$ for the Weyl transformation, the shape of the cutoff brane could be arbitrary. It is remarkable that the Weyl transformations that optimize the path integral exactly reproduce the KR brane configurations in the AdS/BCFT correspondence for the BCFT with one boundary. 
     
     Then it is intriguing to propose that, the cutoff brane configurations from path integral optimization can match to the KR branes in more generic AdS/BCFT setups. If this proposal is right, there should be KR brane configurations as circles \eqref{cutoffbrane2} in the Poincar\'e AdS$_3$ bulk that homologous to the strip BCFT$_2$ with two boundaries. Since the cutoff brane is connected in the bulk, it belongs to the
confined phase \cite{Takayanagi:2011zk,Fujita:2011fp,Nozaki:2012qd}, rather than the deconfined phase where the brane is disconnected in the bulk\footnote{The deconfined configurations were used to derive the co-dimension two Wedge holography \cite{Akal:2020wfl}.}. Unfortunately, such configurations in AdS/BCFT were not confirmed in the literature yet. There are AdS/BCFT configurations where the KR brane on a time slice are circles, but the BCFT is not a strip. For example, in \cite{Fujita:2011fp} the authors considered a BCFT on a round disk: $\tau^2+x^2\leq L^2$ in Euclidean spacetime, and find that the KR brane in AdS bulk dual is given by a sphere,
    \begin{align}
        \tau^2+x^2+(z-\sinh(\rho_0)L)^2-L^2\cosh^2(\rho_0)=0,
    \end{align}
   where $\rho_0=\text{arctanh}\, T$. For any fixed $\tau$, the KR brane is also a portion of a circle. Also in \cite{Geng:2022tfc}, the KR branes connecting the two asymptotic boundaries in the eternal black hole are circles on a time slice when mapping to the Poincar\'e patch (see Fig. 6 in \cite{Geng:2022tfc}). Similar configurations or KR branes were also proposed in \cite{Miao:2017gyt}.

\section{The path-integral-optimized purification as a state in island phase}\label{sec4}
{  In \cite{Takayanagi:2017knl,Nguyen:2017yqw}, it was claimed that the entanglement of purification (EoP) is the holographic dual of the EWCS in holographic theories. The optimization for the path integral that computes the reduced density matrix for an interval was first studied in \cite{Caputa:2018xuf} as a special purification for the interval, where they can compute the entanglement of purification (EoP), and confirm the claim of  \cite{Takayanagi:2017knl,Nguyen:2017yqw}. In \cite{Caputa:2018xuf}, this special purification is taken as a pure state in a normal quantum system without entanglement islands. However, it was pointed out in \cite{Camargo:2022mme} that, there exist negative mutual information in this purification. On the other hand, following the \cite{Basu:2022crn,Basu:2023wmv,Lin:2023ajt}, as well as the discussion in the previous section, we should understand the holographic CFTs under path-integral-optimized Weyl transformation as a simulation of the AdS/BCFT correspondence, hence we should take this purification as a pure state in island phase. In this section, we will compare between these two perspectives and show that, taking this path-integral-optimized purification as a state in island phase solves the puzzle of negative mutual information.}

Let us consider a bipartite region in the boundary field theory with a partition $AB\equiv A\cup B$, and $\rho_{AB}$ is the corresponding reduced density matrix. On a time slice, the entanglement wedge $\mathcal{W}_{AB}$ of $AB$ is the region enclosed by $AB$ and the RT surface $\mathcal{E}_{AB}$. Given the entanglement wedge we can define the EWCS $\Sigma_{AB}$ as the minimal area cross-section of $\mathcal{W}_{AB}$ separating the regions $A$ and $B$. Let $\ket{\Psi}\in \mathcal{H}_{AA'}\otimes \mathcal{H}_{BB'}$ be any purification of $\rho_{AB}$, the EoP between $A$ and $B$ is defined as \cite{Terhal:2002riz}
\begin{align}
	E_{p}(A:B)=\mathop{\text{min}}\limits_{\ket{\Psi},A'}S_{AA'}\,,
\end{align}
where we take the minimization over all possible purifications of $AB$ and over all possible partitions of $A'B'$. The EoP is then given by the minimal value of $S_{AA'}$. The authors in \cite{Caputa:2018xuf} start from the vacuum state of the holographic CFT$_2$ which duels to Poinca\'e AdS$_3$ and $AB$ is an interval in this state, then they perform different Weyl transformations for the complement of $AB$ (or $A'B'$) to get a class of pure states computed by path integral since the Weyl transformations change the boundary condition for the path integral. As $\rho_{AB}$ is preserved by these Weyl transformations, the resulting pure states can be considered as different purification for $\rho_{AB}$. The key observations of \cite{Caputa:2018xuf} are that, 1) the purification $\ket{\Psi}$ where $S_{AA'}$ is minimized is exactly the one given by the Weyl transformation that optimizes the path-integral computation of $\rho_{AB}$, i.e. the one characterized by \eqref{varphi2}. And 2), in the path-integral-optimized $\ket{\Psi}$, the minimized $S_{AA'}$ over all the possible partition of the purifying system $A'B'$ exactly matches the length of the EWCS, upon an additional choice of $\kappa=0$. 

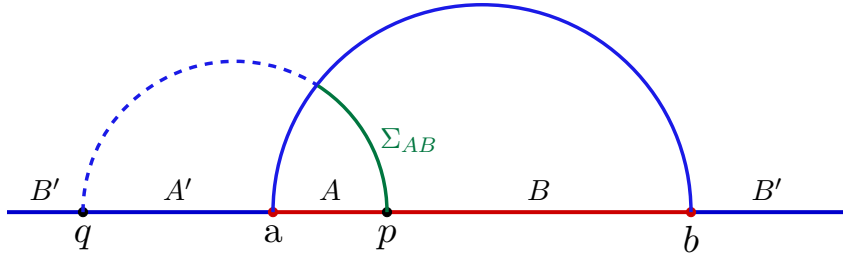
\begin{figure}[t] 
	\begin{center}
		\begin{tikzpicture}
			\draw[color=red!80!black,line width=1.5pt] (-1.5,0)--(0,0)node[midway,above]{$ \textcolor{black}{A} $};
			\draw[color=red!80!black,line width=1.5pt] (0,0)--(4,0)node[midway,above]{$ \textcolor{black}{B} $};			
			\draw[color=blue!80!black,line width=1.5pt] (-4,0)--(-1.5,0)node[midway,above]{$ \textcolor{black}{A'} $};
			\draw[color=blue!80!black,line width=1.5pt] (-5,0)--(-4,0)node[midway,above]{$ \textcolor{black}{B'} $};			
			\draw[color=blue!80!black,line width=1.5pt] (4,0)--(6,0)node[midway,above]{$ \textcolor{black}{B'} $};			
			\draw[fill](0,0)circle(.06)node[below,scale=1.3]{$ p $};	
			\draw[color=red!80!black,fill](4,0)circle(.06)node[below,scale=1.3]{\textcolor{black}{$ b $}};
			\draw[color=red!80!black,fill](-1.5,0)circle(.06)node[below,scale=1.3]{\textcolor{black}{$\text{a}$}};
			\draw[fill](-4,0)circle(.06)node[below,scale=1.3]{$ q $};
			\draw[color=blue!90!yellow,line width=1.3pt] (4,0) arc(0:180:2.75);  
			\draw[color=LouisColor1,line width=1.3pt] (0,0) arc(0:57:2.)node[right,midway,scale=1]{$ \Sigma_{AB} $};    
			\draw[color=blue!90!yellow,dashed,line width=1.3pt] (-4,0) arc(180:57:2);    
		\end{tikzpicture}
	\end{center}   
	\caption{The blue circles are the RT surfaces for $AB$ and $AA'$. The Weyl transformation \eqref{varphi2} with $\kappa=0$ is performed on $A'B'$. When calculating $S_{AA'}$, we should subject a constant term $\frac{c}{6}\abs{\phi(q)}$, which is exactly the length of the dashed portion. Then $S_{AA'}$ exactly matches the length of the EWCS $\Sigma_{AB}$.}       
	\label{noisland}	
\end{figure}

More explicitly, let us consider the configuration shown in \figref{noisland}, where the purification of the mixed state $\rho_{AB}$ is path-integral optimized, hence the Weyl transformation applied on $A'B'$ is characterized by \eqref{varphi2} with $\kappa=0$. Then we minimize the entanglement entropy\footnote{Note that, in \cite{Caputa:2018xuf} the entanglement entropy is calculated following \eqref{twopointf} with $\kappa=0$.}
\begin{align}
S_{AA'}=\frac{c}{3}\log \frac{p-q}{\epsilon}-\frac{c}{6}\log \frac{2(a-q)(b-q)}{\epsilon(b-a)}
\end{align}
by adjusting the partition point $x=q$ and found that  $S_{AA'}$ is minimized when
\begin{align}\label{Oppar}
	q=\frac{2ab-(a+b)p}{a+b-2p}\,,
\end{align}
hence,
\begin{align}
	E_{p}(A:B)=\mathop{\text{min}}\limits_{\ket{\Psi},A'}S_{AA'}=\frac{c}{6}\log(\frac{2(p-a)(b-p)}{\epsilon(b-a)})\,.
\end{align}

{  The authors of \cite{Caputa:2018xuf} studied the path-integral-optimized state $\ket{\Psi}$ in no-island phase\footnote{By the time when \cite{Caputa:2018xuf} was published, there was no concept of entanglement islands.}. 
This indicates that the Hilbert space of the system factorizes as all the space-like separated degrees of freedom are independent, and the entanglement entropy for any interval $[a,b]$ can be defined in a standard way and calculated following the formula \eqref{twopointf}. Nevertheless, it was pointed out in \cite{Camargo:2022mme} that, under such a setup there exists negative mutual information in the path-integral-optimized state $\ket{\Psi}$. Again, let us consider the path-integral-optimized purification of Fig.\ref{noisland} with the partition point settled by \eqref{Oppar}, and assume that the purification is in the no-island phase. Then we can use the formula \eqref{twopointf} to calculate the entanglement entropy for the following intervals,
\begin{align}
S_{A'}=\frac{c}{3}\log \frac{a-q}{\epsilon}+\frac{c}{6}\phi(a)+\frac{c}{6}\phi(q)\,,
\cr
S_{B'}=\frac{c}{3}\log \frac{b-q}{\epsilon}+\frac{c}{6}\phi(b)+\frac{c}{6}\phi(q)\,,
\cr
S_{A'B'}=\frac{c}{3}\log \frac{b-a}{\epsilon}+\frac{c}{6}\phi(a)+\frac{c}{6}\phi(b)\,,
\end{align}
where, according to \eqref{varphi2}, we have $\phi(a)=\phi(b)=0$. Also we can perform a straightforward calculation for the mutual information $I(A':B')$,
\begin{align}\label{mutualinformation}
	I(A':B')&=S_{A'}+S_{B'}-S_{A'B'}\nonumber\\
	&=\frac{c}{3}\log \frac{(a-q)(b-q)}{(b-a)\epsilon}+\frac{c}{3}\phi(q)\nonumber\\
	&=-\frac{c}{3}\log2\,,
\end{align}
where in the last line we used \eqref{Oppar} and \eqref{varphi2}. As we can see, the mutual information $I(A':B')$ is negative, which indicates that the path-integral-optimized purification $\ket{\Psi}$ is not a physical state.

Following our previous discussion on the Weyl transformed CFT$_2$, we turn to the perspective of taking the path-integral-optimized purification as a state in island phase. From this perspective, soon we will show that $A'B'$ is the island region of $AB$, hence the state of $A'B'$ is encoded in $AB$, which means the degrees of freedom in $A'B'$ are not independent. This implies that the strong sub-additivity, which guarantees the positivity for mutual information, does not apply. So the new perspective straightforwardly solves the above puzzle of negative mutual information. {Of course there are other possibilities\footnote{{One may consider other possibilities to avoid the negative mutual information by modifying the way we compute the entanglement entropy instead of using \eqref{twopointf} and \eqref{varphi2}. For example, one may consider the physical degrees of freedom all locate on the boundary interval and the bulk brane, and consider the state/surface correspondence \cite{Miyaji:2015yva}. In this context, $A'$ and $B'$ are sub-regions of the brane and the entanglement entropies are calculated by bulk geodesics connecting their endpoints, which exactly coincide with $A'$ and $B'$ in the case of Fig.\ref{noisland}. So we have $S_{A'B'}=S_{A'}+S_{B'}$, and the mutual information we considered is zero. Nevertheless, the surface/state correspondence is not a context where we study the configurations with entanglement islands. 	Another possible solution is to keep using \eqref{Oppar} and \eqref{varphi2} but consider a constant shift for the scalar field, such that the mutual information \eqref{mutualinformation} satisfies $I(A',B')\geq 0$. This equals to confining the constant term $\kappa$ in \eqref{varphi2} to satisfy $\kappa\geq \log 2$. Also the cutoff brane changes, see section \ref{subsectionA3} for more discussion on the case of $\kappa\neq 0$. One shortcoming of this solution is that, the entanglement entropy $S_{AA'}$ on longer reproduces the EWCS for the entanglement wedge of $\mathcal{W}_{AB}$ due to the shift of a constant, see the discussion section of \cite{Camargo:2022mme} for details. Also it is not obvious that this confinement can guarantee the positivity of an arbitrary mutual information.}} to avoid the negative mutual information, which we will not discuss further in this paper.} }

Let us give a more detailed comparison between the two perspectives on the path-integral-optimized purification. For simplicity, we set $\kappa=0$ hence the Weyl transformation matches with the one used in \cite{Caputa:2018xuf}, and the cutoff brane \eqref{cutoffbrane2} coincide with the RT surface of $AB$ (see the left figure in Fig.\ref{numerical1}). In the island perspective, the RT surfaces for subregions in $AB$ are allowed to anchor on the cutoff brane, which results in a totally new understanding on the entanglement structure of this purification and the corresponding geometric dual on the gravity side. 

{  Firstly, let us consider the configuration depicted by the upper figure in Fig.\ref{illistration}, On the gravity side, consider a sub-interval $\alpha$ in $AB$, the RT surface of $\alpha$ is allowed to anchor on the cutoff brane (see the blue line), hence has two phases which are represented by the dashed and solid green lines. When $\alpha$ approaches $AB$, the solid green line has smaller length hence is the RT surface, and when we take the limit $\alpha\to AB$, the RT surface vanishes. This indicates that $S_{AB}=0$, hence $AB$ is in a pure state. On the field theory side, one can apply the island formula and find that, when $\alpha\to AB$ we have $\text{Is}(\alpha)\to A'B'$, which is consistent with the RT formula.
	
	\begin{figure}[t] 
		\begin{center}
			\begin{tikzpicture}
				\draw[color=red!80!black,line width=1.5pt] (-1.5,0)--(1,0); \draw(0,0)node[below]{$ \textcolor{black}{A} $};
				\draw[color=red!80!black,line width=1.5pt] (1,0)--(4,0);
				\draw (2.2,0)node[below]{$ \textcolor{black}{B} $};			
				\draw[color=blue!80!black,line width=1.5pt] (-3,0)--(-1.5,0)node[midway,above]{$ \textcolor{black}{A'} $};			
				\draw[color=blue!80!black,line width=1.5pt] (4,0)--(5,0)node[midway,above]{$ \textcolor{black}{B'} $};			
				\node[rotate = 180] at (1.25, 0.1) {$\underbrace{\hspace{3.5cm}}$} ;
				\draw (1.25,0.09) node[above,scale=1.3]{$ \alpha $};	
				\draw[color=red!80!black,fill](4,0)circle(.06)node[below,scale=1.3]{\textcolor{black}{$ b $}};
				\draw[color=red!80!black,fill](-1.5,0)circle(.06)node[below,scale=1.3]{\textcolor{black}{$\text{a}$}};
				\draw[color=blue!90!yellow,line width=1.3pt] (4,0) arc(0:180:2.75);  
				\draw[color=LouisColor1,line width=1.3pt] (-0.5,0) arc(0:75:1.1)node[left,midway,scale=1.3]{$ $};    
				\draw[color=LouisColor1,line width=1.3pt] (3,0) arc(180:105:1.1);
				\draw[color=LouisColor1,dashed,line width=1.3pt] (3,0) arc(0:180:1.75);    
				\draw[fill](1,0)circle(.06)node[below,scale=1.2]{$ p $};	
			\end{tikzpicture}
			\begin{tikzpicture}
				\draw[color=red!80!black,line width=1.5pt] (-1.5,0)--(1,0)node[midway,above]{$ \textcolor{black}{A} $};
				\draw[color=red!80!black,line width=1.5pt] (1,0)--(4,0)node[midway,above]{$ \textcolor{black}{B} $};			
				\draw[color=blue!80!black,line width=1.5pt] (-3,0)--(-1.5,0)node[midway,above]{$ \textcolor{black}{A'} $};
				\draw[color=blue!80!black,line width=1.5pt] (4,0)--(5,0)node[midway,above]{$ \textcolor{black}{B'} $};			
				\draw[fill](1,0)circle(.06)node[below,scale=1.3]{$ p $};	
				\draw[color=red!80!black,fill](4,0)circle(.06)node[below,scale=1.3]{\textcolor{black}{$ b $}};
				\draw[color=red!80!black,fill](-1.5,0)circle(.06)node[below,scale=1.3]{\textcolor{black}{$\text{a}$}};
				\draw[color=blue!90!yellow,line width=1.3pt] (4,0) arc(0:180:2.75);  
				\draw[color=LouisColor1,line width=1.3pt] (1,0) arc(0:48:3.25)node[left,midway,scale=1.1]{$ \mathcal{E}_{A} $};    
			\end{tikzpicture}
		\end{center}   
		\caption{Upper figure: RT surfaces for a sub-interval $\alpha$ in $AB$. Lower figure: RT surface for $A$.}       
		\label{illistration}	
	\end{figure}
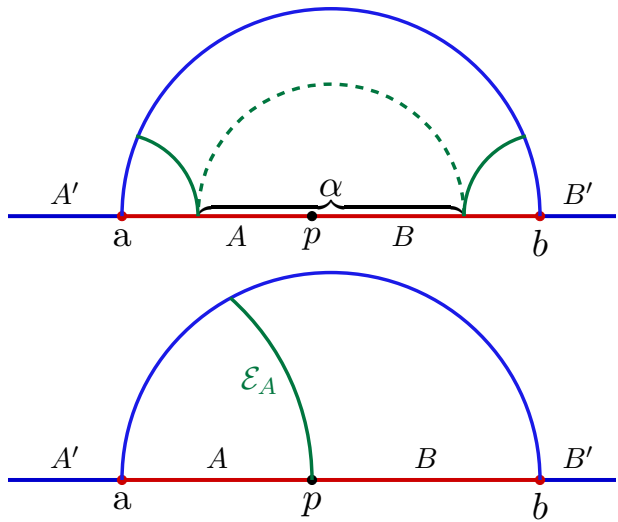

Secondly, from the island perspective the EWCS in $\mathcal{W}_{AB}$ has a new interpretation. On the field theory side we should apply the island formula \eqref{islandf1} to calculate $S_{A}$,
\begin{align}
	S_{A}=&\mathop{\text{min ext}}\limits_{A'} \tilde{S}_{AA'}
	\cr
	=& \mathop{\text{min}}\limits_{q}\frac{c}{3}\log \frac{p-q}{\epsilon}+\frac{c}{6}\phi(q)
	\cr
	=&  \frac{c}{6}\log\frac{2(p-a)(b-p)}{\epsilon(b-a)}\,,
\end{align}
where $A'=[q,a)$ and we have plugged in \eqref{Oppar} in the third equation, which is the solution of the optimization. This solution of Is$(A)$ coincides with $\eqref{Oppar}$, which the minimizes $S_{AA'}$ when calculating the EoP in \cite{Caputa:2018xuf}. It was also pointed out in \cite{Wen:2021qgx,Camargo:2022mme} that, the point $x=q$ satisfying \eqref{Oppar} is just the intersection point between the boundary and the geodesic (see the blue dashed line in Fig.\ref{noisland}) extended from the EWCS $\Sigma_{AB}$. On the gravity side, the RT surface $\mathcal{E}_{A}$ of $A$ is just the geodesic emanating from the partition point $x=p$ and anchored on the cutoff brane vertically (see the green line in the lower figure of Fig.\ref{illistration}), which coincide with the EWCS $\Sigma_{AB}$ in Fig.\ref{noisland}. This is consistent with the expectation that, when $AB$ is in a pure state, the EWCS and its holographic dual quantites (like the EoP, the reflected entropy \cite{Dutta:2019gen} and BPE), should coincide with $S_{A}$.

Thirdly, when we take $AB$ is a sub-region of the interval, $\rho_{AB}$ is in a mixed state. We can also study the EWCS in $\mathcal{W}_{AB}$ and its quantum information dual. From the island perspective, the cutoff brane plays the same role as the KR brane, then both of the RT surface $\mathcal{E}_{AB}$ and the corresponding EWCS $\Sigma_{AB}$ undergo phase transitions since they are allowed to anchor on the cutoff brane. See Fig.\ref{figA2b} for example. On the field theory side, it was proposed in \cite{Wen:2021qgx} that the so-called balance partial entanglement entropy (BPE), which is a partial entanglement entropy (PEE) satisfying certain balance conditions, corresponds to the EWCS on the gravity side. More interestingly, this correspondence was generalized to the holographic configurations with entanglement islands \cite{Basu:2023wmv}. We can calculate the BPE in island phases following the steps in \cite{Basu:2023wmv}, and check whether the BPE coincide with the EWCS. This is a highly non-trivial test for our proposal that the path-integral-optimized state $\ket{\Psi}$ should be understood as in the island phase. Since the analysis of the BPE is complicated and relatively independent from the main topic of this paper, we put it in the appendix. Readers who are interested in the BPE and EWCS for the path-integral-optimized purification can go through the background papers \cite{Wen:2021qgx,Basu:2023wmv,Lin:2023ajt} and the appendix in detail.}


\section{Summary and Discussion}\label{secdis}
In this paper, we considered the special Weyl transformation for a holographic CFT$_2$, which optimizes the path integral computation for the reduced density matrix of an interval. Under such Weyl transformations, the cutoff branes are circles in the AdS bulk passing through the endpoints of the interval at the boundary. When we take the limit that one of the endpoint goes to infinity hence the interval becomes a half line, the cutoff branes coincide with the KR branes in the AdS/BCFT configurations where the BCFT has one boundary. Without taking the above limit, the cutoff brane configurations coincide with a time slice of some AdS/BCFT configurations, where the KR brane are also circles passing through the endpoints on the boundary \cite{Fujita:2011fp,Miao:2017gyt,Geng:2022tfc}. Finding more AdS/BCFT configurations that exactly match with the holographic Weyl transformed CFT we have considered could be an interesting exploration in the future. 

Perhaps it is an even more interesting idea to take the scalar field that characterizes the Weyl transformation dynamical, with its action just being the Liouville action, whose equivalence to the action of 3d gravity has been extensively discussed \cite{Verlinde:1989ua,Carlip:1991zm,Coussaert:1995zp,Martinec:1998wm,Rooman:2000zi,Carlip:2005tz,Skenderis:1999nb,Krasnov:2000zq,Carlip:2005zn}. Then the region under nontrivial Weyl transformation (or the $\phi\neq 0$ region) is naturally coupled to a gravity with background geometry that solves the Liouville equations and the gravitational excitation represented by the perturbation of the scalar field. This may give a justification for our proposal that the island formula applies to the Weyl transformed CFT.

In our previous works, we have assumed that the cutoff branes induced by the Weyl transformations that optimize the path integral play the same role as the KR branes, hence the RT surfaces and the EWCSs can anchor on the brane. Such an assumption also implies that, the corresponding boundary state of the Weyl transformed CFT is in island phase. This implication also extends to the path-integral optimized purification for an interval. The new perspective for the path-integral-optimized purification solves the puzzle of negative mutual information in this state. We calculated the entanglement entropies for subregions of the interval using the island formula, and find them coincide with the area of the RT surfaces that are allowed to anchor on the brane. Furthermore, we calculated the BPE between two arbitrary non-overlapping subregions of the interval in island phases, and find them coincide with the area of the EWCS. This gives non-trivial test for our proposal that the path-integral-optimized purification is a state in island phase.

Our results give a potential new link between the AdS/BCFT correspondence and the path integral optimization. This is quite important for the simulation of the AdS/BCFT configurations via the holographic Weyl transformed CFT, since the Weyl transformation is now determined on the field theory side, rather than adjusted by hand. Our calculations also provide new evidence for the correspondence between the BPE and the EWCS.

\section*{Acknowledgement}
The authors thank Hao Geng, Rongxin Miao and Shan-Ming Ruan for helpful discussions.

\appendix

\section{The BPE/EWCS correspondence in the path-integral-optimized purification}\label{sec5}
In \cite{Wen:2021qgx, Camargo:2022mme}, it was proposed that the so-called balance partial entanglement entropy (BPE) is the quantum information quantity that duals to the EWCS. Furthermore, the concept of BPE was generalized to the island phases in \cite{Basu:2023wmv} and its correspondence to the EWCS was also explicitly checked in the context of AdS/BCFT or the holographic Weyl transformed CFT$_2$ in \cite{Basu:2023wmv,Lin:2023ajt}. See also \cite{Chandrasekaran:2020qtn,Li:2020ceg,Afrasiar:2022fid,Afrasiar:2023jrj,Basu:2023jtf,Basak:2023bnc} for more discussion on the EWCS and its quantum information dual in holographic configurations with entanglement islands. In the appendix, we take the path-integral-optimized purification as a state in island phase, and study the entanglement wedges $\mathcal{W}_{AB}$ for a bipartite subregion $AB$ of the path-integral-optimized interval, with the cutoff brane playing the role of the KR brane in the context of AdS/BCFT. On the gravity side, we classify the phases of the EWCS of $\mathcal{W}_{AB}$ and calculate the area of the EWCS in each phase. On the field theory side, we explicitly calculate the BPE between $A$ and $B$. As expected, we find the agreement between the EWCS and the BPE in all the configurations. 

\subsection{Brief review for the PEE and BPE}
For self-consistency, let us first introduce the basic concept of the partial entanglement entropy (PEE) \cite{Wen:2018whg,Wen:2020ech,Wen:2019iyq}. Some of the texts in this review overlap with section 3 in \cite{Lin:2023ajt}. The PEE is a measure of two-body correlation between two non-overlapping regions $\mathcal{I}(A,B)$. Note that, we should not mix between the mutual information $I(A,B)$ and the PEE $\mathcal{I}(A,B)$. So far, the fundamental definition for PEE based on the reduced density matrix is still not established. In some scenarios of interest, it can be determined by a set of physical requirements \cite{Casini:2008wt,Wen:2019iyq}, which include all the properties satisfied by the mutual information and the additional key property of additivity:
\begin{enumerate}
	\item
	\textit{Additivity:} $\mathcal{I}(A,B\cup C)=\mathcal{I}(A,B)+\mathcal{I}(A,C)$;
	
	\item
	\textit{Permutation symmetry:} $\mathcal{I}(A,B)=\mathcal{I}(B,A)$;
	
	\item
	\textit{Normalization:} $\mathcal{I}(A,\bar{A})=S_{A}$;
	\item
	\textit{Positivity:} $\mathcal{I}(A,B)>0$;
	\item
	\textit{Upper boundedness:}$\mathcal{I}(A,B)\leq \text{min}\{S_{A},S_{B}\}$;
	\item
	\textit{$\mathcal{I}(A,B)$ should be Invariant under local unitary transformations inside $A$ or $B$};
	\item
	\textit{Symmetry:} For any symmetry transformation $\mathcal T$ under which $\mathcal T A = A'$ and $\mathcal T B = B'$, we have $\mathcal{I}(A,B) = \mathcal{I}(A',B')$.
\end{enumerate}
In the above list, $A$, $B$ and $C$ denote non-overlapping regions. For vacuum states of CFTs on a plane, we can determine the formula for PEE up to a coefficient by imposing the above requirements except the normalization property. Then we can determine the coefficient by imposing the normalization requirements for spherical regions where the relation between the geometric cutoff and the UV cutoff can be explored \cite{Han:2019scu}. 

In this paper, we will use a particular construction for the PEE in generic two-dimensional theories on a line or a circle\footnote{See \cite{Wen:2018whg,Wen:2019iyq,Chen_2014,Han:2019scu,SinghaRoy:2019urc} for other prescriptions to construct PEE, and \cite{Wen:2018mev,Kudler-Flam:2019oru,Ageev:2021ipd,Rolph:2021nan} for applications of the ALC proposal. }. This proposal is referred to as the \emph{additive linear combination} (ALC) proposal \cite{Wen:2018whg,Kudler-Flam:2019oru,Wen:2019iyq}, which claims that the PEE in these scenarios can be written as a linear combination of subset entanglement entropies that satisfy the property of additivity. 
\begin{itemize}
	\item \textit{The ALC proposal \cite{Wen:2018whg,Wen:2020ech,Wen:2019iyq}}: 
	
	Consider a boundary interval $A$ and partition it into three non-overlapping subregions $A=\alpha_L\cup\alpha\cup\alpha_R$, where $\alpha$ is some subregion inside $A$ and $\alpha_{L}$ ($\alpha_{R}$) denotes the regions left (right) to it. On this configuration, the claim of the \textit{ALC proposal} is the following:
	\begin{align}\label{alc}
		s_{A}(\alpha)=\mathcal{I}(\alpha,\bar{A})=\frac{1}{2}\left(S_{ \alpha_L\cup\alpha}+S_{\alpha\cup \alpha_R}-S_{ \alpha_L}-S_{\alpha_R}\right)\,. 
	\end{align}
\end{itemize}
It is easy to check that, the above construction satisfies all the seven requirements in general \cite{Wen:2020ech,Kudler-Flam:2019oru}.

The concept of PEE \cite{Wen:2018whg,Kudler-Flam:2019oru,Wen:2020ech,Wen:2019iyq} originates from the study of the \emph{entanglement contour} \cite{Chen_2014}, which is defined as a function $s_{A}(\textbf{x})$ that characterizes the contribution to the entanglement entropy of $A$ from each site $\textbf{x}\in A$. By definition the entanglement contour function should satisfy,
\begin{equation}
	S_{A}=\int_{A}s_A(\textbf{x})d \sigma_\textbf{x}\,,
\end{equation}
where $\sigma_{\textbf{x}}$ is a infinitesimal area element located at the site $\textbf{x}$. Subsequently, we can also define the contribution from a subset $\alpha$ in $A$ to $S_A$, 
\begin{equation}
	s_{A}(\alpha)=\int_{\alpha}s_A(\textbf{x})d\sigma_\textbf{x}\,.
\end{equation}
The contribution $s_{A}(\alpha)$ is a measure of the correlation between the subregion $\alpha$ and the complement $\bar{A}$ of $A$, which is exactly the two-body correlation we have defined as the PEE,
\begin{align}\label{tworepresentations}
	\mathcal{I}(\alpha,\bar{A})\equiv s_{A}(\alpha)\,.
\end{align}
Following \cite{Basu:2023wmv}, we call the notation on the left hand side of the above equation as the two-body correlation representation for the PEE, while the notation on the right hand side is the contribution representation.

The \emph{balanced partial entanglement entropy} (BPE) is a special PEE that satisfies a set of balance requirements \cite{Wen:2021qgx}. In canonical purification, the definition of the BPE coincide with the reflected entropy \cite{Dutta:2019gen}, and the purification independence of the BPE was explored in \cite{Camargo:2022mme}. The BPE was proposed to be dual to the EWCS in holographic theories, and this proposal has undergone rigorous validation across diverse scenarios including both static \cite{Wen:2021qgx} and covariant \cite{Wen:2022jxr} setups of AdS$_3$/CFT$_2$ with or without gravitational anomalies, and in the context of AdS/BCFT with entanglement islands \cite{Basu:2023wmv,Lin:2023ajt}. More interestingly, in the context of 3-dimensional flat holography \cite{Bagchi:2010zz,Bagchi:2012cy,Jiang:2017ecm}, on the field theory side the BPE was calculated in \cite{Camargo:2022mme,Basu:2022nyl}, which matches the EWCS explored in \cite{Basu:2021awn}. 

For a system $A\cup B$ in a mixed state $\rho_{AB}$, one can introduce an auxiliary system $A'B'$ to purify $AB$ such that the whole system $ABA'B'$ is in a pure state $|\varphi\rangle$ and,
\begin{align}
	\text{Tr}_{A'B'}\ket{\varphi}\bra{\varphi}=\rho_{AB}.    
\end{align}
The pure state on $ABA'B'$ is called a purification of $\rho_{AB}$, which could be highly non-unique. Let us consider the simple examples described in Fig.\ref{fig:bpe}, where the balance requirements are summarized in the following \cite{Wen:2021qgx}:
\begin{figure}
	\centering
	\includegraphics[width=0.6\textwidth]{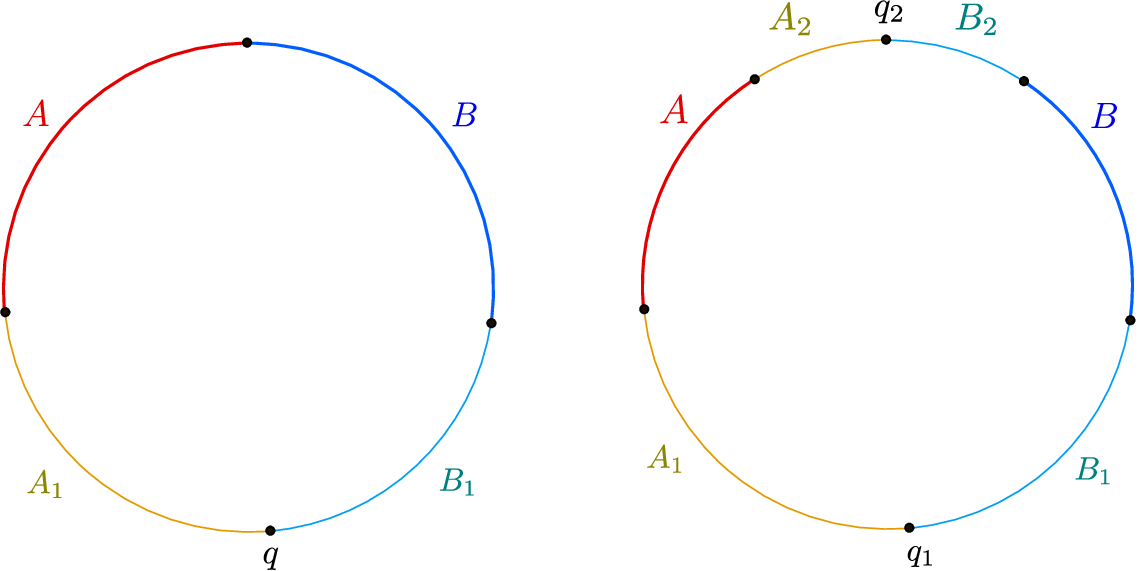}
	\caption{Here we consider the vacuum state of the holographic CFT$_2$ on a circle which duals to the global AdS$_3$, and consider two different partitions, where $A$ and $B$ are adjacent and disjoint respectively. $x=q,q_1,q_2$ are balanced points. This figure is extracted from \cite{Basu:2023wmv}.}
	\label{fig:bpe}
\end{figure}
\subsubsection*{Balance requirements:}
\begin{itemize}
	\item 
	When $A$ and $B$ are adjacent, the contribution from $s_{AA_1}(A)$ and $s_{BB_1}(B)$ should satisfy, 
	\begin{equation}\label{bc1}
		s_{AA_1}(A)=s_{BB_1}(B),\ \text{or}\quad 
		\mathcal{I}(A,BB_1)=\mathcal{I}(B,AA_1).
	\end{equation}
	This balance requirement is sufficient to determine the partition point in the purifying system $A_1B_1$, which we call the balance point. 
	
	\item When $A$ and $B$ are non-adjacent, we have $A'=A_1\cup A_2$ and $B'=B_1\cup B_2$, and the balance requirements become 
	\begin{equation}\label{bc2}
		s_{AA_1A_2}(A_1)=s_{BB_1B_2}(B_1),\ s_{AA_1A_2}(A)=s_{BB_1B_2}(B),
	\end{equation}
	or 
	\begin{equation}  
		\mathcal{I}(A,B_1B_2)=\mathcal{I}(B,A_1A_2),\ \mathcal{I}(A_1,BB_2)=\mathcal I(B_1,AA_2),
	\end{equation}
	which are sufficient to determine the two partition points of the purifying system $A_1B_1A_2B_2$.  Since $S_{AA_1A_2}=S_{BB_1B_2}$, $s_{AA_1A_2}(A_2)=s_{BB_1B_2}(B_2)$ is automatically satisfied provided the satisfaction of the above requirements.  
	\item
	Note that, it is possible that the solution to the balance requirements is non-unique. In such cases, we should choose the solution that minimizes the BPE, which will be defined soon.
\end{itemize}

Provided the balanced requirements are satisfied, the BPE is then defined as 
	\begin{align}\label{BPEd}
		\begin{cases}
			\textit{adjacent cases}:\, &\text{BPE}(A,B)=s_{AA_1}(A)|_{\rm balanced}=s_{BB_1}(B)|_{\rm balanced}\\ \\
			\textit{non-adjacent cases}:\, &\text{BPE}(A,B)=s_{AA_1A_2}(A)|_{\rm balanced}=s_{BB_1B_2}(B)|_{\rm balanced}\,,
		\end{cases} 
	\end{align}
	Since we have $\mathcal{I}(A,B')=\mathcal{I}(A',B)$ at the balance point\footnote{Here $A'=A_1$ ($B'=B_1$) for adjacent configurations, and $A'=A_1A_2$ ($B'=B_1B_2$) for non-adjacent configurations.}, the BPE can also be expressed as
	\begin{equation}
		\text{BPE}(A,B)
		=\mathcal{I}(A,B)+\frac{(\mathcal{I}(A,B')+\mathcal{I}(A',B))|_{\rm balanced}}{2}.
\end{equation}
It is important to note that \cite{Camargo:2022mme}, the summation $\mathcal{I}(A,B')+\mathcal{I}(A',B)$ which we call the crossing PEE is minimized at the balanced point. This indicates that the BPE can be defined via an optimization problem. It is also interesting to note that, when $A$ and $B$ are adjacent, this minimal crossing PEE gives half of the lower bound of Markov gap \cite{Zou:2020bly,Wen:2021qgx,Hayden:2021gno}, which is a universal constant $(c/6)\log 2$.

\subsection{PEE and BPE in island phases}
As described in the article \cite{Basu:2022crn}, a system in the island phase can be understood as a self-encoding system. In other words, the state of certain subsets of the system are totally encoded in the state of their counterpart subsets of the system. Subsequently, the calculation of entanglement entropy should also be modified to the island formulas \eqref{islandf1}. Given the self-encoding property, when we compute the PEE between subregions in island phase, we should also take the contribution from the corresponding island regions into account. Now we generalize our construction of the PEE and BPE to configurations with entanglement islands \cite{Basu:2023wmv}. Here we just list the basic elements we need to carry out the computations. One should consult \cite{Basu:2023wmv} for more details.

\begin{figure}[t]
	\begin{center}
		\begin{tikzpicture}
			\filldraw[lightblue!50] (0,0)--(5,0) arc(0:126.7:5);
			\filldraw[orange!50] (0,0)--(2.5,0) arc(0:126.6:2.5);
			\draw[color=red!80!black,line width=1pt] (0,0)--(2,0)node[midway,below,scale=1]{$ \textcolor{black}{A} $};
			\draw[color=LouisBlue,line width=1pt] (2,0)--(2.5,0)node[midway,below,scale=1]{ \textcolor{black}{$A_2$} };
			\draw[color=yellow!80!black,line width=1pt] (2.5,0)--(3,0)node[midway,below,scale=1]{ \textcolor{black}{$B_2$} };
			\draw[color=blue!80!black,line width=1pt] (3,0)--(5,0)node[midway,below,scale=1]{ \textcolor{black}{$B$} };
			\draw[color=red!80!black,line width=1pt] (0,0)--(-1.2,1.6)node[midway,left,scale=1]{ \textcolor{black}{Is($A$)} };
			\draw[color=black,line width=1pt] (-1.2,1.6)--(-1.8,2.4);
			\node[rotate = -53] at (-1.58, 1.93) {$\underbrace{\hspace{1cm}}$} (-1.58, 1.93) node[below,rotate = -53,sloped,scale=0.7]{ \textcolor{black}{Io($AB$)} };
			\draw[color=blue,line width=1pt] (-1.8,2.4)--(-3,4)node[midway,left,scale=1]{ \textcolor{black}{Is($B$)} };
			\draw[color=LouisColor1,dashed,line width=1pt] (3,0) arc(0:180:0.5) node[midway,above,sloped]{$\text{}$};
			\draw[color=LouisColor2,dashed,line width=1pt] (2.5,0) arc(0:12:2.5);
			\draw[color=LouisColor2,line width=1pt] (2.45,0.5) arc(12:126:2.5) node[midway,above,sloped]{$\text{}$};
			\draw[fill] (-1.8,2.4)circle(0.05);
			\draw[fill] (-1.2,1.6)circle(0.05);
			\draw[fill] (-1.5,2)circle(0.05);
			\draw[fill] (2,0)circle(0.05);
			\draw[fill] (2.5,0)circle(0.05);
			\draw[fill] (3,0)circle(0.05);
			\draw[fill] (2.45,0.5)circle(0.05);
		\end{tikzpicture}
	\end{center}   
	\caption{Ownerless island in a time slice of holographic BCFT with one boundary. Here $A_2\cup B_2$ admits no island, and Is($AB$) covers the whole brane. Nevertheless, the union of Is($A$) and Is($B$) does not cover the whole brane, which means there is an ownerless island Io($AB$) inside Is($AB$) but outside $\text{Is}(A)\cup \text{Is}(B)$.  }        
	\label{one_endpoint}	
\end{figure}
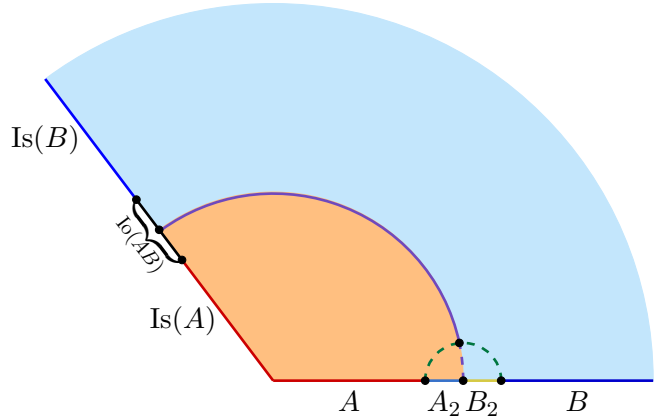

Let us start with the region $AB$ and the island regions Is$(AB)$, Is$(A)$ and Is$(B)$. According to \cite{Basu:2023wmv}, the degrees of freedom in $\is (AB)$ is not independent, and contributes to the entanglement entropy $S_{AB}$ calculated by the island formula. The PEE, for example $s_{AB}(A)$, should contain the contribution from the island region, and it is clear that we should assign the contribution from $\is (A)$ to  $s_{AB}(A)$ and similarly assign $\is (B)$ to  $s_{AB}(B)$. Nevertheless, there are scenarios with regions included in $\is (AB)$ but outside $\is(A)\cup\is(B)$, which should also be assigned to $s_{AB}(A)$ or $s_{AB}(B)$. In other words, when $\is (AB)\neq \emptyset$ and $\is (AB)\supset \is (A)\cup \is (B)$, we define the ownerless island region to be
\begin{align}
	\text{Io}(AB) =\is(AB)/(\is (A)\cup \is (B))\,.
\end{align}
The ownerless island should be further divided into two parts
\begin{align}
	\text{Io}(AB)=\text{Io}(A)\cup \text{Io}(B)\,,
\end{align}
which are assigned to $s_{AB}(A)$ and $s_{AB}(B)$ respectively. We will specify the division of Io$(AB)$ later. All in all, we define the  \emph{generalized islands}
\begin{align}\label{genisland}
	\mathrm{Ir}(A)=\is(A)\cup \text{Io}(A),\qquad \mathrm{Ir}(B)=\is(B)\cup \text{Io}(B)\,,
\end{align}
and assign them to $A$ and $B$ respectively when calculating the PEEs. The generalized islands coincide with the so-called reflected entropy islands defined in \cite{Chandrasekaran:2020qtn}. The assignment can be classified into three classes:
\begin{itemize}
	\item 
	\textbf{Class 1:}
	When Is$(AB)$=Is$(A)$=Is$(B)$=$\emptyset$, we have
	\begin{align}
		\mathrm{Ir}(A)=\mathrm{Ir}(B)=\emptyset\,.
	\end{align}  
	\item
	\textbf{Class 2:}
	When Is$(AB)$=$\is(A)\cup \is (B)\neq\emptyset$, we have 
	\begin{align}
		\text{Io}(A)=\text{Io}(B)=\text{Io}(AB)=\emptyset, \qquad \mathrm{Ir}(A)=\is(A),\qquad \mathrm{Ir}(B)=\is(B)\,.
	\end{align}
	\item 
	\textbf{Class 3:}
	When we have non-trivial ownerless island region, then we have \eqref{genisland}.
\end{itemize}

Now we turn to the computation of the PEEs for a bipartite system $AB$ with entanglement island, and denote $C \equiv\overline{AB\cup\text{Is}(AB)}$ for convenience. After taking into account the contributions from the islands, we should have \cite{Basu:2023wmv},
\begin{equation} \label{pee0}
	s_{AB}(A)=\mathcal{I}(A\cup\text{Ir}(A),C),\quad 
	s_{AB}(B)=\mathcal{I}(B\cup\text{Ir}(B),C),
\end{equation}
where the island contribution has been taken account. The right contribution comes from the generalized island rather than the entanglement island. A key step to compute the PEEs \eqref{pee0} is to write the right hand side of \eqref{pee0} as a linear combination of the PEEs that can be written in this form $\mathcal{I}(\gamma,\bar\gamma)$, i.e. a PEE between a region $\gamma$ and its complement. For example, using the property of additivity, the PEE \eqref{pee0} is just given by
\begin{equation}\label{galc1}
	\begin{split}
		s_{AB}(A)
		=& \mathcal{I}(A\cup\text{Ir}(A),C)\\
		=& \frac{1}{2}\left[
		\mathcal I(A\text{Ir}(A) B \text{Ir}(B), C) + \mathcal I(A\text{Ir}(A), B \text{Ir}(B) C) - \mathcal I(B \text{Ir}(B),A\text{Ir}(A) C)\right]\\
		=& \frac{1}{2}\left[ \tilde S_{A\text{Ir}(A)B\text{Ir}(B)}+\tilde S_{A\text{Ir}(A)}- \tilde S_{B\text{Ir}(B)}\right],
	\end{split}
\end{equation}
where we have used $\text{Ir}(A)\text{Ir}(B)=\text{Is}(AB)$, and the notation $\tilde{S}_{\gamma}\equiv\mathcal{I}(\gamma,\bar{\gamma})$ which will be explained soon.

For the configurations shown Fig.\ref{fig:bpe}, for example the adjacent case in the left figure, we should calculate $s_{AA_1}(A)$ following \eqref{galc1}. For the non-adjacent case in Fig.\ref{fig:bpe} where $A$ is sandwiched by two regions $A_1$ and $A_2$, we have
\begin{equation}\label{galc2}
	s_{A_1AA_2}(A)=\frac{1}{2}\left[ \tilde S_{A_1\text{Ir}(A_1)A\text{Ir}(A)}-\tilde S_{A_1\text{Ir}(A_1)}+\tilde S_{A\text{Ir}(A)A_2\text{Ir}(A_2)}- \tilde S_{A_2\text{Ir}(A_2)}\right].
\end{equation}
We call this formula the \emph{generalized ALC formula} \cite{Basu:2023wmv} for island phases, which is just the ALC formula \eqref{alc} with the replacement $S_{\gamma}\Rightarrow \tilde S_{\gamma\text{Ir}(\gamma)}$ applied to each term. Accordingly, the balance requirements should also be modified to generalized versions, which are given by
\begin{align}\label{gbc1}
	\textit{adjacent cases}:
	\quad \mathcal{I}(A\,\text{Ir}(A),B_1\,\text{Ir}(B_1))=\mathcal{I}(A_1\,\text{Ir}(A_1),B\,\text{Ir}(B)).		
\end{align}
and
\begin{align}\label{gbc2}
	\textit{non-adjacent cases}:
	\begin{cases}
		\mathcal{I}(A_1\text{Ir}(A_1),B\,\text{Ir}(B)\,B_2\text{Ir}(B_2))=\mathcal{I}(B_1\text{Ir}(B_1),A\,\text{Ir}(A)\,A_2\text{Ir}(A_2)), \\ \\
		\mathcal{I}(A\,\text{Ir}(A),B_1\text{Ir}(B_1)\,B_2\text{Ir}(B_2))=\mathcal{I}(B\,\text{Ir}(B),A_1\text{Ir}(A_1)\,A_2\text{Ir}(A_2)).
	\end{cases} 
\end{align} 
In summary, in island phases the BPE are still defined by \eqref{BPEd}, but we need to use the generalized versions of the ALC formula and balanced requirements.

Before we compute the BPE, we need to clarify how to compute $\mathcal{I}(\gamma,\bar{\gamma})$. In this paper we need to deal with two types of $\gamma$, 1) $\gamma=[-a,b]$ a connected interval, 2) $\gamma=A\cup \text{Ir}(A)=[-d,-c]\cup[a,b]$ is consists of two disconnected interval, where $a,b,c,d>0$. For the above two cases, there are two corresponding proposals \cite{Basu:2023wmv} to compute the PEE,
  \begin{align}\label{basicp}
		&\textit{Basic proposal 1}:\quad
		\mathcal{I}(\gamma,\bar\gamma)=\tilde{S}_{\gamma}=\tilde S_{[-a,b]},
	\\\label{basicpp}
		&\textit{Basic proposal 2}:\quad \mathcal{I}(\gamma,\bar\gamma)=\tilde{S}_{[-d,-c]\cup[a,b]}=\tilde{S}_{[-c,a]}+\tilde{S}_{[-d,b]},
	\end{align} 
where, for example, $\tilde S_{[-a,b]}$ is the two-point function of twist operators inserted at $x=-a$ and $x=b$\footnote{Such two-point functions are very subtle in the effective theory of AdS/BCFT correspondence or other doubly holography configurations, since the region $[-a,0]$ may not be the island region of $[0,b]$. In other words, the island formula \eqref{islandf1} only involves $\tilde S_{\gamma}$ where $\gamma=\mathcal{R}\cup \text{Is}(\mathcal{R})$ is the union of a region and its island, and in this case $\tilde S_{\gamma}=S_{\mathcal{R}}$. While the $\gamma$ we deal with usually goes beyond this type, and $\tilde S_{\gamma}$ should not be understood as the entanglement entropy of $\gamma$ or any other region (see \cite{Basu:2023wmv} for more discussions).}.  These two-point functions are well defined in the Weyl transformed CFT, and can be computed by \eqref{twopointf}. 

Though the above two basic proposals have not been proved yet, they have produced highly consistent results between the BPE and the EWCS in various configurations with entanglement islands \cite{Basu:2023wmv}. The \textit{basic proposal 1} looks like a generalization of the $RT$ formula in AdS$_3$/CFT$_2$ for single interval. While the \textit{basic proposal 2} is a generalization of the $RT$ formula for two-intervals with connected entanglement wedge. It looks reasonable as this proposal applies when the region $[a,b]$ admits an island, hence the entanglement wedge looks more like the connected phase of a two-interval. Note that the \textit{basic proposal 2} is not consistent with the normalization and additive property of the PEE, which may be explained by a similar phase transition of the PEE flux in AdS$_3$/CFT$_2$\footnote{This phase transition should originate from the large $c$ limit of the holographic CFT$_2$ \cite{Hartman:2013mia}.} \cite{Lin:2023rbd,Lin:2024dho}. We leave this for future investigation \cite{WenXuZhong}. In summary, our calculations in this paper only involve linear combinations of Weyl transformed two-point functions \eqref{twopointf} of the twist operators.

The remaining problem is the division of the ownerless island region $\text{Io}(AB)=\text{Io}(A)\cup\text{Io}(B)$, which is indeed determined by the balance requirements. In \cite{Basu:2023wmv}, the authors considered the AdS/BCFT set-up and its simulation via a holographic Weyl transformed CFT$_2$. They found that different assignments for the ownerless island region lead to different BPEs, which exactly correspond to different saddles of the EWCS. Then according to the minimum requirement, we should choose the assignment that gives us the minimal BPE. It seems that, the division of the ownerless island depends on the phase structure of the EWCS.

\subsection{A case study for the path-integral-optimized purification}\label{subsectionA3}
Here we give a typical example for the study of the EWCS and BPE in the path-integral-optimized purification of an interval, and assume that the cutoff brane plays the role of the KR brane and the purification is in island phase. We consider the path integral optimization for a single interval [a,b], hence the scalar field that characterizes the Weyl transformation is given by \eqref{varphi2} with $\kappa=0$, and the cutoff brane \eqref{cutoffbrane2} is just a semi-circle. As a case study, we consider the setup shown in Fig.\ref{figA2b}, where we considered two adjacent sub-intervals,
\begin{align}
	A:~~[a_1,p]\qquad B:~~[p,b_1]
\end{align}
whose size and position are adjusted such that, the RT surface $\mathcal{E}_{AB}$ consists of two disconnected pieces of geodesics anchored on the cutoff brane, and the EWCS $\Sigma_{AB}$ also anchors on the brane. 

\begin{figure}[H]
	\begin{center}
		\begin{tikzpicture}
			\filldraw[LouisBlue!30](-2,0) arc (180:0:3.5);
			\begin{scope} 
				\clip (-2,0) arc (180:0:3.5);
				\filldraw[orange,opacity=0.6] (-9,0)--(1.2,0) arc(0:43:4.6);
			\end{scope}
			\begin{scope}
				\clip (-2,0) arc (180:0:3.5);
				\filldraw[yellowr,opacity=1](-0.4,0) arc(0:180:2);
			\end{scope}
			\begin{scope}
				\clip (-2,0) arc (180:0:3.5);
				\filldraw[yellowr,opacity=1](4,0) arc(180:0:1);
			\end{scope}
			\draw[color=red!80!black,line width=1.5pt] (-0.4,0)--(1.2,0)node[midway,above,scale=1]{ \textcolor{black}{$A$} };
			\draw[color=yellow!20!black,very thick] (-2,0)--(-1.2,0)node[midway,above,scale=1]{ \textcolor{black}{$B_1$} };
			\draw[color=blue,line width=1.5pt] (1.2,0)--(4,0)node[midway,above,scale=1]{ \textcolor{black}{$B$} };	
			\draw[color=yellow!20!black,line width=1.5pt] (4,0)--(5,0)node[midway,above,scale=1]{ \textcolor{black}{$B_1$} };		
			\draw[color=blue!80!black,line width=1.5pt] (-1.2,0)--(-0.4,0)node[midway,above,scale=1]{ \textcolor{black}{$A_1$} };
			\draw[very thick] (-8,0)--(-4.4,0)node[midway,above,scale=0.8]{ \textcolor{black}{Ir(A)}};
			\draw[very thick] (-9,0)--(-8,0)node[midway,above,scale=0.8]{ \textcolor{black}{Ir(B)}};
			\draw[line width=1.5pt,LouisBlue] (-4.4,0)--(-2,0)node[midway,below]{$ \textcolor{black}{} $};		    
			\draw[line width=1.5pt] (6.0,0)--(7.6,0)node[midway,above,scale=0.8]{ \textcolor{black}{Ir(B)} };	
			\draw[very thick,LouisBlue] (5,0)--(6,0)node[midway,above]{$ \textcolor{black}{} $};	
			\draw[color=LouisColor1,line width=1pt] (4,0) arc(180:0:1) node[midway,above,sloped]{$\text{}$};
			\draw[color=LouisColor1,line width=1pt] (-1.2,0) arc(0:180:0.9) node[midway,above,sloped]{$\text{}$};
			\draw[color=LouisColor2,line width=1pt] (-0.4,0) arc(0:180:2) node[midway,above,sloped]{$\text{}$};
			\draw[color=purple,line width=1.2pt] (1.2,0) arc(0:43:4.6) node[midway,above,sloped]{$\Sigma_{AB}$};
			\draw[color=purple,dashed,line width=1.5pt] (-8,0) arc(180:43:4.6) node[midway,above,sloped]{$\text{}$};
			\draw[color=LouisBlue,line width=1.5pt] (5,0) arc(0:180:3.5) node[midway,above,sloped,scale=0.8,color=black]{\text{}};
			\draw[fill](-0.4,0)circle(0.05) (-0.4,0) node[below,scale=1]{$a_1$};
			\draw[fill](1.2,0)circle(0.05) (1.2,0) node[below,scale=1]{$p$};
			\draw[fill](4,0)circle(0.05) (4,0) node[below,scale=1]{$b_1$};
			\draw[fill](5,0)circle(0.05) (5,0) node[below,scale=1]{$b$};
			\draw[fill](-1.2,0)circle(0.05) (-1.2,0) node[below,scale=1]{$p_1$};
			\draw[fill](-2,0)circle(0.05) (-2,0) node[below,scale=1]{$a$};
			\draw[fill](-4.4,0)circle(0.05) (-4.4,0) node[below,scale=1]{$q_2$};
			\draw[fill](6,0)circle(0.05) (6,0) node[below,scale=1]{$q_4$};
			\draw[fill] (-8,0)circle(0.05) (-8,0) node(q)[below,scale=1]{$q_1$};	
			\draw[fill](-3,0)circle(0.05) (-3,0) node[below,scale=1]{$q_3$};			
			\draw (5.45,0.3) node[yellow!45!black,scale=0.8,color=black]{Ir($B_1$)};
			\draw (-2.4,0.3) node[yellow!45!black,scale=0.8,color=black]{Ir($B_1$)};
			\draw (-3.6,0.3) node[yellow!45!black,scale=0.8,color=black]{Ir($A_1$)};
			\draw (-2.2,1.3) node[yellow!45!black,scale=0.8,color=black]{Ir($A_1$)};
			\draw (-1.2,2.7) node(IrA)[yellow!45!black,scale=0.8,color=black]{Ir($A$)};
			\draw (4,3) node[yellow!45!black,scale=0.8,color=black]{Ir($B$)};
			\draw[fill] (4.87,0.99)circle(0.05);
			\draw[fill] (-0.03,3.15)circle(0.05);
			\draw[fill] (-1.5,1.78)circle(0.05) node(AA1)[left]{};
			\draw[fill] (-1.89,0.87)circle(0.05);
		\end{tikzpicture}
	\end{center}   
	\caption{Two adjacent intervals $AB$ and their complements $A_1$ and $B_1$, featuring entanglement island regions $\text{Ir}(A)=[q_1,q_2],\quad \text{Ir}(B)=(-\infty,q_1]\cup [q_4,\infty),\quad \text{Ir}(A_1)=[q_2,q_3]\,\,\, \text{and}\,\,\,\, \text{Ir}(B_1)=[q_3,a]\cup [b,q_4]$ on the entire KR brane, are configured schematically. Here $\text{Ir}(B_1)$ indicate disconnected entanglement island.}       
	\label{figA2b}	
\end{figure}

The lengths of the geodesics anchored on the cutoff brane are classified in \cref{geodesics length}. The cross-section of the entanglement wedge $\mathcal{W}_{AB}$ has three saddle points, which anchors on either of the two disconnected pieces of the RT surface, or the cutoff brane, and the EWCS is the saddle point with minimal area. In the case shown in \figref{figA2b}, the EWCS anchors on the cutoff brane.

Firstly, we check the self-consistency of the simulation by calculating the entanglement entropy on both sides of holography. In the gravity side, $S_{AB}$ is calculated by the RT formula, which is the area of the two pieces of geodesic anchored on the brane described as
\begin{align}
	S_{AB}=\frac{c}{6}\log\left[\frac{2(a_1-a)(b-a_2)}{(b-a)\epsilon}\right]+\frac{c}{6}\log\left[\frac{2(b_1-a)(b-b_2)}{(b-a)\epsilon}\right]
\end{align}
On the field theory side, we calculate $S_{AB}$ by island formula and denoting the island region as $\is (AB)=(-\infty,q_2]\cup [q_4,\infty)$. Then, following \eqref{basicpp} we have
\begin{align}
	S_{AB}=&\min_{\is (AB)}\tilde{S}_{\is (AB) AB}=\min_{\is (AB)} \left(\tilde{S}_{[q_2, a_1]}+\tilde{S}_{[b_1,q_4]}\right) 
	\cr
	=&\frac{c}{6}\log\left[\frac{2(a_1-a)(b-a_2)}{(b-a)\epsilon}\right]+\frac{c}{6}\log\left[\frac{2(b_1-a)(b-b_2)}{(b-a)\epsilon}\right]
\end{align}
which coincide with the RT formula. In the second equation we have plugged in the saddle points for $q_1$ and $q_4$ expressed as
\begin{align}
	q_2=\frac{2ab-(a+b)a_1}{a+b-2a_1} ,\qquad q_4= \frac{2ab-(a+b)b_1}{a+b-2b_1}\,.
\end{align}
As was shown in \figref{figA2b}, these saddles are just the intersecting points between the boundary and the extended RT surface.

Then we check the correspondence between the EWCS and the BPE. On the gravity side, the area of the EWCS as shown in \figref{figA2b} for this case can be easily calculated as follows
\begin{align}\label{EWA2b}
	\frac{\text{Area}[\Sigma_{AB}]}{4 G}=\frac{c}{6}\log\left[\frac{2(p-a)(b-p)}{(b-a)\epsilon}\right]\,.
\end{align}
Prior to calculating the BPE for this phase, we should give an assignment for the island regions, $\text{Is}(AB)=\text{Ir}(A)\cup \text{Ir}(B)$, where
\begin{align}
	\text{Ir}(A)=[q_1,q_2]\,,\qquad \text{Ir}(B)=(-\infty,q_1]\cup [q_4,\infty)
\end{align}
where $x=q_1$ is the partition point of $\is (AB)$, which will be determined by the balance requirements \eqref{gbc1}. Now we calculate the PEE on both sides of \eqref{gbc1}, which are given by
\begin{align}
	\mathcal{I}(A\text{Ir}(A),B\text{Ir}(B)B_1\text{Ir}(B_1))&=\frac12\Big(\tilde{S}_{A\text{Ir}(A)A_1\text{Ir}(A_1)}+\tilde{S}_{A\text{Ir}(A)}-\tilde{S}_{A_1\text{Ir}(A_1)}\Big)\nonumber\\
	&=\frac12(\tilde{S}_{[q_1,q_3]\cup [p_1,p]}+\tilde{S}_{[q_1,q_2]\cup [a_1,p]}-\tilde{S}_{[q_2,q_3]\cup [p_1,a_1]})\nonumber\\
	&=\frac12(\tilde{S}_{[q_1,p]}+\tilde{S}_{[q_3,p_1]}+\tilde{S}_{[q_1,p]}+\tilde{S}_{[q_2,a_1]}-\tilde{S}_{[q_2,a_1]}-\tilde{S}_{[q_3,p_1]})\nonumber\\
	&=\frac{c}{6}\log\left[\frac{(p-q_1)^2}{\epsilon^2}\right]+\frac{c}{6}\phi[q_1]\,, \label{bcbpeA2b1}\\
	\mathcal{I}(B\text{Ir}(B),A\text{Ir}(A)A_1\text{Ir}(A_1)) &=\frac12\Big(\tilde{S}_{B\text{Ir}(B)B_1\text{Ir}(B_1)}+\tilde{S}_{B\text{Ir}(B)}-\tilde{S}_{B_1\text{Ir}(B_1)}\Big)\nonumber\\ 
	&=\frac12(\tilde{S}_{(-\infty,q_1]\cup[q_3,p_1]\cup [p,\infty)}+\tilde{S}_{(-\infty,q_1]\cup[p,b_1]\cup [q_4,\infty)}-\tilde{S}_{[q_3,p_1]\cup [b_1,q_4] })\nonumber\\
	&=\frac12(\tilde{S}_{[q_1,p]}+\tilde{S}_{[q_3,p_1]}+\tilde{S}_{[q_1,p]}+\tilde{S}_{[b_1,q_4]}-\tilde{S}_{[q_3,p_1]}-\tilde{S}_{[b_1,q_4]})\nonumber\\
	&=\frac{c}{6}\log\left[\frac{(p-q_1)^2}{\epsilon^2}\right]+\frac{c}{6}\phi[q_1]\,,\label{bcbpeA2b2}
\end{align}
where the scalar field is given by \eqref{varphi2}. Interestingly for this configuration, the balanced condition is satisfied trivially. However, we need to further impose the implicit minimal requirement which settles $q_1$ to be
\begin{align}\label{pointA2bq1}
	q_1&=\frac{2ab-(a+b)p}{a+b-2p}\,.
\end{align}
Again, this point is also the intersection between the boundary and the extended EWCS $\Sigma_{AB}$. Finally, we obtain the BPE for this phase as follows
\begin{align}\label{BPEA2b}
	\text{BPE}=\mathcal{I}(A\text{Ir}(A),B\text{Ir}(B)B_1\text{Ir}(B_1))|_{balanced}=\frac{c}{6}\log\left[\frac{2(p-a)(b-p)}{(b-a)\epsilon}\right]\,,
\end{align}
where we have plugged in \eqref{varphi2} and \eqref{pointA2bq1} in the above equation. As expected, the BPE \eqref{BPEA2b} precisely match with the area of the EWCS \eqref{EWA2b}.

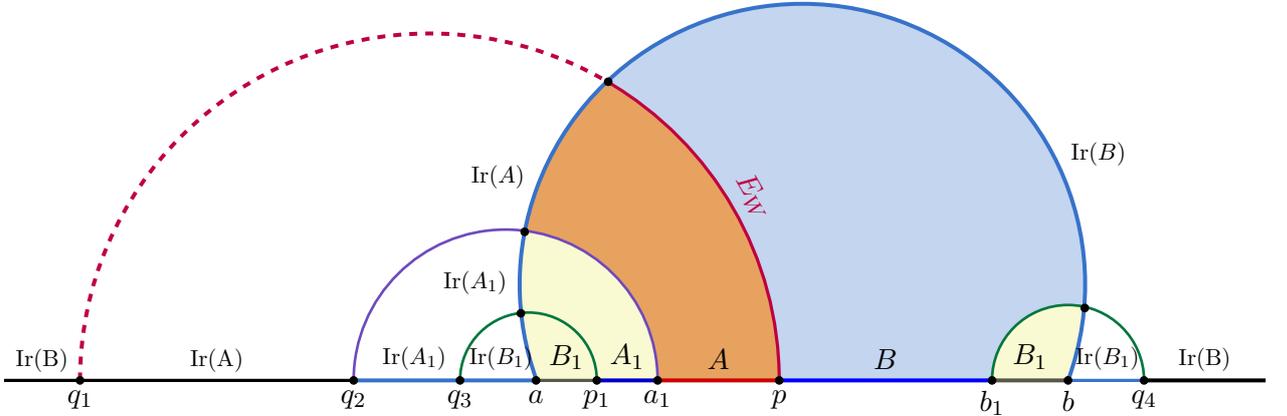
\begin{figure}[H]
	\begin{center}
		\begin{tikzpicture}
			\filldraw[LouisBlue!30](5,0) arc(-20:200:3.72);
			\begin{scope} 
				\clip (5,0) arc(-20:200:3.72);
				\filldraw[orange,opacity=0.6] (-9,0)--(1.2,0) arc(0:59:4.6);
			\end{scope}
			\begin{scope}
				\clip (5,0) arc(-20:200:3.72);
				\filldraw[yellowr,opacity=1](-0.4,0) arc(0:180:2);
			\end{scope}
			\begin{scope}
				\clip (5,0) arc(-20:200:3.72);
				\filldraw[yellowr,opacity=1](4,0) arc(180:0:1);
			\end{scope}
			\draw[color=red!80!black,line width=1.5pt] (-0.4,0)--(1.2,0)node[midway,above,scale=1]{ \textcolor{black}{$A$} };
			\draw[color=yellow!20!black,very thick] (-2,0)--(-1.2,0)node[midway,above,scale=1]{ \textcolor{black}{$B_1$} };
			\draw[color=blue,line width=1.5pt] (1.2,0)--(4,0)node[midway,above,scale=1]{ \textcolor{black}{$B$} };	
			\draw[color=yellow!20!black,line width=1.5pt] (4,0)--(5,0)node[midway,above,scale=1]{ \textcolor{black}{$B_1$} };		
			\draw[color=blue!80!black,line width=1.5pt] (-1.2,0)--(-0.4,0)node[midway,above,scale=1]{ \textcolor{black}{$A_1$} };
			\draw[very thick] (-8,0)--(-4.4,0)node[midway,above,scale=0.8]{ \textcolor{black}{Ir(A)}};
			\draw[very thick] (-9,0)--(-8,0)node[midway,above,scale=0.8]{ \textcolor{black}{Ir(B)}};
			\draw[line width=1.5pt,LouisBlue] (-4.4,0)--(-2,0)node[midway,below]{$ \textcolor{black}{} $};		    
			\draw[line width=1.5pt] (6.0,0)--(7.6,0)node[midway,above,scale=0.8]{ \textcolor{black}{Ir(B)} };	
			\draw[very thick,LouisBlue] (5,0)--(6,0)node[midway,above]{$ \textcolor{black}{} $};	
			\draw[color=LouisColor1,line width=1pt] (4,0) arc(180:0:1) node[midway,above,sloped]{$\text{}$};
			\draw[color=LouisColor1,line width=1pt] (-1.2,0) arc(0:180:0.9) node[midway,above,sloped]{$\text{}$};
			\draw[color=LouisColor2,line width=1pt] (-0.4,0) arc(0:180:2) node[midway,above,sloped]{$\text{}$};
			\draw[color=purple,line width=1.2pt] (1.2,0) arc(0:59.5:4.6) node[midway,above,sloped]{$E_W$};
			\draw[color=purple,dashed,line width=1.5pt] (-8,0) arc(180:59.5:4.6) node[midway,above,sloped]{$\text{}$};
			\draw[color=LouisBlue,line width=1.5pt] (5,0) arc(-20:200:3.72) node[midway,above,sloped,scale=0.8,color=black]{\text{}};
			\draw[fill](-0.4,0)circle(0.05) (-0.4,0) node[below,scale=1]{$a_1$};
			\draw[fill](1.2,0)circle(0.05) (1.2,0) node[below,scale=1]{$p$};
			\draw[fill](4,0)circle(0.05) (4,0) node[below,scale=1]{$b_1$};
			\draw[fill](5,0)circle(0.05) (5,0) node[below,scale=1]{$b$};
			\draw[fill](-1.2,0)circle(0.05) (-1.2,0) node[below,scale=1]{$p_1$};
			\draw[fill](-2,0)circle(0.05) (-2,0) node[below,scale=1]{$a$};
			\draw[fill](-4.4,0)circle(0.05) (-4.4,0) node[below,scale=1]{$q_2$};
			\draw[fill](6,0)circle(0.05) (6,0) node[below,scale=1]{$q_4$};
			\draw[fill] (-8,0)circle(0.05) (-8,0) node(q)[below,scale=1]{$q_1$};	
			\draw[fill](-3,0)circle(0.05) (-3,0) node[below,scale=1]{$q_3$};			
			\draw (5.52,0.3) node[yellow!45!black,scale=0.8,color=black]{Ir($B_1$)};
			\draw (-2.46,0.3) node[yellow!45!black,scale=0.8,color=black]{Ir($B_1$)};
			\draw (-3.6,0.3) node[yellow!45!black,scale=0.8,color=black]{Ir($A_1$)};
			\draw (-2.8,1.3) node[yellow!45!black,scale=0.8,color=black]{Ir($A_1$)};
			\draw (-2.5,2.7) node(IrA)[yellow!45!black,scale=0.8,color=black]{Ir($A$)};
			\draw (5.4,3) node[yellow!45!black,scale=0.8,color=black]{Ir($B$)};
			\draw[fill] (5.22,0.96)circle(0.05);
			\draw[fill] (-1.05,3.96)circle(0.05);
			\draw[fill] (-2.15,1.97)circle(0.05) node(AA1)[left]{};
			\draw[fill] (-2.2,0.89)circle(0.05);
		\end{tikzpicture}
	\end{center}   
	\caption{Two adjacent intervals $AB$ and their complements $A_1$ and $B_1$, featuring entanglement island regions $\text{Ir}(A)=[q_1,q_2],\quad \text{Ir}(B)=(-\infty,q_1]\cup [q_4,\infty),\quad \text{Ir}(A_1)=[q_2,q_3]\,\,\, \text{and}\,\,\,\, \text{Ir}(B_1)=[q_3,a]\cup [b,q_4]$ on the entire KR brane, are configured schematically. Here $\text{Ir}(B_1)$ indicate disconnected entanglement island.}       
	\label{figA2bp}	
\end{figure}

Then we consider the scenarios where $\kappa\neq 0$ in \eqref{varphi2}, such that the cutoff brane is described by \eqref{cutoffbrane2}, which is no longer a semi-circle (see \figref{figA2bp}). The size, position and the partition of $AB$ are adjusted such that, the RT surface $\mathcal{E}_{AB}$ and the EWCS $\Sigma_{AB}$ all anchor on the cutoff brane. The lengths of the geodesics anchored on the cutoff brane are classified in \cref{geodesics length}. On the gravity side, the area of the EWCS can be easily calculated as follows
\begin{align}\label{EWA2bp}
	\frac{\text{Area}[\Sigma_{AB}]}{4 G}&=\frac{c}{6}\log\left[\frac{2(p-a)(b-p)(\cosh (\kappa)+\sinh (\kappa))}{(b-a)\epsilon}\right]
	\cr
	&=\frac{c}{6}\log\left[\frac{2(p-a)(b-p)}{(b-a)\epsilon}\right]+\frac{c}{6}\kappa\,.
\end{align}
Then we turn to the calculation for the BPE on the field theory side. Firstly we give an assignment for the island regions, $\text{Is}(AB)=\text{Ir}(A)\cup \text{Ir}(B)$, where
\begin{align}
	\text{Ir}(A)=[q_1,q_2]\,,\qquad \text{Ir}(B)=(-\infty,q_1]\cup [q_4,\infty)
\end{align}
where $x=q_1$ is the partition point of $\is (AB)$, which will be determined by the balance requirements \eqref{gbc1}. Now we calculate the PEE on both sides of \eqref{gbc1}, which have the same expression as \eqref{bcbpeA2b1} and \eqref{bcbpeA2b2}, with the only difference that, the $\kappa$ in the scalar field \eqref{varphi2} is non-zero. Again the balanced condition is satisfied trivially and the implicit minimal requirement which settles $q_1$ to be
\begin{align}\label{pointA2bq1p}
	q_1&=\frac{2ab-(a+b)p}{a+b-2p}\,.
\end{align}
Finally, we may obtain the BPE for this phase as follows
\begin{align}\label{BPEA2bp}
	\text{BPE}=\mathcal{I}(A\text{Ir}(A),B\text{Ir}(B)B_1\text{Ir}(B_1))|_{balanced}=\frac{c}{6}\log\left[\frac{2(p-a)(b-p)}{(b-a)\epsilon}\right]+\frac{c}{6}k\,,
\end{align}
where we have plugged in \eqref{varphi2} and \eqref{pointA2bq1p} in the above equation. As expected, the BPE expressed in \eqref{BPEA2bp} precisely match with the area of the EWCS given by \eqref{EWA2bp}.

\section{Geodesic Length from Geometrical Analysis}\label{geodesics length}
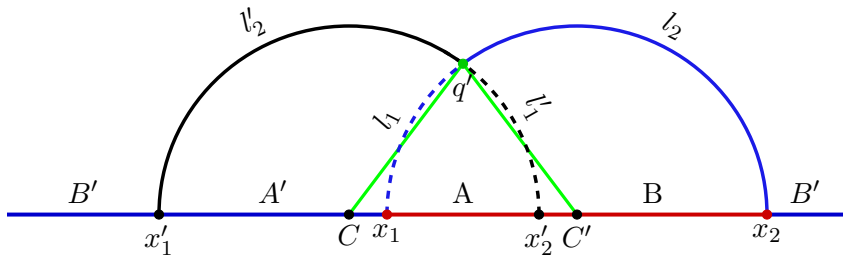
\begin{figure}[H]
		\begin{center}
			\begin{tikzpicture}
				\draw[color=red!80!black,line width=1.5pt] (0,0)--(2,0)node[midway,above]{$ \textcolor{black}{\text{A}} $};
				\draw[color=red!80!black,line width=1.5pt] (2,0)--(5,0)node[midway,above]{$ \textcolor{black}{\text{B}} $};	
				\draw[color=blue!80!black,line width=1.5pt] (-3,0)--(0,0)node[midway,above]{$ \textcolor{black}{A'} $};
				\draw[color=blue!80!black,line width=1.5pt] (-5,0)--(-3,0)node[midway,above]{$ \textcolor{black}{B'} $};			
				\draw[color=blue!80!black,line width=1.5pt] (5,0)--(6,0)node[midway,above]{$ \textcolor{black}{B'} $};	
				\draw[color=green,line width=1.2pt] (-0.5,0)--(1,2);
				\draw[color=green,line width=1.2pt] (2.5,0)--(1,2);		
				\draw[color=blue!90!yellow,line width=1.3pt] (5,0) arc(0:126:2.5) node[pos=0.5,above,sloped]{\textcolor{black}{$l_2$}};
				\draw[color=blue!90!yellow,dashed,line width=1.3pt] (0,0) arc(180:126:2.5) node[pos=0.5,above,sloped]{\textcolor{black}{$l_1$}};
				\draw[color=black,dashed,line width=1.3pt] (2,0) arc(0:60:2.5) node[midway,above,sloped]{\text{$l_1'$}};
				\draw[color=black,line width=1.3pt] (-3,0) arc(180:53:2.5) node[midway,above,sloped]{\text{$l_2'$}};
				\draw (0,0) node[below,scale=1]{\text{$x_1$}};
				\draw (2,0) node[below,scale=1]{\text{$x_2'$}};
				\draw (5,0) node[below,scale=1]{\text{$x_2$}};
				\draw (-3,0) node[below,scale=1]{\text{$x_1'$}};
				\draw (1,2) node[below,scale=1]{$ q' $};
				\draw[fill](-0.5,0)circle(0.06) node[below,scale=1]{\text{$C$}};
				\draw[fill](2.5,0)circle(0.06) node[below,scale=1]{\text{$C'$}};
				\draw[fill](2,0)circle(0.06);
				\draw[color=red!80!black,fill](5,0)circle(0.06);
				\draw[color=red!80!black,fill](0,0)circle(0.06);
				\draw[fill](-3,0)circle(0.06);
				\draw[color=green!80!black,fill](1,2)circle(0.06);
			\end{tikzpicture}
		\end{center}   
		\caption{Schematics shows the purification of the adjacent intervals $A=[x_1,x'_2]$ and $B=[x'_2,x_2]$ located in the vacuum $CFT_2$. The geodesics lengths homologous to the intervals $A\cup A'$ and $A\cup B$ are depicted as $l'_1+l'_2$ (black) and $l_1+l_2$ (blue) respectively and the intersection point of the corresponding geodesics is labelled as $q'$.}       
		\label{basic}	
	\end{figure}

In this section, we give a generic way to derive the lengths for geodesic chords anchored on the boundary and terminated in the bulk, which we frequently encounter in this paper. Let us consider two adjacent intervals $A=[x_1,x'_2]$ and $B=[x'_2,x_2]$ located in the vacuum $CFT_2$ which constitutes a pure state with the intervals $A'=[x'_1,x_1]$ and $B'=[x_2,x'_1]$, see \figref{basic}. When the RT surface of $A'A$ and $AB$ are normal to each other, we can explicitly derive the lengths for the geodesic chords $l_1$, $l_2$, $l'_1$ and $l'_2$ in \figref{basic}. 
	To obtain the location of the point $q'$ as indicated in \figref{basic}, we consider a Euclidean triangle in the $x$-$z$ plane and utilize trivial Euclidean identities. Consequently, the coordinate of the point $q'$ are described as $x=\rho_2 \cos(\theta')$ and $z=\rho_2 \sin(\theta')$ or $\rho_1 \sin(\theta)$ using the parameters in Fig. \ref{basic1}.  We can also re-express the coordinates $\rho_1$, $\rho_2$ and $y$ in terms of the interval endpoints and applying cosine law in the triangle results in following relations,
	\begin{align}
		\rho_1&=\frac{x_2-x_1}{2},\quad \rho_2=\frac{x_2'-x_1'}{2},\quad y=\frac{x_1+x_2-x_1'-x_2'}{2}\label{rou}\,, \nonumber \\
		&\cos\theta=\frac{\rho_1^2+y^2-\rho_2^2}{2\rho_1y},\quad \cos\theta'= \frac{\rho_2^2+y^2-\rho_1^2}{2\rho_2y}\,.
	\end{align}
	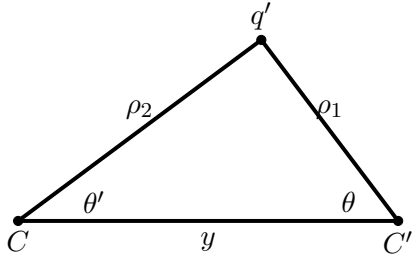
\begin{figure}[H]
		\begin{center}
			\begin{tikzpicture}
				\draw[color=black,line width=1.5pt] (0,0)--(5,0)node[midway,below]{ \textcolor{black}{$y$} };
				\draw[color=black,line width=1.5pt] (0,0)--(3.2,2.4)node[midway,above]{ \textcolor{black}{\text{$\rho_2$}} };
				\draw[color=black,line width=1.5pt] (5,0)--(3.2,2.4)node[midway,above]{ \textcolor{black}{\text{$\rho_1$}} };
				\draw[fill](0,0)circle(0.06) node[below,scale=1]{\text{$C$}};
				\draw[fill](5,0)circle(0.06) node[below,scale=1]{\text{$C'$}};
				\draw[fill](3.2,2.4)circle(0.06) node[above,scale=1]{\text{$q'$}};
				\draw (4.35,0.25) node[scale=1]{\text{$\theta$}};
				\draw (1,0.25) node[scale=1]{\text{$\theta'$}};
			\end{tikzpicture}
		\end{center}
		\caption{Euclidean triangle is considered in the $x$-$z$ plane where $\rho_1$ and $\rho_2$ are the distances from the point $q'$ and the center of the RT surfaces associated with the intervals $A\cup B$ and $A'\cup A$ respectively. The point $y$ is the distance between the corresponding centers. }       
		\label{basic1}	
	\end{figure}
	Note that, the two semi-circles depicted in \figref{basic} are normal to each other and therefore we can determine a constraint equation between the endpoint of the intervals $x_1, x_2, x_1'$ and $x_2'$ by utilizing right angle triangle identity $\rho_1^2+\rho_2^2=y^2$ as follows 
	\begin{align}\label{relation}	-2x_1'x_2'+x_2(x_1'+x_2')+x_1(x_1'+x_2'-2x_2)=0\,.
	\end{align}
	Furthermore, we establish a framework to compute the geodesic length $l'_1$, then we further generalize this analysis to other geodesic segments. In this context, the geodesics equation for the segment $l'_1$ is described by $(x-C)^2+(z-\epsilon)^2=(\rho_2)^2$, where $C$ is the center for the semi-circle $l'_1+l'_2$. Now we utilize the geodesic length formula in pure $AdS_3$ geometry at constant time slice and integrate over the bulk coordinate $z$ with limits involving endpoints of the geodesic segment $l'_1$ as 
	\begin{align}\label{geodesic_length}
		\mathcal{L}_{l'_1}&=\int_{\epsilon}^{\rho_2 \sin(\theta') }\dd{z}\frac{\sqrt{1+\frac{\dd{x}}{\dd{z}}^2}}{z}\, \nonumber\\
		&=\frac12 \log\left[{\frac{\rho_2-\sqrt{\rho_2^2-z^2}}{\rho_2+\sqrt{\rho_2^2-z^2}}}\right]\Bigg|_{\epsilon}^{\rho_2 \sin(\theta')}\,\nonumber\\
		&=\frac12 \log\left[{\frac{1-\cos\theta'}{1+\cos\theta'}\frac{4\rho_2^2}{\epsilon^2}}\right]\,.
	\end{align}
	Note that similar analysis can also be followed for the length of the geodesic segments $l_1$, $l_2$, and $l'_2$. For each of the geodesic length segments shown in \figref{basic}, the holographic proposal of the entanglement entropy discussed in \cite{Ryu:2006bv} may be used to determine the holographic entanglement entropy from the \eqref{geodesic_length} in a following way
	\begin{align}
		\mathcal{L}_{l_1}&=\frac{1}{2}\log\left[{\frac{(x_1-x_1')(x_1-x_2')(x_2-x_1)^2}{(x_1'-x_2)(x_2-x_2')\epsilon^2}}\right]\label{Sl1}\,,\\
		\mathcal{L}_{l_2}&=\frac{1}{2}\log\left[{\frac{(x_2-x_1')(x_2-x_2')(x_2-x_1)^2}{(x_1'-x_1)(x_1-x_2')\epsilon^2}}\right]\label{Sl2}\,,\\
		\mathcal{L}_{l_1'}&=\frac{1}{2}\log\left[{\frac{(x_2-x_2')(x_1-x_2')(x_2'-x_1')^2}{(x_1'-x_1)(x_2-x_1')\epsilon^2}}\right]\label{Sl1'}\,,\\
		\mathcal{L}_{l_2'}&=\frac{1}{2}\log\left[{\frac{(x_1-x_1')(x_1'-x_2)(x_2'-x_1')^2}{(x_2-x_2')(x_1-x_2')\epsilon^2}}\right]\label{Sl2'}\,.
	\end{align}
In the above equations, we have written the lengths for the geodesic chords in terms of the coordinates $\{ x'_1, x'_2, x_1, x_2\}$ of the boundary endpoints, which satisfy \eqref{relation}. In this paper, we utilize the expression of the corresponding geodesic segments to compute the lengths for the EWCS and other geodesic chords in various scenarios.

\section{The BPE/EWCS correspondence for adjacent $AB$ with entanglement islands} \label{appendixB}
Here we consider the configuration of two adjacent intervals $A=[a_1,p]$ and $B=[p,b_1]$ where $AB$ admits an island region. Again there are three possible assignments for the island regions:
		\begin{enumerate}
		\item 
		A2a: $\text{Ir}(A)=\text{Ir}(A_1)=\emptyset,\quad \text{Ir}(B)=(-\infty,q_1]\cup[q_2,\infty),\quad \text{Ir}(B_1)=[q_1,a]\cup[b,q_2]$,
		\item 
		A2b: $ \text{Ir}(A)=[q_1,q_2],\quad \text{Ir}(B)=(-\infty,q_1]\cup[q_4,\infty),\quad \text{Ir}(A_1)=[q_2,q_3],\quad \text{Ir}(B_1)=[q_3,a]\cup [b,q_4]$,
		\item
		A2c:
		$\text{Ir}(B)=\text{Ir}(B_1)=\emptyset,\quad \text{Ir}(A)=(-\infty,q_1]\cup[q_2,\infty),\quad \text{Ir}(A_1)=[q_1,a]\cup[b,q_2]$
			\end{enumerate}
In the phase-A2a, the interval $AA_1$ do not admit an island as shown in \figref{figA2a}. The phase-A2c is symmetric to the phase-A2a. In the phase-A2b, both $AA_1$ and $BB_1$ admit an island, and the analysis is given in the main text.

\subsubsection*{Phase-A2a and phase-A2c}
We now discuss the computation of the EWCS in the scenario depicted in \figref{figA2a}. Here we have divided the complement of the adjacent intervals $A\cup B$ into $A_1\cup B_1$ with partition point $p_1$. Utilizing the constraint \eqref{relation} for the RT surfaces homologous to the intervals $[q_1,a_1]$ and $[p_1,p]$, we can obtain location of the point $p_1$ as
\begin{equation}\label{pointA2ap1}
	\begin{aligned}
		p_1&=\frac{2q_1a_1-p(q_1+a_1)}{q_1+a_1-2p}\,.
	\end{aligned}
\end{equation}
In this scenario, the EWCS may be directly obtained from the length of the geodesic segment described in \cref{geodesics length} as follows
\begin{equation}\label{EWCSA2ac}
	\begin{aligned}
 E_W&=\frac{\text{Area}[\text{EWCS}]}{4 G}\\ 
		&=\frac{c}{6}\log\left[{\frac{2(p-a_1)(p-q_1)}{(a_1-q_1)\epsilon}}\right],
	\end{aligned}
\end{equation}
Note that the value of $q_1$ can also be obtained in terms of the points $a$, $a_1$ and $b$ by using the constraint relation in \eqref{relation} as
\begin{equation}\label{pointA2aq1}
	\begin{aligned}
		q_1&=\frac{2ab-(a+b)a_1}{a+b-2a_1}\,. \\ 
	\end{aligned}
\end{equation}
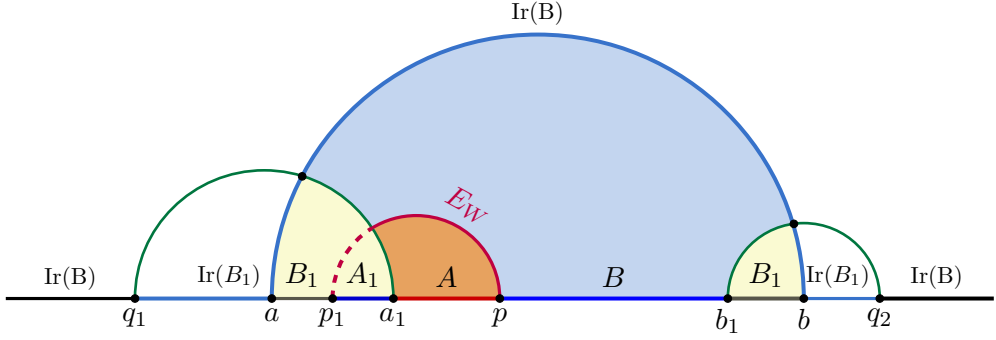
\begin{figure}[H]
		\begin{center}
			\begin{tikzpicture}
                \filldraw[LouisBlue!30](-2,0) arc (180:0:3.5);
                 \begin{scope}
                     \clip (1,0) arc(0:180:1.1);
                     \filldraw[orange,opacity=0.6] (-2,0) arc (180:0:3.5);
                 \end{scope}
                 \begin{scope}
                     \clip (-2,0) arc (180:0:3.5);
                     \filldraw[yellowr,opacity=1](-0.4,0) arc(0:180:1.7);
                 \end{scope}
                 \begin{scope}
                     \clip (-2,0) arc (180:0:3.5);
                     \filldraw[yellowr,opacity=1](4,0) arc(180:0:1);
                 \end{scope}
				\draw[color=red!80!black,line width=1.5pt] (-0.4,0)--(1,0)node[midway,above,scale=1]{$ \textcolor{black}{A} $};
				\draw[color=yellow!20!black,very thick] (-2,0)--(-1.2,0)node[midway,above,scale=1]{$ \textcolor{black}{B_1} $};
				\draw[color=blue,line width=1.5pt] (1,0)--(4,0)node[midway,above,scale=1]{$ \textcolor{black}{B} $};	
				\draw[color=yellow!20!black,line width=1.5pt] (4,0)--(5,0)node[midway,above,scale=1]{$ \textcolor{black}{B_1} $};		
				\draw[color=blue!80!black,line width=1.5pt] (-1.2,0)--(-0.4,0)node[midway,above,scale=1]{$ \textcolor{black}{A_1} $};
				\draw[very thick] (-5.5,0)--(-3.8,0)node[midway,above,scale=0.8]{ \textcolor{black}{Ir(B)}};
                \draw[line width=1.5pt,LouisBlue] (-3.8,0)--(-2,0)node[midway,below]{$ \textcolor{black}{} $};		    
				\draw[line width=1.5pt] (6.0,0)--(7.5,0)node[midway,above,scale=0.8]{ \textcolor{black}{Ir(B)} };	
                \draw[very thick,LouisBlue] (5,0)--(6,0)node[midway,above]{$ \textcolor{black}{} $};	
				\draw[color=LouisColor1,line width=1pt] (4,0) arc(180:0:1) node[midway,above,sloped]{$\text{}$};
				\draw[color=LouisColor1,line width=1pt] (-0.4,0) arc(0:180:1.7) node[midway,above,sloped]{$\text{}$};
				\draw[color=purple,line width=1.2pt] (1,0) arc(0:122:1.1) node[midway,above,sloped]{$E_W$};
				\draw[color=purple,dashed,line width=1.5pt] (-1.2,0) arc(180:122:1.1) node[midway,above,sloped]{$\text{}$};
				\draw[color=LouisBlue,line width=1.5pt] (5,0) arc(0:180:3.5) node[midway,above,sloped,scale=0.8,color=black]{\text{Ir(B)}};
				\draw[fill](-0.4,0)circle(0.05) (-0.4,0) node[below,scale=1]{$a_1$};
				\draw[fill](1,0)circle(0.05) (1,0) node[below,scale=1]{$p$};
				\draw[fill](4,0)circle(0.05) (4,0) node[below,scale=1]{$b_1$};
				\draw[fill](5,0)circle(0.05) (5,0) node[below,scale=1]{$b$};
				\draw[fill](-1.2,0)circle(0.05) (-1.2,0) node[below,scale=1]{$p_1$};
				\draw[fill](-2,0)circle(0.05) (-2,0) node[below,scale=1]{$a$};
				\draw[fill](-3.8,0)circle(0.05) (-3.8,0) node[below,scale=1]{$q_1$};
				\draw[fill](6,0)circle(0.05) (6,0) node[below,scale=1]{$q_2$};
                \draw (5.45,0.3) node[yellow!45!black,scale=0.8,color=black]{Ir($B_1$)};
                \draw (-2.56,0.3) node[yellow!45!black,scale=0.8,color=black]{Ir($B_1$)};
                \draw[fill] (4.87,0.99)circle(0.05);
                \draw[fill] (-1.6,1.62)circle(0.05);
			\end{tikzpicture}
		\end{center}   
		\caption{Schematics shows the configuration of two adjacent intervals considered in Weyl $CFT_2$ where the intervals $B$ and $B_1$ admit entanglement island region on the KR brane. }       
	\label{figA2a}
	\end{figure}

Now we carry out the BPE analysis for this phase, where the island assignment $\text{Ir}(B_1)=[q_1,a]\cup [b,q_2]$ and $\text{Ir}(B)=(-\infty,q_1]\cup [q_2,\infty)$ is shown in \figref{figA2a}. In this phase, the balance requirement is given by
  \begin{equation}\label{bcA2a}
	\begin{aligned}
		\mathcal{I}(A,B\text{Ir}(B)B_1\text{Ir}(B_1))=& \mathcal{I}(B\text{Ir}(B),AA_1) \,.
	\end{aligned}
\end{equation}
We now utilize the generalized ALC proposal to compute PEEs mentioned in the above balance requirement, 
 \begin{align}
\mathcal{I}(A,B\text{Ir}(B)B_1\text{Ir}(B_1))&=\frac12\Big(\tilde{S}_{AA_1}+\tilde{S}_A-\tilde{S}_{A_1}\Big)\nonumber\\
	&=\frac12(\tilde{S}_{[p_1,p]}+\tilde{S}_{[a_1,p]}-\tilde{S}_{[p_1,a_1]})\nonumber\\
	&=\frac{c}{6}\log\left[\frac{(p-p_1)(p-a_1)}{(a_1-p_1)\epsilon}\right]\,, \label{bcbpeA2a}\\
\mathcal{I}(B\text{Ir}(B),AA_1) &=\frac12\Big(\tilde{S}_{B\text{Ir}(B)B_1\text{Ir}(B_1)}+\tilde{S}_{B\text{Ir}(B)}-\tilde{S}_{B_1\text{Ir}(B_1)}\Big)\nonumber\\ 
	&=\frac12(\tilde{S}_{(-\infty,p_1]\cup [p,\infty)  }+\tilde{S}_{(-\infty,q_1]\cup[p,b_1]\cup [q_2,\infty)}-\tilde{S}_{[q_1,p_1]\cup [b_1,q_2] })\nonumber\\
	&=\frac12(\tilde{S}_{[p_1,p]}+\tilde{S}_{[q_1,p]}+\tilde{S}_{[b_1,q_2]}-\tilde{S}_{[q_1,p_1]}-\tilde{S}_{[b_1,q_2]})\nonumber\\
	&=\frac{c}{6}\log\left[\frac{(p-p_1)(p-q_1)}{(p_1-q_1)\epsilon}\right]\,.
\end{align}
Solving the balance requirement \eqref{bcA2a} we get the partition point $p _1$, which is same as \eqref{pointA2ap1}.
Finally, we can obtain the BPE using \eqref{bcbpeA2a} and \eqref{pointA2ap1},
	\begin{align}
		\text{BPE}=\mathcal{I}(A,B\text{Ir}(B)B_1\text{Ir}(B_1))|_{balanced}=\frac{c}{6}\log\left[{\frac{2(p-a_1)(p-q_1)}{(a_1-q_1) \epsilon}}\right]\,,
	\end{align}
	which coincide with the EWCS \eqref{EWCSA2ac}.
	
	The discussion for the phase-A2c is similar.
	
\section{The BPE/EWCS correspondence for non-adjacent intervals} 
\subsection{Disjoint $AB$ with no island}
	We first investigate the simplest case of two disjoint intervals $A=[a_1,a_2]$ and $B=[b_1,b_2]$ with no island region. Here the complement of $AB$ is described by intervals $A_1\cup B_1$ and $A_2\cup B_2$ with partition points located in Weyl $CFT_2$. In this case, we observe that $A_1\cup B_1$ may incorporate an entanglement island region, which can be classified into two distinct phases given by
\begin{enumerate}
	\item 
	D1a: $\text{Ir}(AB)=\emptyset,\quad \text{Ir}(B_1)=\left(-\infty, a\right] \cup \left[b, \infty \right)$,
	\item
	D1b:
	$\text{Ir}(AB)=\emptyset,\quad \text{Ir}(A_1)=[q_1,q_2], \quad \text{Ir}(B_1)=(-\infty,q_1]\cup\left[q_2,a\right]\cup [b,\infty ) $.
\end{enumerate}
In the initial phase-D1a, $\text{Ir}(B_1)$ spans entire entanglement island region. However, the other phases-D1a incorporates the entanglement island region $\text{Ir}(A_1)$ and $\text{Ir}(B_1)$ for $A_1\cup B_1$. In both of the above phases, the entanglement wedge of $AB$ is connected indicated from the RT surface homologous to $AB$.


\subsubsection*{Phase-D1a}
To proceed, we need to determine the location of EWCS endpoints situated on the RT surfaces homologous to the intervals $[a_1,b_2]$ and $[a_2,b_1]$. These two endpoints may be obtained by constructing Euclidean triangles as discussed in \cref{geodesics length} for the RT surfaces homologous to the intervals $[p_1,p_2]$ and $[a_1,b_2]$ for $y_2$ subsequently $[p_1,p_2]$ and $[a_2,b_1]$ for $y_1$. Therefore the location of these endpoints in terms of boundary coordinates are given by 
\begin{equation}
    \begin{aligned}\label{endpointD1a}
	y_1&= \frac{\sqrt{a_2-p_1} \sqrt{p_1-b_1} \sqrt{a_2-p_2} \sqrt{b_1-p_2}}{p_1+p_2-a_2-b_1}\,,\\
 y_2&= \frac{\sqrt{a_1-p_1} \sqrt{a_1-p_2} \sqrt{p_1-b_2} \sqrt{b_2-p_2}}{p_1+p_2-a_1-b_2}\,.
\end{aligned}
\end{equation}
The geodesic length connecting the above bulk points can be computed using the following length formula 
\begin{equation}
\begin{aligned}\label{lengthD1a}
	\mathcal{L}&=\int_{y_1}^{y_2 }\dd{z}\frac{\sqrt{1+\frac{\dd{x}}{\dd{z}}^2}}{z}\, \\
	&=\frac12 \log\left[{\frac{(p_2-p_1)/2-\sqrt{((p_2-p_1)/2)^2-z^2}}{(p_2-p_1)/2+\sqrt{((p_2-p_1)/2)^2-z^2}}}\right]\Bigg|_{y_1}^{y_2}\,.
\end{aligned}
\end{equation}
Note that, the aforementioned length formula is minimal since, in determining the EWCS endpoints, we took into consideration that the RT surface homologous to $[p_1,p_2]$ is perpendicular to the RT surfaces homologous to $[a_1,b_2]$ and $[a_2,b_1]$. In this context, the EWCS in the phase can be obtained by utilizing \eqref{lengthD1a} and \eqref{endpointD1a} as follows 
\begin{equation}
\begin{aligned}\label{ewcsD1a}
E_W&=\frac{\mathcal{L}}{4 G}\\ 
	&=\frac{c}{12}\log \left[\frac{\left(a_2-p_1\right) \left(a_1-p_2\right) \left(p_1-b_1\right) \left(b_2-p_2\right)}{\left(a_1-p_1\right) \left(a_2-p_2\right) \left(p_1-b_2\right) \left(b_1-p_2\right)}\right]\,.
\end{aligned}
\end{equation}
where $p_1$ and $p_2$ are given as follow through solving \eqref{relation} for the RT surfaces homologous to $[p_1,p_2]$ and $[a_1,b_2]$, and the RT surfaces homologous to $[p_1,p_2]$ and $[a_2,b_1]$.
	
\begin{equation}
	\begin{aligned}\label{partitionD1a}
		p_1&=\frac{a_1 b_2-a_2 b_1-\sqrt{\left(a_1-a_2\right) \left(b_1-b_2\right) \left(a_1-b_1\right) \left(a_2-b_2\right)}}{a_1-a_2-b_1+b_2}\,,\\
		p_2	&=\frac{a_1 b_2-a_2 b_1+\sqrt{\left(a_1-a_2\right) \left(b_1-b_2\right) \left(a_1-b_1\right) \left(a_2-b_2\right)}}{a_1-a_2-b_1+b_2}\,.
	\end{aligned}
\end{equation}

\begin{figure}[H]
	\begin{center}
		\begin{tikzpicture}
			\filldraw[yellowr!80](-2,0) arc (180:0:3.5);
			\begin{scope} 
				\clip (0,0) arc(180:0:2);
				\filldraw[LouisBlue!30] (-2,0) arc (180:0:3.5);
			\end{scope}
			\begin{scope} 
				\clip (0,0) arc(180:0:2);
				\filldraw[orange!60] (-1.4,0) arc(180:0:1.3);
			\end{scope}
			\draw[color=red!80!black,line width=1.5pt] (0,0)--(0.6,0)node[midway,above,scale=1]{$ \textcolor{black}{A} $};
			\draw[color=blue!60!black,line width=1.5pt] (0.6,0)--(1.2,0)node[midway,above,scale=1]{ \textcolor{black}{$A_2$} };
			\draw[color=yellow!20!black,very thick] (-2,0)--(-1.4,0)node[midway,above,scale=1]{$ \textcolor{black}{B_1} $};
			\draw[color=blue!60!black,line width=1.5pt] (1.2,0)--(2.2,0)node[midway,above,scale=1]{ \textcolor{black}{$B_2$} };	
			\draw[color=blue,line width=1.5pt] (2.2,0)--(4,0)node[midway,above,scale=1]{$ \textcolor{black}{B} $};
			\draw[color=yellow!20!black,line width=1.5pt] (4,0)--(5,0)node[midway,above,scale=1]{$ \textcolor{black}{B_1} $};		
			\draw[color=blue!80!black,line width=1.5pt] (-1.4,0)--(0,0)node[midway,above,scale=1]{$ \textcolor{black}{A_1} $};
			\draw[line width=1.5pt,LouisBlue] (-4.5,0)--(-2,0)node[midway,below]{$ \textcolor{black}{} $};
			\draw[very thick,LouisBlue] (5,0)--(7,0)node[midway,above]{$ \textcolor{black}{} $};	
			\draw[color=purple,dashed,line width=1.2pt] (1.2,0) arc(0:32:1.3) node[midway,above,sloped]{$\text{}$};
			\draw[color=purple,line width=1.2pt] (1,0.7) arc(32.5:65:1.3) node[midway,above,sloped,scale=0.8]{$E_W$};
			\draw[color=purple,dashed,line width=1.2pt] (-1.4,0) arc(180:64:1.3) node[midway,above,sloped]{$\text{}$};
			\draw[color=LouisColor2,dashed,line width=1.5pt] (0,0) arc(180:0:2) node[midway,above,sloped]{$\text{}$};
			\draw[color=LouisColor1,dashed,line width=1pt] (0.6,0) arc(180:0:0.8) node[midway,above,sloped]{$\text{}$};
			\draw[color=LouisBlue,line width=1.5pt] (5,0) arc(0:180:3.5) node(IrB1)[midway,above,sloped,scale=0.8,color=black]{\text{Ir($B_1$)}};
			\draw[fill](0,0)circle(0.05) (0,0) node[below,scale=1]{$a_1$};
			\draw[fill](1.2,0)circle(0.05) (1.2,0) node[below,scale=1]{$p_2$};
			\draw[fill](4,0)circle(0.05) (4,0) node[below,scale=1]{$b_2$};
			\draw[fill](5,0)circle(0.05) (5,0) node(b)[below,scale=1]{$b$};
			\draw[fill](-1.4,0)circle(0.05) (-1.4,0) node[below,scale=1]{$p_1$};
			\draw[fill](-2,0)circle(0.05) (-2,0) node(a)[below,scale=1]{$a$};
			\draw[fill](0.6,0)circle(0.05) (0.6,0) node[below,scale=1]{$a_2$};
			\draw[fill](2.2,0)circle(0.05) (2.2,0) node[below,scale=1]{$b_1$};
			\draw[fill](1,0.68)circle(0.05) (1,0.68) node[right,scale=1]{$y_1$};
			\draw[fill](0.4,1.21)circle(0.05) (0.4,1.21) node[above,scale=1]{$y_2$};
			\draw (5.9,0.3) node[yellow!45!black,scale=0.8,color=black]{Ir($B_1$)};
			\draw (-3.4,0.3) node[yellow!45!black,scale=0.8,color=black]{Ir($B_1$)};
		\end{tikzpicture}
	\end{center}   
\caption{The diagram depicts phases-D1a of two disjoint intervals $A=[a_1,a_2]$ and $B=[b_1,b_2]$ which are sandwiched by $A_2\cup B_2=[a_2,b_1]$ with an entanglement island region spanned by $\text{Ir}(B_1)=(-\infty,a]\cup[b,\infty)$ on the full KR brane.}       
\label{figD1a}	
\end{figure}
	
 We now compute the BPE.  As indicated in \figref{figD1a}, where $\text{Ir}(B_1)=(-\infty,a]\cup[b,\infty)$ occupies the whole island region. We need to determine the corresponding partition point via the balance requirement. In this context, the balance requirements are given by the following two equations
 \begin{equation}\label{bcD1a}
 \begin{aligned}
 	\mathcal{I}(A,BB_1\text{Ir}(B_1)B_2)=& \mathcal{I}(B,A_1A_2A) \,,
 	\cr
 \mathcal{I}(A_1,BB_1\text{Ir}(B_1)B_2)	=& \mathcal{I}(B_1\text{Ir}(B_1),A_1A_2A)\,.
 \end{aligned}
 \end{equation}
 The above PEEs can be computed using the generalized ALC proposal as follows
	\begin{align}
	\mathcal{I}(A,BB_1\text{Ir}(B_1)B_2)	&=\frac12(\tilde{S}_{AA_1}+\tilde{S}_{AA_2}-\tilde{S}_{A_1}-\tilde{S}_{A_2})\nonumber\\
	&=\frac{1}{2} \left(\tilde{S}_{\left[p_1, a_2\right]}+\tilde{S}_{\left[a_1, p_2\right]}-\tilde{S}_{\left[p_1, a_1\right]}-\tilde{S}_{\left[a_2,  p_2\right]}\right)\nonumber\\
		&=\frac{c}{6}\log\left[\frac{(a_2-p_1)(p_2-a_1)}{(a_1-p_1)(p_2-a_2)}\right]\,,\label{bpeD1a}\\
 \mathcal{I}(B,A_1A_2A)	&=\frac12(\tilde{S}_{BB_1\text{Ir}(B_1)}+\tilde{S}_{BB_2}-\tilde{S}_{B_1\text{Ir}(B_1)}-\tilde{S}_{B_2})\nonumber\\
 &=\frac{1}{2}\left(\tilde{S}_{\left(-\infty, p_1\right] \cup \left[b_1, \infty \right)}+\tilde{S}_{\left[p_2, b_2\right]}-\tilde{S}_{\left(-\infty, p_1\right] \cup \left[b_2, \infty \right)}-\tilde{S}_{\left[p_2,b_1\right]}\right)\nonumber\,\\
&=\frac{1}{2} \left(\tilde{S}_{\left[p_1, b_1\right]}+\tilde{S}_{\left[p_2, b_2\right]}-\tilde{S}_{\left[p_1, b_2\right]}-\tilde{S}_{\left[p_2,b_1\right]}\right)\nonumber\\
&=\frac{c}{6}\log\left[\frac{(b_1-p_1)(b_2-p_2)}{(b_2-p_1)(b_1-p_2)}\right]\,,
\end{align}
and
\begin{align}
 \mathcal{I}(A_1,BB_1\text{Ir}(B_1)B_2)	 &=\frac12(\tilde{S}_{AA_1A_2}+\tilde{S}_{A_1}-\tilde{S}_{AA_2})\nonumber\\
  &=\frac{1}{2} \left(\tilde{S}_{\left[p_1, p_2\right]}+\tilde{S}_{\left[p_1, a_1\right]}-\tilde{S}_{\left[a_1,  p_2\right]}\right)\nonumber\\
		&=\frac{c}{6}\log\left[\frac{(p_2-p_1)(a_1-p_1)}{(p_2-a_1)\epsilon}\right]\,,\\
	 \mathcal{I}(B_1\text{Ir}(B_1),A_1A_2A)
  &=\frac12(\tilde{S}_{BB_1\text{Ir}(B_1)B_2}+\tilde{S}_{B_1\text{Ir}(B_1)}-\tilde{S}_{BB_2})\nonumber\\
&=\frac{1}{2}\left(\tilde{S}_{\left(-\infty, p_1\right] \cup \left[p_2, \infty \right)}+\tilde{S}_{\left(-\infty, p_1\right] \cup \left[b_2, \infty \right)}-\tilde{S}_{\left[p_2,b_2\right]}\right)\nonumber\,\\
&=\frac{1}{2} \left(\tilde{S}_{\left[p_1, p_2\right]}+\tilde{S}_{\left[p_1, b_2\right]}-\tilde{S}_{\left[p_2, b_2\right]}\right)\nonumber\\
&=\frac{c}{6}\log\left[\frac{(b_2-p_1)(p_2-p_1)}{(b_2-p_2)\epsilon}\right]\,.
	\end{align}
From the balance requirement \eqref{bcD1a}, we can compute the partition points $p_1$ and $p_2$, which are the same as \eqref{partitionD1a}. 
The BPE for this phase-D1a may be obtained utilizing \eqref{partitionD1a} and \eqref{bpeD1a} as
\begin{align}
	\text{BPE}&=\mathcal{I}(A,BB_1\text{Ir}(B_1)B_2)|_{balanced} \nonumber\\
	&=\frac{c}{6}\log\left[\frac{b_2 \left(a_2+b_1\right)+a_1 \left(a_2+b_1-2 b_2\right)-2 a_2 b_1+2 \sqrt{\left(a_1-a_2\right) \left(b_1-b_2\right) \left(a_1-b_1\right) \left(a_2-b_2\right)}}{\left(a_2-b_1\right) \left(a_1-b_2\right)}\right]\,.
\end{align}
Note that the above result of BPE can be exactly matched with the EWCS \eqref{ewcsD1a}.
	
	
\subsubsection*{Phase-D1b}
\begin{figure}[H]
	\begin{center}
		\begin{tikzpicture}
			\filldraw[yellowr!80](-2,0) arc (180:0:3.5);
			\begin{scope} 
				\clip (-0.2,0) arc(180:0:1.7);
				\filldraw[LouisBlue!30] (-2,0) arc (180:0:3.5);
			\end{scope}
			\begin{scope} 
				\clip (-2,0) arc (180:0:3.5);
				\filldraw[yellowr!80] (1.2,0) arc(0:180:4.6);
			\end{scope}
			\begin{scope} 
				\clip (-0.2,0) arc(180:0:1.7);
				\filldraw[orange!60] (1.2,0) arc(0:180:4.6);
			\end{scope}
			\draw[color=red!80!black,line width=1.5pt] (-0.2,0)--(0.6,0)node[midway,above,scale=1]{$ \textcolor{black}{A} $};
			\draw[color=blue!60!black,line width=1.5pt] (0.6,0)--(1.2,0)node[midway,above,scale=1]{ \textcolor{black}{$A_2$} };
			\draw[color=yellow!20!black,very thick] (-2,0)--(-1.2,0)node[midway,above,scale=1]{ \textcolor{black}{$B_1$} };
			\draw[color=blue!60!black,line width=1.5pt] (1.2,0)--(1.8,0)node[midway,above,scale=1]{ \textcolor{black}{$B_2$} };
			\draw[color=blue,line width=1.5pt] (1.8,0)--(3.2,0)node[midway,above,scale=1]{$ \textcolor{black}{B} $};	
			\draw[color=yellow!20!black,line width=1.5pt] (3.2,0)--(5,0)node[midway,above,scale=1]{ \textcolor{black}{$B_1$} };		
			\draw[color=blue!80!black,line width=1.5pt] (-1.2,0)--(-0.2,0)node[midway,above,scale=1]{ \textcolor{black}{$A_1$} };
			\draw[line width=1.5pt,LouisBlue] (-9,0)--(-2,0);		    
			\draw[very thick,LouisBlue] (5,0)--(7,0)node[midway,above]{$ \textcolor{black}{} $};	
			\draw[color=LouisColor1,line width=1pt] (-1.2,0) arc(0:180:0.9) node[midway,above,sloped]{$\text{}$};
			\draw[color=LouisColor2,dashed,line width=1pt] (0.6,0) arc(180:0:0.6) node[midway,above,sloped]{$\text{}$};
			\draw[color=purple,dashed,line width=1.2pt] (1.2,0) arc(0:8:4.6) node[midway,above,sloped]{$\text{}$};
			\draw[color=purple,line width=1.2pt] (1.15,0.63) arc(8:20:4.6) node[midway,above,sloped]{$E_W$};
			\draw[color=purple,dashed,line width=1.5pt] (-8,0) arc(180:20:4.6) node[midway,above,sloped]{$\text{}$};
			\draw[color=purple,dashed,line width=1.5pt] (-0.2,0) arc(180:0:1.7);
			\draw[color=LouisBlue,line width=1.5pt] (5,0) arc(0:180:3.5) node[midway,above,sloped,scale=0.8,color=black]{\text{}};
			\draw[fill](-0.2,0)circle(0.05) (-0.2,0) node[below,scale=1]{$a_1$};
			\draw[fill](1.2,0)circle(0.05) (1.2,0) node[below,scale=1]{$p_2$};
			\draw[fill](3.2,0)circle(0.05) (3.2,0) node[below,scale=1]{$b_2$};
			\draw[fill](5,0)circle(0.05) (5,0) node(b)[below,scale=1]{$b$};
			\draw[fill](-1.2,0)circle(0.05) (-1.2,0) node[below,scale=1]{$p_1$};
			\draw[fill](-2,0)circle(0.05) (-2,0) node[below,scale=1]{$a$};
			\draw[fill](0.6,0)circle(0.05) (0.6,0) node[below,scale=1]{$a_2$};
			\draw[fill](1.8,0)circle(0.05) (1.8,0) node[below,scale=1]{$b_1$};
			\draw (5.9,0.3) node[yellow!45!black,scale=0.8,color=black]{Ir($B_1$)};
			\draw (-2.4,0.3) node[yellow!45!black,scale=0.8,color=black]{Ir($B_1$)};
			\draw (-8.5,0.3) node[yellow!45!black,scale=0.8,color=black]{Ir($B_1$)};
			\draw (-4.5,0.3) node[yellow!45!black,scale=0.8,color=black]{Ir($A_1$)};
			\draw (-1.1,2.9) node(IrA1)[,scale=0.8,color=black]{Ir($A_1$)};
			\draw (4.6,2.3) node(IrB1)[yellow!45!black,scale=0.8,color=black]{Ir($B_1$)};
			\draw[fill](-3,0)circle(0.05) (-3,0) node[below,scale=1]{$q_2$};
			\draw[fill] (-8,0)circle(0.05) (-8,0) node[below,scale=1]{$q_1$};
			\draw[fill] (1.15,0.62)circle(0.05) (1.15,0.63) node[left,scale=1]{$y_1$};
			\draw[fill] (0.91,1.59)circle(0.05)(0.91,1.59) node[left,scale=1]{$y_2$};
			\draw[fill] (-0.03,3.15)circle(0.05);
			\draw[fill] (-1.89,0.87)circle(0.05);
		\end{tikzpicture}
	\end{center}   
	\caption{Schematics shows the configuration of two disjoint intervals $AB$ with entanglement island regions $\text{Ir}(A_1)=[q_1,q_2]$ and $\text{Ir}(B_1)=(-\infty,q_1]\cup\left[q_2,a\right]\cup [b,\infty )$  where the partition points of $A_2\cup B_2$ and $A_1\cup B_1$ are labeled as $p_2$ and $p_1$ respectively. }       
\label{figD1b}	
\end{figure}
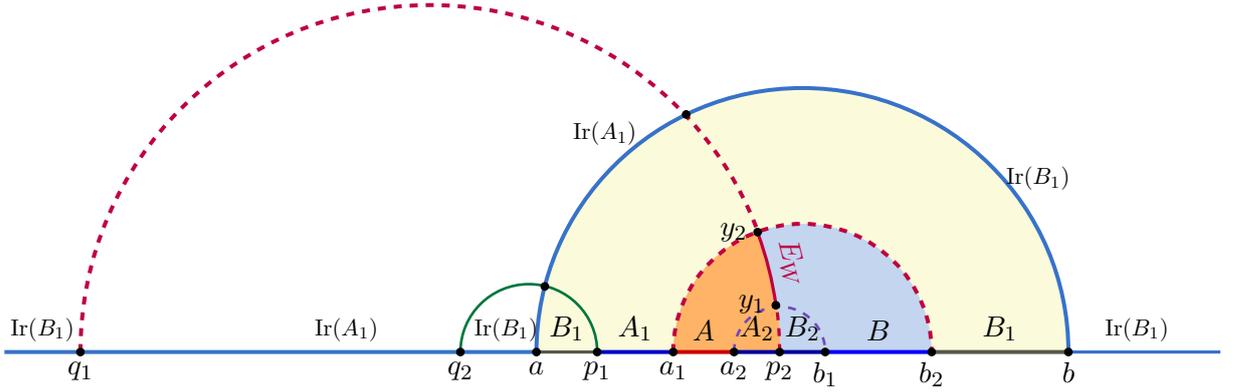

In this phase-D1b, we follow similar analysis provided in the earlier phase-D1a to compute the EWCS. Consequently, we proceed to obtain the endpoints of the EWCS which are given by $y_1$ and $y_2$ as indicated in \figref{figD1b},
\begin{equation}
	\begin{aligned}\label{endpointD1b}
		y_1&= \frac{\sqrt{a_1-q_1} \sqrt{a_1-p_2} \sqrt{q_1-b_2} \sqrt{b_2-p_2}}{q_1+p_2-a_1-b_2}\,,\\
		y_2&= \frac{\sqrt{a_2-q_1} \sqrt{a_2-p_2} \sqrt{q_1-b_1} \sqrt{b_1-p_2}}{q_1+p_2-a_2-b_1}\,,
	\end{aligned}
\end{equation}
where we used Euclidean triangle construction as discussed in \cref{geodesics length} for the RT surfaces homologous to the intervals $[q_1,p_2]$, $[a_1,b_2]$ and $[q_1,p_2]$, $[a_2,b_1]$ for calculating the $y_2$ and $y_1$ respectively. The minimal length at a constant time slice associated to these endpoints is given by
\begin{equation}
\begin{aligned}\label{lengthD1b}
	\mathcal{L}&=\int_{y_1}^{y_2 }\dd{z}\frac{\sqrt{1+\frac{\dd{x}}{\dd{z}}^2}}{z}\, \\
	&=\frac12 \log\left[{\frac{(p_2-q_1)/2-\sqrt{((p_2-q_1)/2)^2-z^2}}{(p_2-q_1)/2+\sqrt{((p_2-q_1)/2)^2-z^2}}}\right]\Bigg|_{y_1}^{y_2}\,.
\end{aligned}
\end{equation}
Utilizing \eqref{endpointD1b} and \eqref{lengthD1b}, we may the compute the EWCS for this phase as follows
\begin{equation}
\begin{aligned}\label{ewcsD1b}
E_W&=\frac{\mathcal{L}}{4 G}\\ 
	&=\frac{c}{12}\log\left[\frac{\left(a_2-q_1\right) \left(a_1-p_2\right) \left(q_1-b_1\right) \left(b_2-p_2\right)}{\left(a_1-q_1\right) \left(a_2-p_2\right) \left(q_1-b_2\right) \left(b_1-p_2\right)}\right]\,.
\end{aligned}
\end{equation}
where $q_1$ and $p_2$ are given by:
\begin{equation}
	\begin{aligned}\label{partitionD1b}
		q_1&=\frac{a_1 b_2-a_2 b_1-\sqrt{\left(a_1-a_2\right) \left(b_1-b_2\right) \left(a_1-b_1\right) \left(a_2-b_2\right)}}{a_1-a_2-b_1+b_2}\,,\\
		p_2	&=\frac{a_1 b_2-a_2 b_1+\sqrt{\left(a_1-a_2\right) \left(b_1-b_2\right) \left(a_1-b_1\right) \left(a_2-b_2\right)}}{a_1-a_2-b_1+b_2}\,.
	\end{aligned}
\end{equation}

Note that $q_1$ can also be written in terms of the points $a_2$, $b_1$ and $p_2$ using the constraint condition in \eqref{relation} for the RT surfaces homologous to $[q_1,p_2]$ and $[a_2,b_1]$ as
\begin{equation}\label{q2D1b}
q_1=\frac{2 a_2 b_1-p_2 \left(a_2+b_1\right)}{a_2+b_1-2 p_2}\,.
\end{equation}

In this phase-D1b, the computation of BPE involves entanglement island regions $\text{Ir}(A_1)=[q_1,q_2]$ and $\text{Ir}(B_1)=(-\infty,q_1]\cup\left[q_2,a\right]\cup [b,\infty)$ shown in \figref{figD1b}. We now calculate the location of partition points $p_1$ and $p_2$ from balance requirements
\begin{equation}\label{bcD1b}
	\begin{aligned}
		\mathcal{I}(A,BB_1\text{Ir}(B_1)B_2)=& \mathcal{I}(B,A_1\text{Ir}(A_1)A_2A) \,,
		\cr
		\mathcal{I}(A_1\text{Ir}(A_1),BB_1\text{Ir}(B_1)B_2)	=& \mathcal{I}(B_1\text{Ir}(B_1),A_1\text{Ir}(A_1)A_2A)\,.
	\end{aligned}
\end{equation}
The above PEEs may be computed using the generalized ALC proposal as follows
\begin{align}
		\mathcal{I}(A,BB_1\text{Ir}(B_1)B_2)&=\frac12(\tilde{S}_{AA_1\text{Ir}(A_1)}+\tilde{S}_{AA_2}-\tilde{S}_{A_1\text{Ir}(A_1)}-\tilde{S}_{A_2})\nonumber\\
		 &=\frac{1}{2}\left(\tilde{S}_{\left[q_1, q_2\right] \cup \left[p_1,a_2 \right]}+\tilde{S}_{\left[a_1, p_2\right]}-\tilde{S}_{\left[q_1, q_2\right] \cup \left[p_1, a_1 \right]}-\tilde{S}_{\left[a_2,p_2\right]}\right)\nonumber\,\\
		&=\frac{1}{2} \left( \tilde{S}_{\left[q_1, a_2\right]}+\tilde{S}_{\left[q_2, p_1\right]}+\tilde{S}_{\left[a_1, p_2\right]}-\tilde{S}_{\left[q_1,a_1\right]}-\tilde{S}_{\left[q_2,p_1\right]}-\tilde{S}_{\left[a_2, p_2\right]}\right)\nonumber\\
	&=\frac{c}{6}\log\left[\frac{\left(a_2-q_1\right) \left(p_2-a_1\right)}{\left(a_1-q_1\right) \left(p_2-a_2\right)}\right]\,,\\
	\mathcal{I}(B,A_1\text{Ir}(A_1)A_2A)&=\frac12(\tilde{S}_{BB_1\text{Ir}(B_1)}+\tilde{S}_{BB_2}-\tilde{S}_{B_1\text{Ir}(B_1)}-\tilde{S}_{B_2})\nonumber\\
	 &=\frac{1}{2}\left(\tilde{S}_{\left(-\infty, q_1\right]\cup[q_2,p_1] \cup \left[b_1, \infty \right)}+\tilde{S}_{\left[p_2, b_2\right]}-\tilde{S}_{\left(-\infty, q_1\right] \cup[q_2,p_1]\cup \left[b_2, \infty\right)}-\tilde{S}_{\left[p_2,b_1\right]}\right)\nonumber\,\\
	&=\frac{1}{2} \left(\tilde{S}_{\left[q_1, b_1\right]}+\tilde{S}_{\left[q_2,p_1\right]}+\tilde{S}_{\left[p_2, b_2\right]}-\tilde{S}_{\left[q_1, b_2\right]}-\tilde{S}_{\left[q_2, p_1\right]}-\tilde{S}_{\left[p_2,b_1\right]}\right)\nonumber\\
	&=\frac{c}{6}\log\left[\frac{\left(b_1-q_1\right) \left(b_2-p_2\right)}{\left(b_2-q_1\right) \left(b_1-p_2\right)}\right]\,,
	\end{align}
and
\begin{align}
	\mathcal{I}(A_1\text{Ir}(A_1),BB_1\text{Ir}(B_1)B_2)&=\frac12(\tilde{S}_{AA_1\text{Ir}(A_1)A_2}+\tilde{S}_{A_1\text{Ir}(A_1)}-\tilde{S}_{AA_2})\nonumber\\
	 &=\frac{1}{2}\left(\tilde{S}_{\left[q_1, q_2\right] \cup \left[p_1, p_2 \right]}+\tilde{S}_{\left[q_1, q_2\right]\cup \left[p_1, a_1 \right]}-\tilde{S}_{\left[a_1,p_2\right]}\right)\nonumber\,\\
	&=\frac{1}{2} \left(\tilde{S}_{\left[q_1, p_2\right]}+\tilde{S}_{\left[q_2, p_1\right]}+\tilde{S}_{\left[q_1, a_1\right]}+\tilde{S}_{\left[q_2, p_1\right]}-\tilde{S}_{\left[a_1, p_2\right]}\right)\nonumber\\
 &=\frac{c}{6}\log\left[\frac{(p_1-q_2)^2(p_2-q_1)(a_1-q_1)}{\epsilon^3(p_2-a_1)}\right]+\frac{c}{6}\phi(q_1)+\frac{c}{6}\phi(q_2)\,,\\
	\mathcal{I}(B_1\text{Ir}(B_1),A_1\text{Ir}(A_1)A_2A)&=\frac12(\tilde{S}_{BB_1\text{Ir}(B_1)B_2}+\tilde{S}_{B_1\text{Ir}(B_1)}-\tilde{S}_{BB_2})\nonumber\\
	 &=\frac{1}{2}\left(\tilde{S}_{\left(-\infty, q_1\right] \cup[q_2,p_1]\cup \left[p_2, \infty \right)}+\tilde{S}_{\left(-\infty, q_1\right] \cup[q_2,p_1] \cup\left[b_2, \infty \right]}-\tilde{S}_{\left[p_2,b_2\right]}\right)\nonumber\,\\
	&=\frac{1}{2} \left(\tilde{S}_{\left[q_1, p_2\right]}+\tilde{S}_{\left[q_2, p_1\right]}+\tilde{S}_{\left[q_1, b_2\right]}+\tilde{S}_{\left[q_2,p_1\right]}-\tilde{S}_{\left[p_2,b_2\right]}\right)\nonumber\\
 &=\frac{c}{6}\log\left[\frac{(p_1-q_2)^2(p_2-q_1)(b_2-q_1)}{\epsilon^3(b_2-p_2)}\right]+\frac{c}{6}\phi(q_1)+\frac{c}{6}\phi(q_2)\,.
\end{align}
From the balance requirement \eqref{bcD1b}, the partition points $q_1$ and $p_2$ are the same as \eqref{partitionD1b}.
Finally, we may obtain BPE for this phase as follows 
\begin{align}\label{bpeD1b}
	\text{BPE}&=\mathcal{I}(A,BB_1\text{Ir}(B_1)B_2)|_{balanced} \nonumber\\
	&=\frac{c}{6}\log\left[\frac{b_2 \left(a_2+b_1\right)+a_1 \left(a_2+b_1-2 b_2\right)-2 a_2 b_1+2 \sqrt{\left(a_1-a_2\right) \left(b_1-b_2\right) \left(a_1-b_1\right) \left(a_2-b_2\right)}}{\left(a_2-b_1\right) \left(a_1-b_2\right)}\right]\,.
\end{align}
The above result of BPE exactly matches with the EWCS in \eqref{ewcsD1b} using the value of partition point $p_2$ and $q_1$ given in \eqref{q2D1b} and \eqref{partitionD1b}.

	\subsection{Disjoint $AB$ admits island}
	In this subsection, we consider the configuration of two disjoint intervals $A=[a_1,a_2]$ and $B=[b_1,b_2]$ where $AB$ admits entanglement island region. We study the following assignments for the island region,
\begin{enumerate}
	\item 
	D2a: $\text{Ir}(AA_1A_2)=\emptyset,\quad \text{Ir}(B_1)=[q_1,a]\cup [b,q_2], \quad \text{Ir}(B)=(-\infty,q_1]\cup[q_2,\infty)$\,,
 \item D2b: $ \text{Ir}(A)=[q_1,q_2],\quad \text{Ir}(B)=(-\infty,q_1]\cup[q_4,\infty),\quad \text{Ir}(A_1)=[q_2,q_3],\quad \text{Ir}(B_1)=[q_3,a]\cup [b,q_4]$\,,
	\item D2c:
	$\text{Ir}(A)=[q_2,q_3],\quad \text{Ir}(B)=[q_5,q_6],\quad \text{Ir}(A_1)=[q_3,q_4],\quad \text{Ir}(B_1)=[q_4,a]\cup [b,q_5],\quad \text{Ir}(A_2)=[q_1,q_2],\quad \text{Ir}(B_2)=(-\infty,q_1]\cup [q_6,\infty)$\,.
\end{enumerate}
The intervals $A$, $A_1$ and $A_2$ does not receive any contributions from the island region in the phase-D2a as shown in \figref{figD2}. 
The last phase-D2c involves all entanglement island regions for the all the intervals $A$, $B$, $A_1$, $B_1$, $A_2$ and $B_2$. However in this phase, the entanglement wedge gets disconnected consequently the EWCS and BPE become zero.

	
	\subsubsection*{Phase-D2a}
\begin{figure}[H]
	\begin{center}
		\begin{tikzpicture}
			\filldraw[LouisBlue!30](-2,0) arc (180:0:3.5);
			\begin{scope}
				\clip (1,0) arc(0:180:1.1);
				\filldraw[orange,opacity=0.6] (-2,0) arc (180:0:3.5);
			\end{scope}
			\begin{scope}
				\clip (-2,0) arc (180:0:3.5);
				\filldraw[yellowr,opacity=1](-0.4,0) arc(0:180:1.7);
			\end{scope}
			\begin{scope}
				\clip (-2,0) arc (180:0:3.5);
				\filldraw[yellowr,opacity=1](4,0) arc(180:0:1);
			\end{scope}
			\draw[color=red!80!black,line width=1.5pt] (-0.4,0)--(0.3,0)node[midway,above,scale=1]{$ \textcolor{black}{A} $};
			\draw[color=blue!80!black,line width=1.5pt] (0.3,0)--(1,0)node[midway,above,scale=1]{ \textcolor{black}{$A_2$} };
			\draw[color=yellow!20!black,very thick] (-2,0)--(-1.2,0)node[midway,above,scale=1]{$ \textcolor{black}{B_1} $};
			\draw[color=blue!80,line width=1.5pt] (2,0)--(4,0)node[midway,above,scale=1]{$ \textcolor{black}{B} $};	
			\draw[color=blue!80!black,line width=1.5pt] (1,0)--(2,0)node[midway,above,scale=1]{ \textcolor{black}{$B_2$} };	
			\draw[color=yellow!20!black,line width=1.5pt] (4,0)--(5,0)node[midway,above,scale=1]{$ \textcolor{black}{B_1} $};		
			\draw[color=blue!80!black,line width=1.5pt] (-1.2,0)--(-0.4,0)node[midway,above,scale=1]{$ \textcolor{black}{A_1} $};
			\draw[very thick] (-5.5,0)--(-3.8,0)node[midway,above,scale=0.8]{ \textcolor{black}{Ir(B)}};
			\draw[line width=1.5pt,LouisBlue] (-3.8,0)--(-2,0)node[midway,below]{$ \textcolor{black}{} $};		    
			\draw[line width=1.5pt] (6.0,0)--(7.5,0)node[midway,above,scale=0.8]{ \textcolor{black}{Ir(B)} };	
			\draw[very thick,LouisBlue] (5,0)--(6,0)node[midway,above]{$ \textcolor{black}{} $};	
			\draw[color=LouisColor1,line width=1pt] (4,0) arc(180:0:1) node[midway,above,sloped]{$\text{}$};
			\draw[color=LouisColor1,line width=1pt] (-0.4,0) arc(0:180:1.7) node[midway,above,sloped]{$\text{}$};
			\draw[color=LouisColor2,dashed,line width=1pt] (2,0) arc(0:180:0.85) node[midway,above,sloped]{$\text{}$};
			\draw[color=purple,dashed,line width=1.2pt] (1,0) arc(0:43:1.1) node[midway,above,sloped]{$\text{}$};
			\draw[color=purple,line width=1.2pt] (0.73,0.73) arc(42:122:1.1) node[midway,above,sloped]{$E_W$};
			\draw[color=purple,dashed,line width=1.5pt] (-1.2,0) arc(180:122:1.1) node[midway,above,sloped]{$\text{}$};
			\draw[color=LouisBlue,line width=1.5pt] (5,0) arc(0:180:3.5) node(IrB)[midway,above,sloped,scale=0.8,color=black]{\text{Ir(B)}};
			\draw[fill](-0.4,0)circle(0.05) (-0.4,0) node[below,scale=1]{$a_1$};
			\draw[fill](0.3,0)circle(0.05) (0.3,0) node[below,scale=1]{$a_2$};
			\draw[fill](1,0)circle(0.05) (1,0) node[below,scale=1]{$p_2$};
			\draw[fill](4,0)circle(0.05) (4,0) node[below,scale=1]{$b_2$};
			\draw[fill](5,0)circle(0.05) (5,0) node[below,scale=1]{$b$};
			\draw[fill](2,0)circle(0.05) (2,0) node[below,scale=1]{$b_1$};
			\draw[fill](-1.2,0)circle(0.05) (-1.2,0) node[below,scale=1]{$p_1$};
			\draw[fill](-2,0)circle(0.05) (-2,0) node[below,scale=1]{$a$};
			\draw[fill](-3.8,0)circle(0.05) (-3.8,0) node[below,scale=1]{$q_1$};
			\draw[fill](6,0)circle(0.05) (6,0) node[below,scale=1]{$q_2$};
			\draw[fill](0.73,0.73)circle(0.05) (0.73,0.73) node[above,scale=1]{$y_1$};
			\draw[fill](-0.68,0.93)circle(0.05) (-0.68,0.93) node[above,scale=1]{$y_2$};
			\draw (5.45,0.3) node[yellow!45!black,scale=0.8,color=black]{Ir($B_1$)};
			\draw (-2.56,0.3) node[yellow!45!black,scale=0.8,color=black]{Ir($B_1$)};
			\draw[fill] (4.87,0.99)circle(0.05) node(IrBB1){};
			\draw[fill] (-1.6,1.62)circle(0.05) node(IrBB12){};
		\end{tikzpicture}
	\end{center}   
	\caption{The diagram shows the configuration of $AB$ considered in Weyl $CFT_2$ where the intervals $B$ and $B_1$ admit entanglement island regions $\text{Ir}(B_1)=[q_1,a_1]\cup [b,q_2]$ and $\text{Ir}(B)=(-\infty,q_1]\cup [q_2,\infty)$. }       
		\label{figD2}	       
\end{figure}
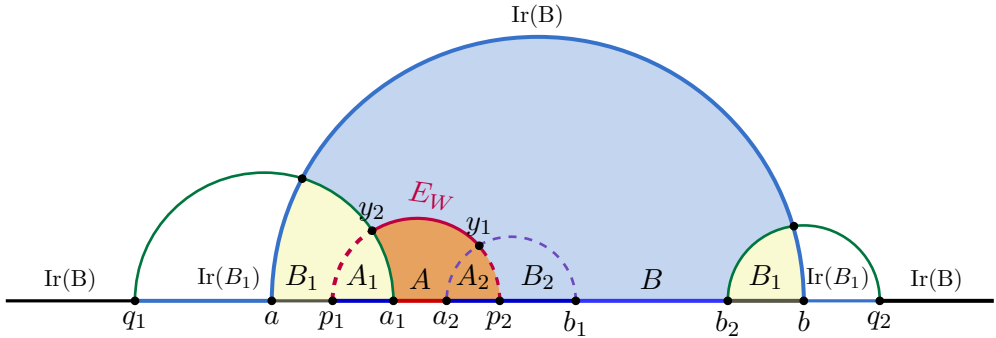

 We use the same approach as in phase-D1a to calculate the EWCS in this phase. Correspondingly the endpoints $y_1$ and $y_2$ of the EWCS as shown in \figref{figD2}, are thus obtained as

\begin{equation}
    \begin{aligned}\label{endpointD2}
	y_1&= \frac{\sqrt{a_2-p_1} \sqrt{a_2-p_2} \sqrt{p_1-b_1} \sqrt{b_1-p_2}}{p_1+p_2-a_2-b_1}\,,\\
 y_2&= \frac{\sqrt{a_1-p_1} \sqrt{a_1-p_2} \sqrt{p_1-q_1} \sqrt{q_1-p_2}}{p_1+p_2-a_1-q_1}\,.
\end{aligned}
\end{equation}
At a constant time slice, the minimal length between the endpoints $y_1$ and $y_2$ located on the RT surfaces homologous to  $[q_1,a_1]$ and $[a_2,b_1]$ can be computed as follows
\begin{equation}
\begin{aligned}\label{lengthD2}
	\mathcal{L}&=\int_{y_1}^{y_2 }\dd{z}\frac{\sqrt{1+\frac{\dd{x}}{\dd{z}}^2}}{z}\, \\
	&=\frac12 \log\left[{\frac{(p_2-p_1)/2-\sqrt{((p_2-p_1)/2)^2-z^2}}{(p_2-p_1)/2+\sqrt{((p_2-p_1)/2)^2-z^2}}}\right]\Bigg|_{y_1}^{y_2}\,.
\end{aligned}
\end{equation}
Finally in this phase, the EWCS can be determined utilizing \eqref{endpointD2} and \eqref{lengthD2} in the following proposal of the EWCS in $AdS_3$ geometries
\begin{equation}
\begin{aligned}\label{ewcsD2}
E_W&=\frac{\mathcal{L}}{4 G}\\ 
	&=\frac{c}{12}\log\left[\frac{\left(a_2-p_1\right) \left(a_1-p_2\right) \left(p_1-b_1\right) \left(q_1-p_2\right)}{\left(a_1-p_1\right) \left(a_2-p_2\right) \left(b_1-p_2\right) \left(p_1-q_1\right)}\right]\,.
\end{aligned}
\end{equation}
where $p_1$ and $p_2$ are given by:
\begin{equation}
	\begin{aligned}\label{partitionD2}
		p_1&=\frac{a_1 q_1-a_2 b_1-\sqrt{\left(a_1-a_2\right) \left(b_1-q_1\right) \left(a_1-b_1\right) \left(a_2-q_1\right)}}{a_1-a_2-b_1+q_1}\,,\\
		p_2	&=\frac{a_1 q_1-a_2 b_1+\sqrt{\left(a_1-a_2\right) \left(b_1-q_1\right) \left(a_1-b_1\right) \left(a_2-q_1\right)}}{a_1-a_2-b_1+q_1}\,.
	\end{aligned}
\end{equation}

We now compute the BPE in \figref{figD2}. Consequently, the balance requirements are  given by
\begin{equation}\label{bcD2}
	\begin{aligned}
		\mathcal{I}(A,B\text{Ir}(B)B_1\text{Ir}(B_1)B_2)=& \mathcal{I}(B\text{Ir}(B),A_1A_2A) \,,
		\cr
		\mathcal{I}(A_1,B\text{Ir}(B)B_1\text{Ir}(B_1)B_2)	=& \mathcal{I}(B_1\text{Ir}(B_1),A_1A_2A)\,.
	\end{aligned}
\end{equation}
Utilizing the generalized ALC proposal, the above PEEs may be described as follows
	\begin{align}
	\mathcal{I}(A,B\text{Ir}(B)B_1\text{Ir}(B_1)B_2)&=\frac12(\tilde{S}_{AA_1}+\tilde{S}_{AA_2}-\tilde{S}_{A_1}-\tilde{S}_{A_2})\nonumber\\
	 &=\frac{1}{2}\left(\tilde{S}_{\left[p_1,a_2 \right]}+\tilde{S}_{\left[a_1, p_2\right]}-\tilde{S}_{\left[p_1,a_1\right]}-\tilde{S}_{\left[a_2,p_2\right]}\right)\nonumber\,\\
	&=\frac{c}{6}\log\left[\frac{\left(a_2-p_1\right) \left(p_2-a_1\right)}{\left(a_1-p_1\right) \left(p_2-a_2\right)}\right]\,,\\
	\mathcal{I}(B\text{Ir}(B),A_1A_2A)&=\frac12(\tilde{S}_{B\text{Ir}(B)B_1\text{Ir}(B_1)}+\tilde{S}_{B\text{Ir}(B)B_2}-\tilde{S}_{B_1\text{Ir}(B_1)}-\tilde{S}_{B_2})\nonumber\\
	 &=\frac{1}{2}\left(\tilde{S}_{(-\infty, p_1] \cup \left[b_1, \infty \right)}+\tilde{S}_{\left(-\infty, q_1\right] \cup[p_2,b_2]\cup \left[q_2, \infty\right)}-\tilde{S}_{[q_1, p_1] \cup[b_2,q_2]}-\tilde{S}_{\left[p_2,b_1\right]}\right)\nonumber\,\\
	&=\frac{1}{2} \left(\tilde{S}_{\left[p_1, b_1\right]}+\tilde{S}_{\left[q_1,p_2\right]}+\tilde{S}_{\left[b_2, q_2\right]}-\tilde{S}_{\left[q_1, p_1\right]}-\tilde{S}_{\left[b_2, q_2\right]}-\tilde{S}_{\left[p_2,b_1\right]}\right)\nonumber\\
	&=\frac{c}{6}\log\left[\frac{\left(b_1-p_1\right) \left(p_2-q_1\right)}{\left(b_1-p_2\right) \left(p_1-q_1\right)}\right]\,,
\end{align}
and
	\begin{align}	
	\mathcal{I}(A_1,B\text{Ir}(B)B_1\text{Ir}(B_1)B_2)&=\frac12(\tilde{S}_{AA_1A_2}+\tilde{S}_{A_1}-\tilde{S}_{AA_2})\nonumber\\
	&=\frac{1}{2}\left(\tilde{S}_{\left[p_1, p_2\right]}+\tilde{S}_{\left[p_1, a_1 \right]}-\tilde{S}_{\left[a_1,p_2\right]}\right)\nonumber\,\\
	&=\frac{c}{6}\log\left[\frac{\left(p_2-p_1\right) \left(a_1-p_1\right)}{\epsilon  \left(p_2-a_1\right)}\right]\,,\\
	\mathcal{I}(B_1\text{Ir}(B_1),A_1A_2A)&=\frac12(\tilde{S}_{B\text{Ir}(B)B_1\text{Ir}(B_1)B_2}+\tilde{S}_{B_1\text{Ir}(B_1)}-\tilde{S}_{B\text{Ir}(B)B_2})\nonumber\\
	&=\frac{1}{2}\left(\tilde{S}_{\left(-\infty, p_1\right] \cup \left[p_2, \infty \right)}+\tilde{S}_{[q_1, p_1] \cup[b_2,q_2] }-\tilde{S}_{\left(-\infty, q_1\right] \cup[p_2,b_2]\cup \left[q_2, \infty \right)}\right)\nonumber\,\\
	&=\frac{1}{2} \left(\tilde{S}_{\left[p_1, p_2\right]}+\tilde{S}_{\left[q_1, p_1\right]}+\tilde{S}_{\left[b_2, q_2\right]}-\tilde{S}_{\left[q_1, p_2\right]}-\tilde{S}_{\left[b_2,q_2\right]}\right)\nonumber\\
	&=\frac{c}{6}\log\left[\frac{\left(p_2-p_1\right) \left(p_1-q_1\right)}{\epsilon  \left(p_2-q_1\right)}\right]\,.
\end{align} 
Utilizing the balance requirements in \eqref{bcD2}, we get the solutions of $p_1$ and $p_2$ that are the same as \eqref{partitionD2}.
Finally we obtain the BPE for this phase-D2 satisfying the balance conditions as follows
\begin{align}
	\text{BPE}&=	\mathcal{I}(A,B\text{Ir}(B)B_1\text{Ir}(B_1)B_2)|_{balanced}\\
	&=\frac{c}{6}\log\left[\frac{q_1 \left(a_2+b_1\right)+a_1 \left(a_2+b_1-2 q_1\right)-2 a_2 b_1+2 \sqrt{\left(a_1-a_2\right) \left(a_1-b_1\right) \left(a_2-q_1\right) \left(b_1-q_1\right)}}{\left(a_2-b_1\right) \left(a_1-q_1\right)}\right]\,.
\end{align}
The above result of BPE exactly coincide with the EWCS by putting the value of $p_2$ in \eqref{ewcsD2}.


\subsubsection*{Phase-D2b}
\begin{figure}[H]
	{
		\centering
		\begin{tikzpicture}
			\filldraw[LouisBlue!30](-2,0) arc (180:0:3.5);
			\begin{scope} 
				\clip (-2,0) arc (180:0:3.5);
				\filldraw[orange,opacity=0.6] (-9,0)--(1.2,0) arc(0:43:4.6);
			\end{scope}
			\begin{scope}
				\clip (-2,0) arc (180:0:3.5);
				\filldraw[yellowr,opacity=1](-0.4,0) arc(0:180:2);
			\end{scope}
			\begin{scope}
				\clip (-2,0) arc (180:0:3.5);
				\filldraw[yellowr,opacity=1](4,0) arc(180:0:1);
			\end{scope}
			\draw[color=red!80!black,line width=1.5pt] (-0.4,0)--(0.5,0)node[midway,above,scale=1]{ \textcolor{black}{$A$} };
			\draw[color=yellow!20!black,very thick] (-2,0)--(-1.2,0)node[midway,above,scale=1]{ \textcolor{black}{$B_1$} };
			\draw[color=yellow!20!black,line width=1.5pt] (4,0)--(5,0)node[midway,above,scale=1]{ \textcolor{black}{$B_1$} };		
			\draw[color=blue!80!black,line width=1.5pt] (-1.2,0)--(-0.4,0)node[midway,above,scale=1]{ \textcolor{black}{$A_1$} };
			\draw[color=blue!80!black,line width=1.5pt] (0.5,0)--(1.2,0)node[midway,above,scale=1]{ \textcolor{black}{$A_2$} };
			\draw[color=blue!80!black,line width=1.5pt] (1.2,0)--(1.9,0)node[midway,above,scale=1]{ \textcolor{black}{$B_2$} };
			\draw[color=blue!80,line width=1.5pt] (1.9,0)--(4,0)node[midway,above,scale=1]{$ \textcolor{black}{B} $};
			\draw[very thick] (-8,0)--(-4.4,0)node[midway,above,scale=0.8]{ \textcolor{black}{Ir(A)}};
			\draw[very thick] (-9,0)--(-8,0)node[midway,above,scale=0.8]{ \textcolor{black}{Ir(B)}};
			\draw[line width=1.5pt,LouisBlue] (-4.4,0)--(-2,0)node[midway,below]{$ \textcolor{black}{} $};		    
			\draw[line width=1.5pt] (6.0,0)--(7.6,0)node[midway,above,scale=0.8]{ \textcolor{black}{Ir(B)} };	
			\draw[very thick,LouisBlue] (5,0)--(6,0)node[midway,above]{$ \textcolor{black}{} $};	
			\draw[color=LouisColor1,line width=1pt] (4,0) arc(180:0:1) node[midway,above,sloped]{$\text{}$};
			\draw[color=LouisColor1,line width=1pt] (-1.2,0) arc(0:180:0.9) node[midway,above,sloped]{$\text{}$};
			\draw[color=LouisColor2,line width=1pt] (-0.4,0) arc(0:180:2) node[midway,above,sloped]{$\text{}$};
			\draw[color=purple,dashed,line width=1.2pt] (1.2,0) arc(0:9:4.6) node[midway,above,sloped]{$\text{}$};
			\draw[color=purple,line width=1.2pt] (1.15,0.71) arc(9:43:4.6) node[midway,above,sloped]{$E_W$};
			\draw[color=purple,dashed,line width=1.5pt] (-8,0) arc(180:43:4.6) node[midway,above,sloped]{$\text{}$};
			\draw[color=LouisColor2,dashed,line width=1.2pt] (0.5,0) arc(180:0:0.7);
			\draw[color=LouisBlue,line width=1.5pt] (5,0) arc(0:180:3.5) node[midway,above,sloped,scale=0.8,color=black]{\text{}};
			\draw[fill](-0.4,0)circle(0.05) (-0.4,0) node[below,scale=1]{$a_1$};
			\draw[fill](0.5,0)circle(0.05) (0.5,0) node[below,scale=1]{$a_2$};
			\draw[fill](1.2,0)circle(0.05) (1.2,0) node[below,scale=1]{$p_2$};
			\draw[fill](1.9,0)circle(0.05) (1.9,0) node[below,scale=1]{$b_1$};
			\draw[fill](4,0)circle(0.05) (4,0) node[below,scale=1]{$b_2$};
			\draw[fill](5,0)circle(0.05) (5,0) node[below,scale=1]{$b$};
			\draw[fill](-1.2,0)circle(0.05) (-1.2,0) node[below,scale=1]{$p_1$};
			\draw[fill](-2,0)circle(0.05) (-2,0) node[below,scale=1]{$a$};
			\draw[fill](-4.4,0)circle(0.05) (-4.4,0) node[below,scale=1]{$q_2$};
			\draw[fill](6,0)circle(0.05) (6,0) node[below,scale=1]{$q_4$};
			\draw[fill](-3,0)circle(0.05) (-3,0) node[below,scale=1]{$q_3$};
			\draw[fill] (-8,0)circle(0.05) (-8,0) node[below,scale=1]{$q_1$};
			\draw[fill] (1.15,0.71)circle(0.05) (1.15,0.71) node[right,scale=1]{$y_1$};
			\draw[fill] (-0.03,3.15)circle(0.05) (-0.03,3.15)node[above,scale=1]{$y_2$};
			\draw (5.45,0.3) node[yellow!45!black,scale=0.8,color=black]{Ir($B_1$)};
			\draw (-2.4,0.3) node[yellow!45!black,scale=0.8,color=black]{Ir($B_1$)};
			\draw (-3.6,0.3) node[yellow!45!black,scale=0.8,color=black]{Ir($A_1$)};
			\draw (-2.2,1.3) node[yellow!45!black,scale=0.8,color=black]{Ir($A_1$)};
			\draw (-1.2,2.7) node(IrA)[yellow!45!black,scale=0.8,color=black]{Ir($A$)};
			\draw (4,3) node[yellow!45!black,scale=0.8,color=black]{Ir($B$)};
			\draw[fill] (4.87,0.99)circle(0.05);
			\draw[fill] (-1.5,1.78)circle(0.05);
			\draw[fill] (-1.89,0.87)circle(0.05);
		\end{tikzpicture}
	} 
	\caption{The BPE for this configuration receives contribution from entanglement island of $A$ and $B$ and other island regions are described as $\text{Ir}(A_1)=[q_2,q_3]$ and $\text{Ir}(B_1)=[q_3,a]\cup [b,q_4]$. Here the $B_1$ has disconnected entanglement island region.}       
	\label{figD3a}
\end{figure}
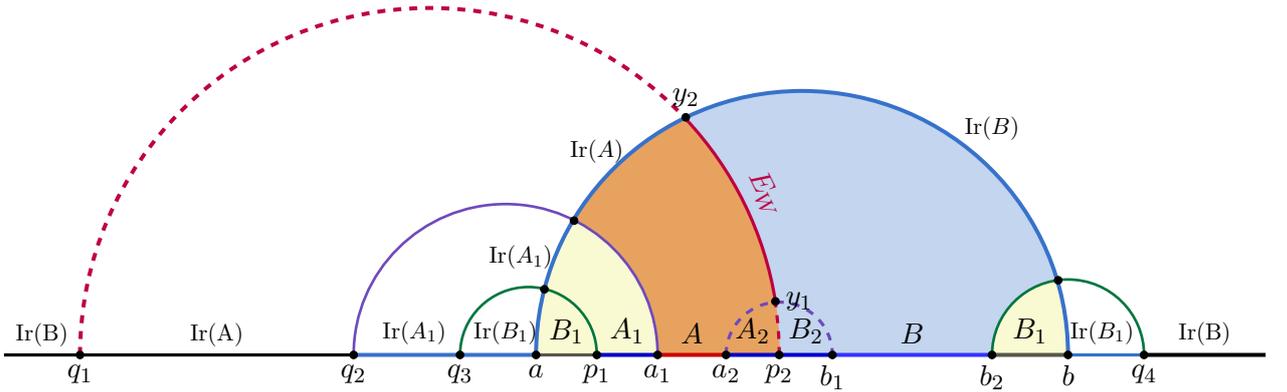

As depicted in \figref{figD3a}, the computation of the bulk EWCS involves endpoints $y_2$ and $y_1$ where $y_2$ is located on the KR brane. Here $y_1$ is situated on the RT surface homologous to the interval $[a_2,b_1]$. Note that the minimal length between the corresponding endpoints can be calculated using the following length formula in the $AdS_3$ geometry at a constant time slice as
\begin{equation}
\begin{aligned}\label{lengthD2a}
	\mathcal{L}&=\int_{y_1}^{y_2 }\dd{z}\frac{\sqrt{1+\frac{\dd{x}}{\dd{z}}^2}}{z}\, \\
	&=\frac12 \log\left[{\frac{(p_2-q_1)/2-\sqrt{((p_2-q_1)/2)^2-z^2}}{(p_2-q_1)/2+\sqrt{((p_2-q_1)/2)^2-z^2}}}\right]\Bigg|_{y_1}^{y_2}\,,
\end{aligned}
\end{equation}
where the endpoints of the EWCS $y_1$ and $y_2$ are given by 
\begin{equation}
	\begin{aligned}\label{endpointD2a}
		y_1&= \frac{\sqrt{a_2-q_1} \sqrt{a_2-p_2} \sqrt{q_1-b_1} \sqrt{b_1-p_2}}{q_1+p_2-a_2-b_1}\,,\\
		y_2&= \frac{\sqrt{a-q_1} \sqrt{a-p_2} \sqrt{q_1-b} \sqrt{b-p_2}}{q_1+p_2-a-b}\,.
	\end{aligned}
\end{equation}
Utilizing \eqref{endpointD2a} and \eqref{lengthD2a}, we may compute the EWCS for this phase as follows
\begin{equation}
\begin{aligned}\label{ewcsD2a}
E_W&=\frac{\mathcal{L}}{4 G}\\ 
	&=\frac{c}{12}\log\left[\frac{\left(a_2-q_1\right) \left(a-p_2\right) \left(q_1-b_1\right) \left(b-p_2\right)}{\left(a-q_1\right) \left(a_2-p_2\right) \left(q_1-b\right) \left(b_1-p_2\right)}\right]\,.
\end{aligned}
\end{equation}
where $q_1$ and $p_2$ are given by:
\begin{equation}
	\begin{aligned}\label{partitionD2a}
		q_1=&\frac{2 a_2 b_1-a_2 p_2-b_1 p_2}{a_2+b_1-2 p_2}\,,\\
		p_2=&\frac{a b-a_2 b_1+\sqrt{\left(a-a_2\right) \left(b-b_1\right) \left(a_2-b\right) \left(b_1-a\right)}}{a-a_2+b-b_1}\,.
	\end{aligned}
\end{equation}

We now computate the BPE in \figref{figD3a}. The balance requirements are given by, 
\begin{equation}\label{bcD2a}
	\begin{aligned}
		\mathcal{I}(A\text{Ir}(A),B\text{Ir}(B)B_1\text{Ir}(B_1)B_2)=& \mathcal{I}(B\text{Ir}(B),A\text{Ir}(A)A_1\text{Ir}(A_1)A_2) \,,
		\cr
		\mathcal{I}(A_1\text{Ir}(A_1),B\text{Ir}(B)B_1\text{Ir}(B_1)B_2)	=& \mathcal{I}(B_1\text{Ir}(B_1),A\text{Ir}(A)A_1\text{Ir}(A_1)A_2)\,.
	\end{aligned}
 \end{equation}
Above PEEs in the balance conditions can be computed by using generalized ALC proposal
as follows  
\begin{align}
	\mathcal{I}(A\text{Ir}(A),&B\text{Ir}(B)B_1\text{Ir}(B_1)B_2)\nonumber\\
 =&\frac12(\tilde{S}_{A\text{Ir}(A)A_1\text{Ir}(A_1)}+\tilde{S}_{A\text{Ir}(A)A_2}-\tilde{S}_{A_1\text{Ir}(A_1)}-\tilde{S}_{A_2})\nonumber\,\\
	=&\frac{1}{2}\left(\tilde{S}_{\left[p_1,a_2\right]\cup \left[q_1,q_3\right]}+\tilde{S}_{\left[a_1,p_2\right]\cup \left[q_1,q_2\right]}-\tilde{S}_{\left[p_1,a_1\right]\cup \left[q_2,q_3\right]}-\tilde{S}_{\left[a_2,p_2\right]}\right)\nonumber\,\\
	=&\frac{1}{2} \left(\tilde{S}_{\left[q_1, a_2\right]}+\tilde{S}_{\left[q_3,p_1\right]}+\tilde{S}_{\left[q_2,a_1\right]}+\tilde{S}_{\left[q_1,p_2\right]}-\tilde{S}_{\left[q_3,p_1\right]}-\tilde{S}_{\left[q_2,a_1\right]}-\tilde{S}_{\left[a_2,p_2\right]}\right)\nonumber\\
	=&\frac{c}{6}\log\left[\frac{ \left(a_2-q_1\right) \left(p_2-q_1\right)}{\epsilon \left(p_2-a_2\right) }\right]+\frac{c}{6}\phi(q_1)\,,\\
	\mathcal{I}(B\text{Ir}(B),&A\text{Ir}(A)A_1\text{Ir}(A_1)A_2)\nonumber\\
 =&\frac12(\tilde{S}_{B\text{Ir}(B)B_1\text{Ir}(B_1)}+\tilde{S}_{B\text{Ir}(B)B_2}-\tilde{S}_{B_1\text{Ir}(B_1)}-\tilde{S}_{B_2})\nonumber\\	
 =&\frac{1}{2}\left(\tilde{S}_{\left(-\infty,q_1 \right]\cup \left[q_3, p_1\right]\cup \left[b_1, \infty \right)} +\tilde{S}_{\left(-\infty,q_1 \right)\cup \left[p_2,b_2\right]\cup \left[q_4,\infty \right)}-\tilde{S}_{\left[b_2,q_4\right]\cup \left[q_3,p_1\right]}-\tilde{S}_{\left[p_2,b_1\right]}\right)\nonumber\\
	=&\frac{1}{2} \left(\tilde{S}_{\left[q_1,b_1\right]}+\tilde{S}_{\left[q_3,p_1\right]}+\tilde{S}_{\left[b_2,q_4\right]}+\tilde{S}_{\left[q_1,p_2\right]}-\tilde{S}_{\left[q_3,p_1\right]}-\tilde{S}_{\left[b_2,q_4\right]}-\tilde{S}_{\left[p_2,b_1\right]}\right)\nonumber\\
	=&\frac{c}{6}\log\left[\frac{\left(b_1-q_1\right) \left(p_2-q_1\right)}{\epsilon (b_1-p_2)}\right]+\frac{c}{6}\phi(q_1)\,,
  \end{align}
 and
 \begin{align}
	\mathcal{I}(A_1\text{Ir}(A_1),&B\text{Ir}(B)B_1\text{Ir}(B_1)B_2)\nonumber\\
 =&\frac12(\tilde{S}_{A\text{Ir}(A)A_1\text{Ir}(A_1)A_2}+\tilde{S}_{A_1\text{Ir}(A_1)}-\tilde{S}_{A\text{Ir}(A)A_2})\nonumber\\
 =&\frac{1}{2}\left(\tilde{S}_{\left[p_2,p_1\right]\cup \left[q_1,q_3\right]}+\tilde{S}_{\left[p_1,a_1\right]\cup \left[q_2,q_3\right]}-\tilde{S}_{\left[a_1,p_2\right]\cup \left[q_1q_2\right]}\right)\nonumber\\
	=&\frac{1}{2} \left(\tilde{S}_{\left[q_1,p_2\right]}+\tilde{S}_{\left[q_3,p_1\right]}+\tilde{S}_{\left[q_2,a_1\right]}+\tilde{S}_{\left[q_3,p_1\right]}-\tilde{S}_{\left[q_2,a_1\right]}-\tilde{S}_{\left[q_1,p_2\right]}\right)\nonumber\\
 =&\frac{c}{6}\log\left[\frac{(p_1-q_3)^2}{\epsilon^2}\right]+\frac{c}{6}\phi(q_3)\,,\\
	\mathcal{I}(B_1\text{Ir}(B_1),&A\text{Ir}(A)A_1\text{Ir}(A_1)A_2)\nonumber\\
 =&\frac12(\tilde{S}_{B\text{Ir}(B)B_1\text{Ir}(B_1)B_2}+\tilde{S}_{B_1\text{Ir}(B_1)}-\tilde{S}_{B\text{Ir}(B)B_2})\nonumber\\
  =&\frac{1}{2}\left(\tilde{S}_{\left(-\infty,q_1 \right]\cup \left[q_3,p_1\right]\cup [p_2,\infty)}+\tilde{S}_{\left[b_2,q_4\right]\cup \left[q_3,p_1\right]}-\tilde{S}_{\left(-\infty,q_1 \right]\cup \left[p_2,b_2\right]\cup \left[ ,q_4,\infty\right)}\right)\nonumber\\
  =&\frac{1}{2}\left(\tilde{S}_{\left[q_1,p_2\right]}+\tilde{S}_{\left[q_3,p_1\right]}+\tilde{S}_{\left[b_2,q_4\right]}+\tilde{S}_{\left[q_3,p_1\right]}-\tilde{S}_{\left[b_2,q_4\right]}-\tilde{S}_{\left[q_1,p_2\right]}\right)\nonumber\\
 =&\frac{c}{6}\log\left[\frac{(p_1-q_3)^2}{\epsilon^2}\right]+\frac{c}{6}\phi(q_3)\,.
\end{align} 
As we can observe from the second balance constraint from above which are trivially satisfied. We may obtain $q_1$
from the first balance condition and it reduces in terms of the partition point $p_2$. Therefore, we need to minimize the BPE for this phase over the partition point $p_2$ and it provides following value of $q_1$ and $p_2$, which are the same as \eqref{partitionD2a}.
The BPE for this phase is given by
\begin{align}\label{BPED2a}
	\text{BPE}&=	\mathcal{I}(A\text{Ir}(A),B\text{Ir}(B)B_1\text{Ir}(B_1)B_2)|_{balanced}\\
	&=\frac{c}{6}\log\left[\frac{(a-b) \left(a_2-b_1\right){} \left(a_2-p_2\right){} \left(b_1-p_2\right){}}{\left(p_2 \left(a_2+b_1\right)+a \left(a_2+b_1-2 p_2\right)-2 a_2 b_1\right){}^2 \left(a_2 \left(b-2 b_1+p_2\right)+b_1 p_2+b \left(b_1-2 p_2\right)\right){}}\right]\,.
\end{align}
The above result of BPE exactly coincide with the EWCS in \eqref{ewcsD2a} by putting the value of $q_1$.
	
	\subsubsection*{Phase-D2c}
\begin{figure}[H]
	\begin{center}
		\begin{tikzpicture}
			\filldraw[LouisBlue!30](-2,0) arc (180:0:3.5);
			\begin{scope} 
				\clip (-2,0) arc (180:0:3.5);
				\filldraw[orange,opacity=0.6] (-9,0)--(1.2,0) arc(0:43:4.6);
			\end{scope}
			\begin{scope}
				\clip (-2,0) arc (180:0:3.5);
				\filldraw[yellowr,opacity=1](-0.4,0) arc(0:180:2);
			\end{scope}
			\begin{scope}
				\clip (-2,0) arc (180:0:3.5);
				\filldraw[yellowr,opacity=1](4,0) arc(180:0:1);
			\end{scope}
			\draw[color=red!80!black,line width=1.5pt] (-0.4,0)--(0.5,0)node[midway,above,scale=1]{ \textcolor{black}{$A$} };
			\draw[color=yellow!20!black,very thick] (-2,0)--(-1.2,0)node[midway,above,scale=1]{ \textcolor{black}{$B_1$} };
			\draw[color=yellow!20!black,line width=1.5pt] (4,0)--(5,0)node[midway,above,scale=1]{ \textcolor{black}{$B_1$} };		
			\draw[color=blue!80!black,line width=1.5pt] (-1.2,0)--(-0.4,0)node[midway,above,scale=1]{ \textcolor{black}{$A_1$} };
			\draw[color=blue!80!black,line width=1.5pt] (0.5,0)--(1.2,0)node[midway,above,scale=1]{ \textcolor{black}{$A_2$} };
			\draw[color=blue!80!black,line width=1.5pt] (1.2,0)--(3.2,0)node[midway,above,scale=1]{ \textcolor{black}{$B_2$} };
			\draw[color=blue!80,line width=1.5pt] (3.2,0)--(4,0)node[midway,above,scale=1]{$ \textcolor{black}{B} $};
			\draw[very thick] (-8,0)--(-4.4,0)  (-5.5,0)node[above,scale=0.8]{ \textcolor{black}{Ir(A)}};
			\draw[very thick] (-9,0)--(-8,0)node[midway,above,scale=0.8]{ \textcolor{black}{Ir($B_2$)}};
			\draw[line width=1.5pt,LouisBlue] (-4.4,0)--(-2,0)node[midway,below]{$ \textcolor{black}{} $};		    
			\draw[line width=1.5pt] (6.0,0)--(7,0)node[midway,above,scale=0.8]{ \textcolor{black}{Ir(B)} };	
			\draw[line width=1.5pt] (7,0)--(8,0)node[midway,above,scale=0.8]{ \textcolor{black}{Ir($B_2$)} };
			\draw[very thick,LouisBlue] (5,0)--(6,0)node[midway,above]{$ \textcolor{black}{} $};	
			\draw[color=LouisColor1,line width=1pt] (4,0) arc(180:0:1) node[midway,above,sloped]{$\text{}$};
			\draw[color=LouisColor1,line width=1pt] (-1.2,0) arc(0:180:0.9) node[midway,above,sloped]{$\text{}$};
			\draw[color=LouisColor2,line width=1pt] (-0.4,0) arc(0:180:2) node[midway,above,sloped]{$\text{}$};
			\draw[color=purple,line width=1.2pt] (1.2,0) arc(0:43:4.6);
			\draw[color=purple,dashed,line width=1.5pt] (-8,0) arc(180:43:4.6);
			\draw[color=LouisColor2,line width=1pt] (0.5,0) arc(0:180:3.5) node[midway,above,sloped]{$\text{}$};
			\draw[color=LouisColor2,line width=1pt] (3.2,0) arc(180:0:1.95);
			\draw[color=LouisBlue,line width=1.5pt] (5,0) arc(0:180:3.5) node[midway,above,sloped,scale=0.8,color=black]{\text{}};
			\draw[fill](-0.4,0)circle(0.05) (-0.4,0) node[below,scale=1]{$a_1$};
			\draw[fill](0.5,0)circle(0.05) (0.5,0) node[below,scale=1]{$a_2$};
			\draw[fill](1.2,0)circle(0.05) (1.2,0) node[below,scale=1]{$p_2$};
			\draw[fill](3.2,0)circle(0.05) (3.2,0) node[below,scale=1]{$b_1$};
			\draw[fill](4,0)circle(0.05) (4,0) node[below,scale=1]{$b_2$};
			\draw[fill](5,0)circle(0.05) (5,0) node[below,scale=1]{$b$};
			\draw[fill](-1.2,0)circle(0.05) (-1.2,0) node[below,scale=1]{$p_1$};
			\draw[fill](-2,0)circle(0.05) (-2,0) node[below,scale=1]{$a$};
			\draw[fill](-4.4,0)circle(0.05) (-4.4,0) node[below,scale=1]{$q_3$};
			\draw[fill](6,0)circle(0.05) (6,0) node[below,scale=1]{$q_5$};
			\draw[fill](-3,0)circle(0.05) (-3,0) node[below,scale=1]{$q_4$};
			\draw[fill] (-8,0)circle(0.05) (-8,0) node[below,scale=1]{$q_1$};
			\draw[fill] (7.1,0)circle(0.05) (7.1,0) node[below,scale=1]{$q_6$};
			\draw[fill] (-6.5,0)circle(0.05) (-6.5,0) node[below,scale=1]{$q_2$};
			\draw (5.45,0.3) node[yellow!45!black,scale=0.8,color=black]{Ir($B_1$)};
			\draw (-2.4,0.3) node[yellow!45!black,scale=0.8,color=black]{Ir($B_1$)};
			\draw (-3.6,0.3) node[yellow!45!black,scale=0.8,color=black]{Ir($A_1$)};
			\draw (-2.2,1.3) node[yellow!45!black,scale=0.8,color=black]{Ir($A_1$)};
			\draw (-1.5,2.4) node(IrA)[yellow!45!black,scale=0.8,color=black]{Ir($A$)};
			\draw (-0.7,3.1) node[yellow!45!black,scale=0.8,color=black]{Ir($A_2$)};
			\draw (5.1,1.4) node[yellow!45!black,scale=0.8,color=black]{Ir($B$)};
			\draw (3.23,3.45) node[yellow!45!black,scale=0.8,color=black]{Ir($B_2$)};
			\draw (-7.23,0.25) node[yellow!45!black,scale=0.8,color=black]{Ir($A_2$)};
			\draw[fill] (4.87,0.99)circle(0.05);
			\draw[fill] (-0.03,3.15)circle(0.05);
			\draw[fill] (-0.75,2.67)circle(0.05);
			\draw[fill] (-1.5,1.78)circle(0.05);
			\draw[fill] (-1.89,0.87)circle(0.05);
			\draw[fill] (4.5,1.83)circle(0.05);
		\end{tikzpicture}
	\end{center}   
	\caption{The figure illustrates entanglement islands regions for all the intervals located in Weyl $CFT_2$ indicated as $\text{Ir}(A)=[q_2,q_3]$,  $\text{Ir}(B)=[q_5,q_6]$, $\text{Ir}(A_1)=[q_3,q_4]$,  $\text{Ir}(B_1)=[q_4,a]\cup[b,q_5]$, $\text{Ir}(A_2)=[q_1,q_2]$ and $\text{Ir}(B_2)=(-\infty,q_1]\cup[q_6,\infty)$.}
\label{figD3c}
\end{figure}
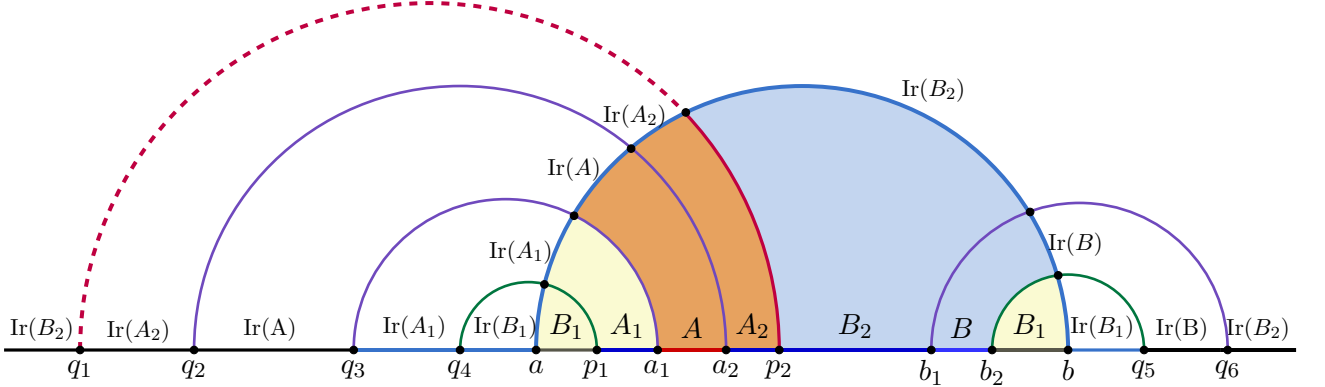

As displayed in \figref{figD3c}, the intervals $A_2$ and $B_2$ also admit entanglement islands. In this context, the entanglement wedge in this phase become disconnected implying the EWCS to be vanished, which indicates a zero BPE. In this phase the balance requirements are described as follows
\begin{equation}\label{bcD3c}
	\begin{aligned}
		\mathcal{I}(A\text{Ir}(A),B\text{Ir}(B)B_1\text{Ir}(B_1)B_2\text{Ir}(B_2))=& \mathcal{I}(B\text{Ir}(B),A_1\text{Ir}(A_1)A_2\text{Ir}(A_2)A\text{Ir}(A)) \,,
		\cr
		\mathcal{I}(A_1\text{Ir}(A_1),B\text{Ir}(B)B_1\text{Ir}(B_1)B_2\text{Ir}(B_2))	=& \mathcal{I}(B_1\text{Ir}(B_1),A_1\text{Ir}(A_1)A_2\text{Ir}(A_2)A\text{Ir}(A))\,.
	\end{aligned}
\end{equation}
Where

\begin{align}
	\mathcal{I}(A\text{Ir}(A)&,B\text{Ir}(B)B_1\text{Ir}(B_1)B_2\text{Ir}(B_2))\nonumber\\
	&=\frac12(\tilde{S}_{A\text{Ir}(A)A_1\text{Ir}(A_1)}+\tilde{S}_{A\text{Ir}(A)A_2\text{Ir}(A_2)}-\tilde{S}_{A_1\text{Ir}(A_1)}-\tilde{S}_{A_2\text{Ir}(A_2)})\nonumber\\
	&=\frac{1}{2}\left(\tilde{S}_{\left[q_2,q_4\right]\cup\left[p_1,a_2\right]}+\tilde{S}_{\left[q_1,q_3\right]\cup\left[a_1,p_2\right]}-\tilde{S}_{\left[q_3,q_4\right]\cup\left[p_1,a_1\right]}-\tilde{S}_{\left[q_1,q_2\right]\cup\left[p_2,a_2\right]}\right)\nonumber\\
	&=\frac{1}{2}\left(\tilde{S}_{\left[q_4,p_1\right]}+\tilde{S}_{\left[q_2,a_2\right]}+\tilde{S}_{\left[q_3,a_1\right]}+\tilde{S}_{\left[q_1,p_2\right]}-\tilde{S}_{\left[q_4,p_1\right]}-\tilde{S}_{\left[q_3,a_1\right]}-\tilde{S}_{\left[q_2,a_2\right]}-\tilde{S}_{\left[q_1,p_2\right]}\right)\nonumber\\
	&=0\,,\\
	\mathcal{I}(B\text{Ir}(B)&,A_1\text{Ir}(A_1)A_2\text{Ir}(A_2)A\text{Ir}(A))\nonumber\\
	&=\frac12(\tilde{S}_{B\text{Ir}(B)B_1\text{Ir}(B_1)}+\tilde{S}_{B\text{Ir}(B)B_2\text{Ir}(B_2)}-\tilde{S}_{B_1\text{Ir}(B_1)}-\tilde{S}_{B_2\text{Ir}(B_2)})\nonumber\\
	&=\frac{1}{2}\left(\tilde{S}_{\left[q_4,p_1\right]\cup\left[b_1,q_6\right]}+\tilde{S}_{\left(-\infty,q_1\right]\cup \left[p_2,b_2\right]\cup\left[q_5,\infty\right) }-\tilde{S}_{\left[b_2,q_5\right]\cup \left[q_4,p_1\right]}-\tilde{S}_{\left(-\infty,q_1\right]\cup\left[p_2,b_1\right]\cup\left(q_6,\infty\right] }\right)\nonumber\\
&=\frac{1}{2}\left(\tilde{S}_{\left[b_1q_6\right]}+\tilde{S}_{\left[q_4,p_1\right]}+\tilde{S}_{\left[b_2,q_5\right]}+\tilde{S}_{\left[q_1,p_2\right]}-\tilde{S}_{\left[b_2,q_5\right]}-\tilde{S}_{\left[q_4,p_1\right]}-\tilde{S}_{\left[b_1,q_6\right]}-\tilde{S}_{\left[q_1,p_2\right]}\right)\nonumber\\
	&=0\,.
\end{align} 
It is obvious that the balanced requirements \eqref{bcD3c} are satisfied and the BPE for this phase is just zero.

\bibliographystyle{JHEP}
\bibliography{Citation}




\end{document}